\documentclass[12pt,english,journal=jcpafh]{achemso}
\pdfoutput=1
\usepackage[T1]{fontenc}
\usepackage[latin9]{inputenc}
\usepackage{geometry}
\usepackage{color}
\usepackage{amsbsy}
\usepackage{amstext}
\usepackage{graphicx}
\usepackage{subscript}

\makeatletter

\title{Systematic First Principles Configuration-Interaction Calculations
of Linear Optical Absorption Spectra in Silicon Hydrides : Si\textsubscript{$2$}H\textsubscript{$2n$}
($n=1-3$)}

\author{Pritam Bhattacharyya}

\affiliation{Department of Physics, Indian Institute of Technology Bombay, Powai,
Mumbai 400076, India}

\email{pritambhattacharyya01@gmail.com}

\author{Deepak Kumar Rai}

\affiliation{Department of Physics, Indian Institute of Technology Bombay, Powai,
Mumbai 400076, India}

\email{dkriitb@gmail.com}

\author{Alok Shukla}

\affiliation{Department of Physics, Indian Institute of Technology Bombay, Powai,
Mumbai 400076, India}

\email{shukla@phy.iitb.ac.in}

\providecommand{\tabularnewline}{\\}

\@ifundefined{showcaptionsetup}{}{%
 \PassOptionsToPackage{caption=false}{subfig}}
\usepackage{subfig}
\makeatother

\usepackage{babel}
\begin{document}
\begin{abstract}
{\normalsize{}We have performed first principles electron-correlated
calculations employing large basis sets to optimize the geometries,
and to compute linear optical absorption spectra of various low-lying
conformers of silicon hydrides: Si$_{2}$H$_{2n}$, $n=1,2,3$. The
geometry optimization for various isomers was carried out at the coupled-cluster
singles-doubles-perturbative-triples {[}CCSD(T){]} level of theory,
while their excited states and absorption spectra were computed using
a large-scale multi-reference singles-doubles configuration-interaction
(MRSDCI) approach, which includes electron-correlation effects at
a sophisticated level. Our calculated spectra are the first ones for
Si$_{2}$H$_{2}$ and Si$_{2}$H$_{4}$ conformers, while for Si$_{2}$H$_{6}$
we obtain excellent agreement with the experimental measurements,
suggesting that our computational approach is reliable. Our calculated
absorption spectra exhibit a strong structure-property relationship,
suggesting the possibility of identifying various conformers based
on their optical absorption fingerprints. Furthermore, we have also
performed geometry optimization for the selected optically excited
states, providing us insights into their character.}{\normalsize\par}
\end{abstract}

\section{Introduction}

Silicon is not only one of the most abundant elements on earth, it
is also technologically one of the most important ones, with almost
the entire semiconductor industry based upon it. Being in the same
group of the periodic table as carbon, it shares many chemical characteristics
with it. However, unlike carbon, it is not known to participate in
$sp^{2}$ hybridization, as a result of which it has no planar graphite-like
allotrope. However, there are indications that a quasi-planar allotrope
of silicon named silicene can be isolated on substrates.\cite{Silicene1,Silicene2,Silicene3,Silicene4}
Silicon and its compounds have fascinated physicists and chemists
alike, over the years, both from a fundamental as well as from the
applied points of view.\cite{Review_1,Review_2,Review_3,Review_4}
There has always been a lot of interest in the field of hydrides of
various substances, and silicon is no exception.\cite{sih_1,sih_2,sih_3,sih_4,sih_5}
The study of hydrides of silicon is important in many ways: (a) during
the formation of silicon thin films using plasma-enhanced chemical
vapor deposition, various types of hydrides of silicon are produced,
which needs to be understood,\cite{doi:10.1021/j100205a046,si-cvd,doi:10.1021/jp015523n}
(b) degradation of silicon based electronic devices happens normally
through hydrogenation of silicon caused by atmospheric moisture,\cite{hydrogenated-a-si-solar-cells-review}
and (c) hydrogenated amorphous silicon is used extensively in thin
film solar cells.\cite{hydrogenated-a-si-solar-cells-review} Functionalization
of silicene films grown on metal substrates by hydrogenation, with
the aim of device applications, is an active area of research these
days.\cite{silicene-hydrogenation1,silicene-hydrogenation2} Furthermore,
the process of hydrogenation in silicon is fascinating from a fundamental
point of view, given that the study of hydrogenated carbon in form
of hydrocarbons is such a mature field. 

Unsaturated silicon hydrides such as SiH, SiH\textsubscript{2}, and
SiH$_{3}$ etc. have been the subject of a number of experimental
investigations because of their importance in astrophysics, due to
their presence in space.\cite{sih_expt_1,sih_expt_2,Wang} Silicon
hydrides of the form Si\textsubscript{2}H\textsubscript{2n}, n=1-3
have also been extensively studied both experimentally,\cite{:/content/aip/journal/jcp/95/4/10.1063/1.460947,PhysRevLett.66.413,Destombes,Thaddeus,si2h2-opt-absorp-exp-2016}
and by means of high-level theory,\cite{Chaeho_Pak,doi:10.1021/j100157a052,Alexander_Sax,soma-si2h4-1986,POIRIER-si2h4-1981,doi:10.1021/j100377a036,LISCHKA1982467,doi:10.1021/ja00386a006,doi:10.1021/ja00360a016,Destombes,Roger_and_Henry,Sannigrahi_Nandi,Jursic,Gottfried_Olbrich,Thaddeus,Monobridged_Si2H4}
because of their ability to form mono- and dibridged hydrides, as
well as multiple bonds of silicon. The carbon analog of Si$_{2}$H$_{2}$
is acetylene, which is a triply-bonded system, while that of Si$_{2}$H$_{4}$
is ethylene, which contains a carbon-carbon double bond. Therefore,
the question arises do Si$_{2}$H$_{4}$ and Si\textsubscript{2}H\textsubscript{2}
similarly have silicon-silicon double, and triple bonds, respectively?
And, what similarities, if any, does Si$_{2}$H$_{6}$ (disilane)
have with C$_{2}$H$_{6}$ (ethane)?

Measurements on Si\textsubscript{2}H\textsubscript{2} have been
performed by several groups.\cite{PhysRevLett.66.413,:/content/aip/journal/jcp/95/4/10.1063/1.460947,si2h2-monobridged-exp,Destombes,si2h2-opt-absorp-exp-2016}
Bogey \emph{et al.}\cite{PhysRevLett.66.413,si2h2-monobridged-exp,Destombes}
performed millimeter- and submillimeter-wave spectroscopy measurements
on Si\textsubscript{2}H\textsubscript{2} produced in silane-argon
plasma. Ruscic and Berkowitz\cite{:/content/aip/journal/jcp/95/4/10.1063/1.460947}
produced several Si$_{2}$H$_{n}$ hydrides, including Si\textsubscript{2}H\textsubscript{2},
by reacting fluorine with Si$_{2}$H$_{6}$, and reported photoionization
mass spectrometric measurements on it. Recently, Mohapatra \emph{et
al.}\cite{si2h2-opt-absorp-exp-2016} synthesized carbene stabilized
Si\textsubscript{2}H\textsubscript{2}, and reported measurements
of its UV-Vis spectrum. Several first-principles quantum chemical
studies of the structure and bonding in Si$_{2}$H$_{2}$ have been
performed by several groups. Lischka and K\"ohler\cite{doi:10.1021/ja00360a016}
studied this molecule at the SCF and CEPA-2 level, and were the first
ones to predict that in the ground state, the molecule has a \emph{dibridged}
structure (Si(H$_{2}$)Si) with $C_{2v}$ symmetry. This prediction
was also verified in the experiments of Bogey \emph{et al}.\cite{PhysRevLett.66.413,Destombes}\textcolor{red}{{}
}People also believed that the second most stable isomer of Si$_{2}$H$_{2}$
is disilavinylidene (H$_{2}$SiSI) having $C_{2v}$ symmetry, while
the \emph{trans}-bent structure (HSiSiH) with $C_{2h}$ symmetry being
the only other minimum.\cite{Roger_and_Henry} Later on, the structure
of Si$_{2}$H$_{2}$ was re-investigated by Colegrove and Schaefer\textcolor{black}{,\cite{doi:10.1021/j100377a036}
and by Grev and Schaefer\cite{Roger_and_Henry} and they found another
higher minimum, corresponding to a monobridged structure (Si(H)SiH)
with }$C_{s}$ symmetry, confirmed in the experiment of Cordonnier
\emph{et al}.\cite{si2h2-monobridged-exp} Sax and Kalcher, using
a first-principles pseudopotential based multi-configuration-self-consistent
field (MCSCF) approach, computed the enthalpies of formation of a
number of hydrides of silicon, including various isomers of Si\textsubscript{2}H\textsubscript{2}.
\cite{doi:10.1021/j100157a052} Sannigrahi and Nandi studied the bonding
properties of various isomers of Si\textsubscript{2}H\textsubscript{2}
using an \emph{ab initio} self-consistent-field (SCF) approach.\cite{Sannigrahi_Nandi}
Jursic performed an extensive first principles study of the potential
energy surface of Si\textsubscript{2}H\textsubscript{2} using density-functional
theory (DFT), as well as by employing second-order Møller Plesset
perturbation theory combined with Gaussian-1/Gaussian-2/Gaussian-3
approaches. More recently, Schaefer and coworkers\cite{Chaeho_Pak}
computed the electron affinities several hydrides of silicon, including\textcolor{red}{{}
}Si\textsubscript{2}H\textsubscript{2}.

The subject of doubly bonded silicon compounds silenes was reviewed
by Raabe and Michl.\cite{Michl_2} In the year 1981, West, Fink and
Michl\cite{Michl} experimentally stabilized the silicon-silicon doubly
bonded compound tetramesityldisilene Si\textsubscript{2}R\textsubscript{4},
by using the bulky 2,4,6- trimethylphenyl (mesityl) group, denoted
here as R. These authors, also measured its UV-Vis absorption spectrum.\cite{Michl}
However, the first synthesis of doubly-bonded molecule disilene (Si$_{2}$H$_{4}$),
was reported by Ruscic and Berkowitz\cite{:/content/aip/journal/jcp/95/4/10.1063/1.460947}
in 1991, who also reported the measurements of its ionization potential.
Using infrared spectroscopy, supported by first principles DFT calculations,
Andrews and Wang concluded that Si$_{2}$H\textsubscript{4} has a
$C_{2h}$ structure.\cite{Wang} Sari \emph{et al}.\cite{Thaddeus},
based upon rotational spectrum measurements of Si\textsubscript{2}H\textsubscript{4}
by means of Fourier Transform Microwave (FTM) spectroscopy, supported
by sophisticated coupled-cluster calculations, concluded that the
molecule has a monobridged structure. Later on, the same group presented
another experimental and theoretical study on Si\textsubscript{2}H\textsubscript{4}
molecule, and concluded that indeed the monobridged isomer was most
abundant. However, several unidentified spectral lines in the experiment
could imply the presence of a dibridged isomer as well.\cite{Monobridged_Si2H4}
As far as theoreticians are concerned, there appears to be a general
agreement that the trans-bent structure of disilene corresponds to
the true ground state, while the monobridged structure is believed
to be the next higher energy isomer.\cite{Chaeho_Pak,Pople,Trinquier,Gottfried_Olbrich,Alexander_Sax,LISCHKA1982467,doi:10.1021/ja00386a006,Wang,doi:10.1021/j100206a004,Dolgonos,Monobridged_Si2H4,Thaddeus}
Silysilylene isomer, with the structure H$_{3}$SiSiH, has been computed
to be the third higher energy structure.\cite{Monobridged_Si2H4,Thaddeus,Dolgonos,Alexander_Sax,doi:10.1021/j100206a004,LISCHKA1982467,doi:10.1021/ja00386a006}

Disilane (Si\textsubscript{2}H\textsubscript{6}), which is the structural
analog of ethane (C$_{2}$H$_{6}$), is a stable compound existing
in gas phase at room temperature, and has been known for a long time.
As a result, a number of experimental and theoretical studies have
been performed on it.  The infrared spectrum of gas phase disilane
was observed by Gutowsky and Stejskal from 350 to 4000 cm\textsuperscript{-1}\cite{Gutowsky},
as well as by Andrews and Wang by reacting laser-ablated silicon atoms
with molecular hydrogen to form the silicon hydrides\cite{Wang}.
Itoh \emph{et al. }experimentally studied the vacuum ultraviolet absorption
cross sections of disilane.\cite{Si2H6_optical_expt} Several authors
experimentally studied the photo-electron spectra of silanes (Si\textsubscript{n}H\textsubscript{2n+2}),
and measured their ionization potentials.\cite{Photoelectron_Si2H6_1,Photoelectron_Si2H6_2}
As far as theoretical calculations on disilane are concerned, in 1976,
geometry optimization was performed by Blustin\cite{Si2H6_GS_2},
and Pople and coworkers\cite{Si2H6_GS_5}. In the same year, the valence
electronic structure and internal rotation barrier of the molecule
was computed by Nicolas, Barthelat, and Durand using a pseudo-potential
method.\cite{Si2H6_GS_6} In 1981, Ratner and coworkers performed
electronic structure calculations employing a Hartree-Fock-Slater
(HFS) procedure, based upon the local-density functional approach.\cite{Si2H6_GS_3}
Photolytic fragmentation was studied by Janoschek and coworkers using
a pseudo-potential method.\cite{Si2H6_GS_4} In 1986, the geometries
and the single point energies of many singly bonded silicon compounds,
including disilane, were computed by Schleyer and coworkers using
HF and MP4 level of theory.\cite{Si2H6_GS_1} Sax performed local
pseudopotential calculations to optimize the ground state geometry
of disilane.\cite{Alexander_Sax} Using \emph{ab initio} propagator
theory, ionization potentials of various silicon hydrides, including
Si\textsubscript{2}H\textsubscript{6}, were theoretically computed
by Ortiz and Mintmire.\cite{Ortiz_Mintmire} Using a first-principles
configuration interaction (CI) approach, Kawai \emph{et al.}\cite{Kawai}
computed the ultraviolet photoabsorption spectrum of disilane. Rohlfing
and Louie computed the optical absorption spectrum of disilane by
solving the Bethe-Salpeter equation, within\emph{ }a first-principles
formalism based upon density-functional theory (DFT), with quasi-particle
effects included using the GW approximation.\cite{Si2H6_optical_theory}

In this work we undertake a comprehensive study of structural stability
and optical properties of three hydrides of silicon dimer, namely,
Si\textsubscript{2}H\textsubscript{2}, Si\textsubscript{2}H\textsubscript{4}
and Si\textsubscript{2}H\textsubscript{6} using state-of-the-art
correlated-electron first principles electronic structure methodology.
Geometry optimization for all the molecules considered was carried
out using the coupled-cluster singles-doubles-triples (CCSD(T)) level
of theory, using large basis sets including polarization functions.
The optical absorption spectra of various clusters were computed using
the multi-reference singles-doubles configuration interaction (MRSDCI)
approach, which has been used in our group to study the optical properties
of a variety of systems such as atomic clusters,\cite{epjd-pradip,epjd-shinde-mg,Shinde_nano_life,Shinde_PCCP}
conjugated polymers,\cite{Aryanpour_Shukla,PhysRevB.65.125204Shukla65,PhysRevB.69.165218Shukla69,Himanshu,himanshu-triplet,Priya_Sony}
and graphene quantum dots.\cite{dkr1,Tista1} We would like to emphasize
that the first-principles electronic structure studies of Si based
systems are more computationally expensive as compared to similar
studies of clusters made up of smaller atoms, simply because Si has
more electrons, thereby requiring larger basis sets, and hence more
computer memory and time. For Si\textsubscript{2}H\textsubscript{2}
and Si\textsubscript{2}H\textsubscript{4}, the linear optical absorption
spectra were computed for a number of isomers, with the aim of understanding
the influence of geometry on the optical properties of these molecules.
The relation between absorption spectra and geometry can be used for
the detection and identification of these molecules in optical experiments.
To achieve better understanding of the nature of optically excited
states, we have also performed geometry optimization on a few selected
excited states. Our calculations reveal that higher energy excited
states of various clusters have significantly different relaxed geometries
as compared to the ground state.

So far, optical absorption experiments on Si\textsubscript{2}H\textsubscript{2}
and Si\textsubscript{2}H\textsubscript{4} have not been performed.
Therefore, our theoretical calculations of their absorption spectra
will be useful in guiding future experimental efforts on these systems.
For the case of disilane, our calculated optical absorption spectrum
is in very good agreement with the experimental measurements of Itoh
\emph{et al.}\cite{Si2H6_optical_expt} as well as with the Bethe-Salpeter
equation based theoretical calculations of Rohlfing and Louie\cite{Si2H6_optical_theory}.
This excellent agreement of our calculations with the experiments
for disilane gives us confidence that our calculations on Si\textsubscript{2}H\textsubscript{2}
and Si\textsubscript{2}H\textsubscript{4} should be equally accurate. 

\section{Theory And Computational Details}

\label{sec:theory}

\subsection{General Methodology}

All the calculations were carried out using a wave-function-based
first-principles methodology, employing the molecular Born-Oppenheimer
Hamiltonian, in which the orbitals of molecules are expressed as linear
combinations of Cartesian-Gaussian-type basis functions. Such a quantum-chemical
electronic structure approach has been implemented in a number of
program packages, and in this work we used the packages PSI4\cite{PSI4}
and MELD.\cite{MELD} The geometries of the silicon hydride (Si\textsubscript{2}H\textsubscript{2n},
n = 1-3) molecules studied in this work were optimized using the coupled-cluster
singles-doubles-perturbative-triples (CCSD(T)) approach, as implemented
in the program package PSI4,\cite{PSI4} utilizing the correlation-consistent
polarized valence-triple-zeta (cc-pVTZ) basis sets. Once the ground
state geometries were determined for each isomer, calculations of
their optical absorption spectra were performed. For the purpose,
calculations of the excited states of the molecules was performed
using the multireference singles-doubles configuration-interaction
(MRSDCI) approach, as implemented in the MELD package.\cite{MELD}
To perform the MRSDCI calculations, first we transform the Hamiltonian
from the atomic orbital (AO) representation consisting of Cartesian
Gaussian basis functions, to the molecular orbital (MO) representation.
This is achieved by obtaining the MOs of the concerned isomer by performing
restricted Hartree-Fock (RHF) calculations on its optimized geometry,
and then transforming the one- and two-electron integrals from the
AO to the MO representation. Next, using the transformed Hamiltonian,
a singles-doubles CI (SDCI) calculation is performed using an appropriate
single reference wave function, both for the ground state, and the
excited states. The many-body wave functions obtained from the SDCI
calculation are used for computing its optical absorption spectrum.
Next, the wave functions contributing to various peaks are analyzed,
and the configuration state functions (CSFs) making significant contributions
to the corresponding excited states are included as references for
the next set of MRSDCI calculation. This procedure is iterated until
the calculated optical absorption spectrum of the isomer converges.
At every stage of the calculation, point-group and spin symmetries
are fully utilized, thus our ground and excited state wave functions
are also eigenfunctions of the corresponding point-group projection
operator, and total-spin operator. For details of the MRSDCI approach
adopted in this work, we refer the reader to our earlier works.\cite{Aryanpour_Shukla,dkr1,epjd-pradip,epjd-shinde-mg,Himanshu,himanshu-triplet,PhysRevB.65.125204Shukla65,PhysRevB.69.165218Shukla69,Priya_Sony,Shinde_PCCP,Shinde_nano_life,Tista1}

The optical absorption spectrum $\sigma(\omega)$, is calculated within
the electric-dipole approximation, using the formula 

\begin{equation}
\sigma(\omega)=4\pi\alpha\sum_{i}\frac{\omega_{i0}|\langle i|\boldsymbol{\hat{e}.r}|0\rangle|^{2}\gamma^{2}}{(\omega_{i0}-\omega)^{2}+\gamma^{2}},\label{eq:sigma}
\end{equation}

where $\omega$, $\boldsymbol{\hat{e}}$, $\boldsymbol{r}$, and $\alpha$,
respectively denote the frequency of the incident light, polarization
direction of the incident light, the position operator, and the fine
structure constant. Furthermore, $\omega_{i0}$ is the energy difference
(in frequency units) between the ground state ($0$) and the $i$\textsuperscript{th}
excited state, while $\gamma$ is the uniform lined width associated
with each excited state energy level. The summation over $i$ in Eq.
\ref{eq:sigma} includes an infinite number of excited states, however,
in our calculations we restricted the sum to those excited states
which are no more than 10 eV higher than the ground state.

The absorption spectrum calculated using the formalism described above
corresponds to vertical excitations, i.e., those optical excitations
during which the geometry of the system has no time to relax. However,
eventually the system undergoes relaxation, and the molecular geometry
of the excited state changes as compared to that of the ground state.
In order to understand this aspect of the excited states of various
isomers, for a few important states, we performed geometry optimization.
For the purpose, CI-singles (CIS)-based excited-state geometry-optimization
approach, as implemented in Gaussian-16 program,\cite{g16} was employed.
For visualizing the molecular orbitals we used GaussView 6\cite{gv6},
and for plotting the molecular geometries, XCrySDen software\cite{XCrySDen}
was employed.

\subsection{Computational parameters}

In this section we discuss our choices of three computational parameters,
namely: (a) Gaussian basis functions, (b) orbital basis set, and (c)
many-particle configurations.

\subsubsection{Choice of the Gaussian basis set}

A wide variety of Gaussian basis functions are available, depending
upon the task at hand. For example, for ground state geometry many
workers prefer Pople basis sets of the type 6-311, which are moderately
sized, and yield good results for ground state properties. However,
we wanted to use a basis set which can be utilized both for ground
state geometry optimization, as well as excited state calculations
needed for computation of the optical absorption spectra. Therefore,
in these calculations we adopted correlation consistent Dunning basis
set cc-pVTZ, which is known to yield good results both for the ground
and excited state calculations. However, in order to ensure the convergence
of our computed optical spectrum, we also computed it by using larger
aug-cc-pVTZ basis set containing several diffuse exponents for an
isomer (see Fig. S9 of the Supporting Information), and found that
except for changes in absolute intensities of the peaks, no significant
changes in peak locations were observed. Thus, cc-pVTZ basis set was
used both for Si and H atoms, during geometry optimization performed
at the CCSD(T) level,\cite{PSI4} and also for excited state calculations
performed using the MRSDCI approach. In an earlier work from our group
dealing with the optical absorption of bare Al clusters, we had also
used the cc-pVTZ basis set for Al atoms.\cite{Shinde_PCCP} Given
the quality of this basis set, we expect our calculations to be fairly
accurate.

\subsubsection{Molecular Orbital basis}

It is a well-known fact that the computational effort at the CI level
scales as $\approx N^{6}$, where $N$ is the total number of molecular
orbitals employed in the CI calculations. Thus, the computation time
increases steeply, with the increasing size of the MO set used in
the CI calculations. Therefore, to keep the calculations tractable,
we employed the frozen-core approximation, thereby not only reducing
the orbital basis size, but also the number of electrons employed
in the CI calculations to just four per atom, corresponding to the
valence electrons of each Si atom. In our earlier works on various
clusters, we carefully studied the influence of frozen core approximation
on the computed optical absorption spectra, and found that the results
were unaffected by it.\cite{Shinde_nano_life,Shinde_PCCP,epjd-pradip,epjd-shinde-mg}
As far as virtual orbitals are concerned, we did not discard any orbital
in MRSDCI calculations for all the molecules except for mono-briged
(Si-H-SiH), disilavinylidene (Si-SiH\textsubscript{2}), and disilane
(H\textsubscript{3}Si-SiH\textsubscript{3}), for which we retained
all those virtual orbitals whose energies were less than one Hartree.
This ``one Hartree'' cutoff is computationally sound because we
are interested in optical excitations whose energy is much smaller.
Nevertheless, we would like to emphasize that during geometry optimization
performed using the CCSD(T) method,\cite{PSI4} each calculation was
carried out at the all-electron level, without truncating the available
MO set.

\subsubsection{Size of the CI expansion.}

\begin{figure}[H]
\begin{centering}
\includegraphics[scale=0.4]{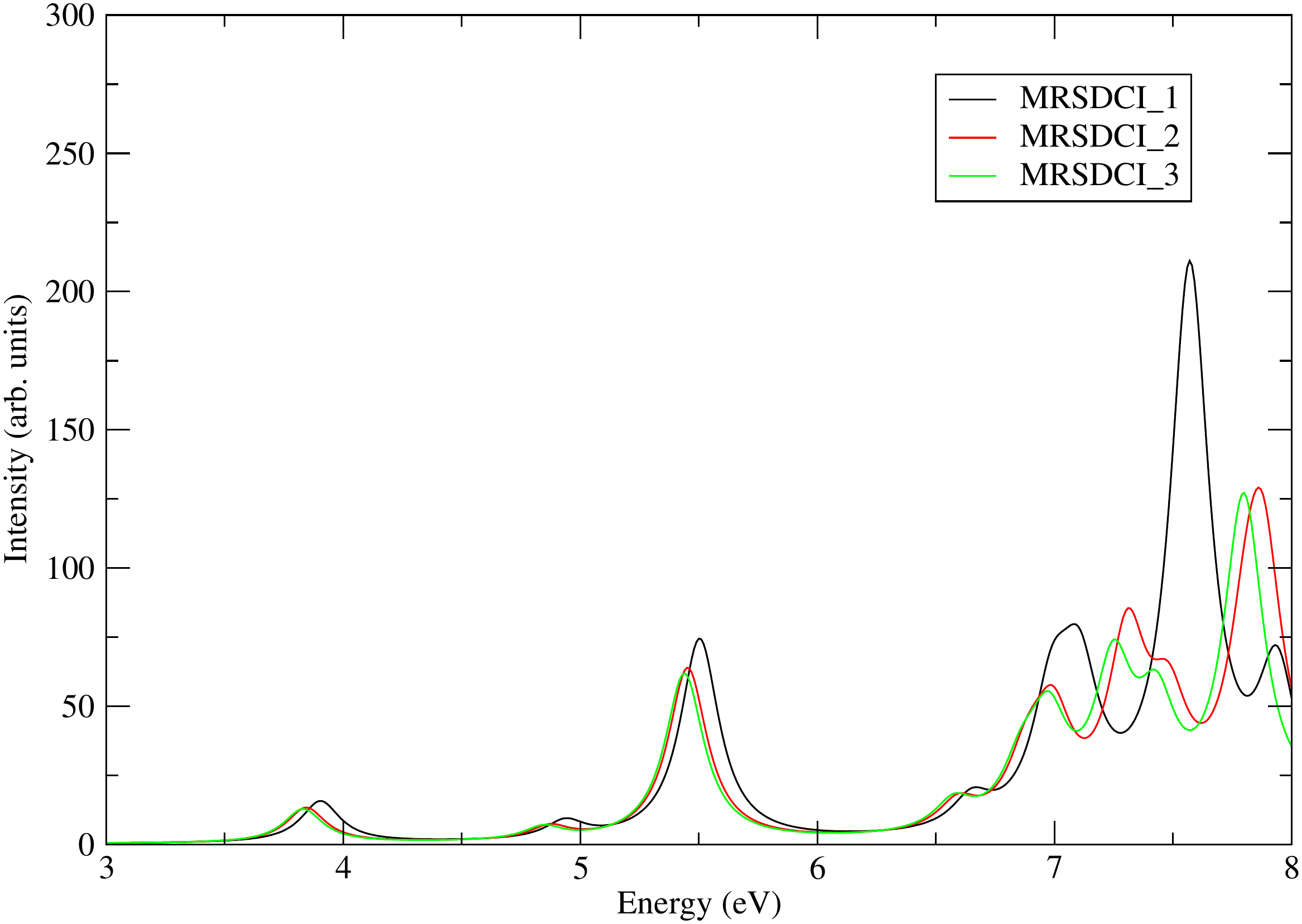}
\par\end{centering}
\caption{Convergence of the MRSDCI optical absorption spectrum of the monobridged
isomer of Si$_{2}$H$_{4}$ molecule, with respect to the increasing
number of reference configurations ($N_{ref})$. Calculations labeled
MRSDCI\_1, MRSDCI\_2, and MRSDCI\_3 were performed using 13, 32, and
41 reference configurations, respectively. \label{fig:spectrum-Convergence}}
\end{figure}

As discussed earlier, we initiate the MRSDCI calculations with a small
number of reference configurations, and compute the optical absorption
spectrum of the system concerned. By analyzing the excited states
contributing to the peaks in the computed spectrum, we increase the
number of reference configurations, and perform next level of MRSDCI
calculation leading to a new optical absorption spectrum. This procedure
is iterated until the calculated absorption spectrum exhibits reasonable
convergence. Whether to include a given configuration in the list
of reference configurations is based upon the magnitude of its coefficient
in the many-particle wave function of an excited state contributing
to a significant peak in the calculated absorption spectrum. In Fig.
\ref{fig:spectrum-Convergence} we demonstrate this procedure for
the case of monobridged isomer of Si$_{2}$H$_{4}$ molecule. Denoting
the total number of reference configurations as $N_{ref}$, and the
total number of CSFs in that CI expansion as $N_{total}$, the three
MRSDCI calculations presented in Fig. \ref{fig:spectrum-Convergence}
were performed using $N_{ref}=13$ ($N_{total}=851933)$, $N_{ref}=32$
($N_{total}=1975358)$, and $N_{ref}=41$$(N_{total}=2506254)$, respectively.
From the plotted spectra it is obvious that MRSDCI\_3 calculation
has converged to an acceptable level, both qualitatively and quantitatively,
when compared to the MRSDCI\_2 calculation. 

\section{Results and Discussion}

\label{sec:Results}

In Table \ref{tab:Ground-state-energies} we present our results on
the energetics corresponding to the optimized geometries of various
molecules, to be discussed in detail later on. An inspection of Table
\ref{tab:Ground-state-energies} reveals that all the isomers studied
in this work have large binding energies in the range 10.62--24.23
eV, implying that they will be bound at the ambient conditions. Furthermore,
for each isomer we also performed vibrational frequency analysis,
and no imaginary frequencies were obtained. This suggests that all
the conformers studied in this work represent stable structures.

In Table \ref{tab:total-configurations} we present the sizes of the
CI matrices involved in the MRSDCI calculations performed on these
molecules, for various irreducible representations (irreps) corresponding
to their ground, and excited state, wave functions. From the sizes
of the CI matrices, it is obvious that our calculations were large
scale, and, therefore, account for electron correlation effects in
an adequate manner. Next, we discuss the calculated ground state geometries,
and the optical absorption spectra, of various isomers of silicon
hydride molecules. We present the results of our calculations on Si$_{2}$H$_{6}$
molecule first, because it is the only molecule studied in this work
for which both the experimental, as well theoretical, results are
available. Therefore, comparing our results with those will allow
us to benchmark our approach, and make it trustworthy for other molecules
for which no earlier data on optical absorption is available.

\begin{table}[H]
\caption{{\small{}Total ground state (GS) energies (in Hartree), point group,
symmetry of the ground state, relative energies, correlation energies,
and the binding energies of the hydrogenated silicon conformers computed
using the CCSD(T) approach, using the cc-pVTZ basis sets. The correlation
energy indicates the difference of energies computed using the CCSD(T)
and HF levels of theory. To obtain the total binding energy of a system
presented in the last column, we have also taken care of the basis
set superposition error (BSSE) using the counterpoise correction.\label{tab:Ground-state-energies}}}

\begin{tabular}{cccccccc}
\hline 
Molecule & Conformer & Point  & Symmetry & Total energy & Relative  & Correlation & Binding\tabularnewline
 &  & group & of the GS  & (Ha) & energy & energy (eV) & energy\tabularnewline
 &  &  & wave function &  & (eV) &  & (eV)\tabularnewline
\hline 
Si\textsubscript{2}H\textsubscript{2} & Dibridged disilyne & C\textsubscript{2v} & \textsuperscript{1}A\textsubscript{1} & -579.33918 & 0.0 & 10.58 & 11.38\tabularnewline
 &  &  &  &  &  &  & \tabularnewline
Si\textsubscript{2}H\textsubscript{2} & Monobridged & C\textsubscript{s} & \textsuperscript{1}A\textsuperscript{$'$}  & -579.32320 & 0.4348 & 10.74 & 10.93\tabularnewline
 &  &  &  &  &  &  & \tabularnewline
Si\textsubscript{2}H\textsubscript{2} & Disilavinylidene & C\textsubscript{2v} & \textsuperscript{1}A\textsubscript{1} & -579.31788 & 0.5796 & 10.24 & 10.86\tabularnewline
 &  &  &  &  &  &  & \tabularnewline
Si\textsubscript{2}H\textsubscript{2} & Trans-bent & C\textsubscript{2h} & \textsuperscript{1}A\textsubscript{g} & -579.30975 & 0.8008 & 10.94 & 10.62\tabularnewline
 &  &  &  &  &  &  & \tabularnewline
 &  &  &  &  &  &  & \tabularnewline
Si\textsubscript{2}H\textsubscript{4} & Disilene & C\textsubscript{2h} & \textsuperscript{1}A\textsubscript{g} & -580.55905 & 0.0 & 11.41 & 17.32\tabularnewline
 &  &  &  &  &  &  & \tabularnewline
Si\textsubscript{2}H\textsubscript{4} & Monobridged & C\textsubscript{1} & \textsuperscript{1}A & -580.54854 & 0.2859 & 11.41 & 17.01\tabularnewline
 &  &  &  &  &  &  & \tabularnewline
Si\textsubscript{2}H\textsubscript{4} & Silylsilylene & C\textsubscript{s} & \textsuperscript{1}A$^{\prime}$ & -580.54823 & 0.2944 & 11.06 & 17.05\tabularnewline
 &  &  &  &  &  &  & \tabularnewline
 &  &  &  &  &  &  & \tabularnewline
Si\textsubscript{2}H\textsubscript{6} & Disilane & D\textsubscript{3d} & \textsuperscript{1}A\textsubscript{1g} & -581.81644 & - & 12.13 & 24.23\tabularnewline
 &  &  &  &  &  &  & \tabularnewline
\hline 
\end{tabular}
\end{table}

\begin{table}[H]
\caption{{\small{}Point group symmetry employed in the calculations, along
with the total number of configurations ($N_{total}$) in the MRSDCI
expansion, aimed at computing the optical absorption spectra of various
hydrogenated silicon conformers. In all the calculations, cc-pVTZ
basis sets were used both for Si and H atoms.\label{tab:total-configurations}}}

\begin{tabular}{ccccccccc}
\hline 
Molecule & \hspace{0.3cm} & Conformer & \hspace{0.3cm} & Point group used  & \hspace{0.3cm} & Symmetry & \hspace{0.3cm} & $N_{total}$\tabularnewline
\hline 
Si\textsubscript{2}H\textsubscript{2} &  & Dibridged disilyne &  & C\textsubscript{2v} &  & \textsuperscript{1}A\textsubscript{1} &  & 1458235\tabularnewline
 &  &  &  &  &  & \textsuperscript{1}B\textsubscript{1} &  & 1951202\tabularnewline
 &  &  &  &  &  & \textsuperscript{1}B\textsubscript{2} &  & 1105180\tabularnewline
 &  &  &  &  &  &  &  & \tabularnewline
Si\textsubscript{2}H\textsubscript{2} &  & Monobridged &  & C\textsubscript{1} &  & \textsuperscript{1}A\textsubscript{} &  & 1681403\tabularnewline
 &  &  &  &  &  &  &  & \tabularnewline
Si\textsubscript{2}H\textsubscript{2} &  & Disilavinylidene &  & C\textsubscript{1} &  & \textsuperscript{1}A &  & 2526917\tabularnewline
 &  &  &  &  &  &  &  & \tabularnewline
Si\textsubscript{2}H\textsubscript{2} &  & Trans-bent &  & C\textsubscript{2h} &  & \textsuperscript{1}A\textsubscript{g} &  & 221709\tabularnewline
 &  &  &  &  &  & \textsuperscript{1}A\textsubscript{u} &  & 2818861\tabularnewline
 &  &  &  &  &  & \textsuperscript{1}B\textsubscript{u} &  & 2643120\tabularnewline
 &  &  &  &  &  &  &  & \tabularnewline
 &  &  &  &  &  &  &  & \tabularnewline
Si\textsubscript{2}H\textsubscript{4} &  & disilene &  & C\textsubscript{2h} &  & \textsuperscript{1}A\textsubscript{g} &  & 42206\tabularnewline
 &  &  &  &  &  & \textsuperscript{1}A\textsubscript{u} &  & 359342\tabularnewline
 &  &  &  &  &  & \textsuperscript{1}B\textsubscript{u} &  & 501900\tabularnewline
 &  &  &  &  &  &  &  & \tabularnewline
Si\textsubscript{2}H\textsubscript{4} &  & monobridged &  & C\textsubscript{1} &  & \textsuperscript{1}A &  & 2506254\tabularnewline
 &  &  &  &  &  &  &  & \tabularnewline
Si\textsubscript{2}H\textsubscript{4} &  & silylsilylene &  & C\textsubscript{1} &  & \textsuperscript{1}A &  & 3404169\tabularnewline
 &  &  &  &  &  &  &  & \tabularnewline
 &  &  &  &  &  &  &  & \tabularnewline
Si\textsubscript{2}H\textsubscript{6} &  & Disilane &  & C\textsubscript{2h} &  & \textsuperscript{1}A\textsubscript{g} &  & 20621\tabularnewline
 &  &  &  &  &  & \textsuperscript{1}A\textsubscript{u} &  & 1634632\tabularnewline
 &  &  &  &  &  & \textsuperscript{1}B\textsubscript{u} &  & 2038895\tabularnewline
\hline 
\end{tabular}
\end{table}

\subsection{Disilane Si$_{2}$H$_{6}$}

\begin{figure}[H]
\begin{centering}
\includegraphics[scale=0.35]{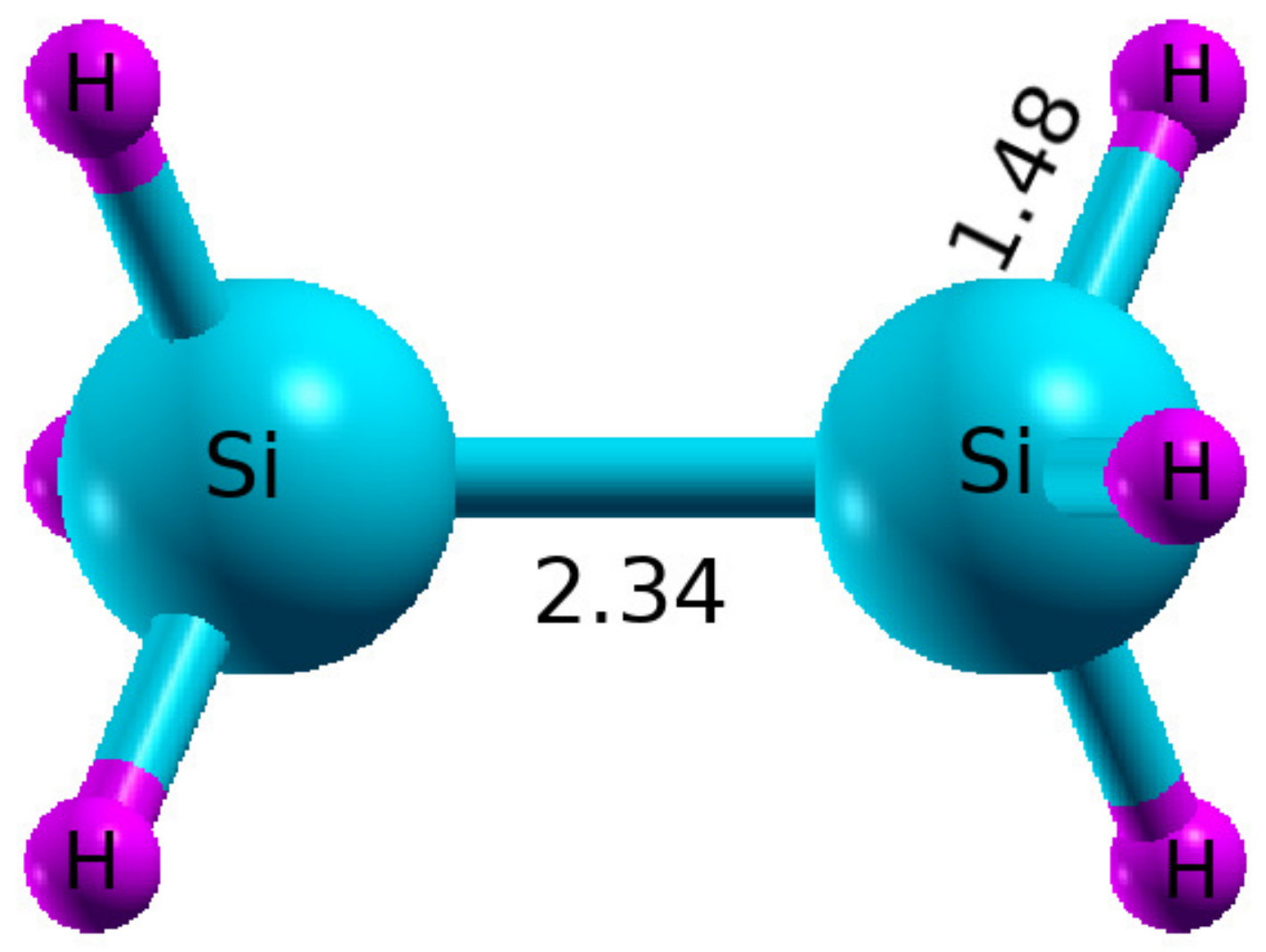}
\par\end{centering}
\centering{}\caption{{\small{}Ground state geometry of}\emph{\small{} }{\small{}disilane
(H}\protect\textsubscript{{\small{}3}}{\small{}Si-SiH}\protect\textsubscript{{\small{}3}}{\small{}),
optimized using the CCSD(T) method, and the cc-pVTZ basis set. All
the bond lengths are in $\textrm{\AA}$ unit.\label{fig:geometry-si2h6}}}
\end{figure}

As far as disilane (Si$_{2}$H$_{6}$) is concerned, it has only one
stable conformer with a three-dimensional sp$^{3}$-hybridized structure,
and the configuration {\small{}H}\textsubscript{{\small{}3}}{\small{}Si-SiH}\textsubscript{{\small{}3}}{\small{},}
similar to the case of ethane{\small{} (C$_{2}$H$_{6}$).} In this
structure, again similar to the case of ethane, the hydrogen atoms
are arranged in a staggered configuration, instead of an eclipsed
one, leading to the ground state point-group symmetry $D_{3d}$. Our
optimized geometry obtained using the CCSD(T) method, and cc-pVTZ
basis set, depicted in Fig. \ref{fig:geometry-si2h6}, has only two
unique bond lengths: (a) Si-Si bond length of 2.34 \AA , and (b) Si-H
bond length 1.48 \AA , along with the H-Si-H and H-Si-Si bond angles
of 108.6$^{o}$ and 110.3$^{o}$. Our calculated geometry parameters
are in good agreement with the experimental,\cite{Si2H6_geo_expt_1,Si2H6_geo_expt_2}
as well as theoretical values computed by other authors.\cite{Si2H6_GS_6,Alexander_Sax,Ortiz_Mintmire}
The atomic coordinates corresponding to our optimized ground state
geometries are presented in Table S16 of the Supporting Information.

\begin{figure}[H]
\begin{centering}
\includegraphics[scale=0.4]{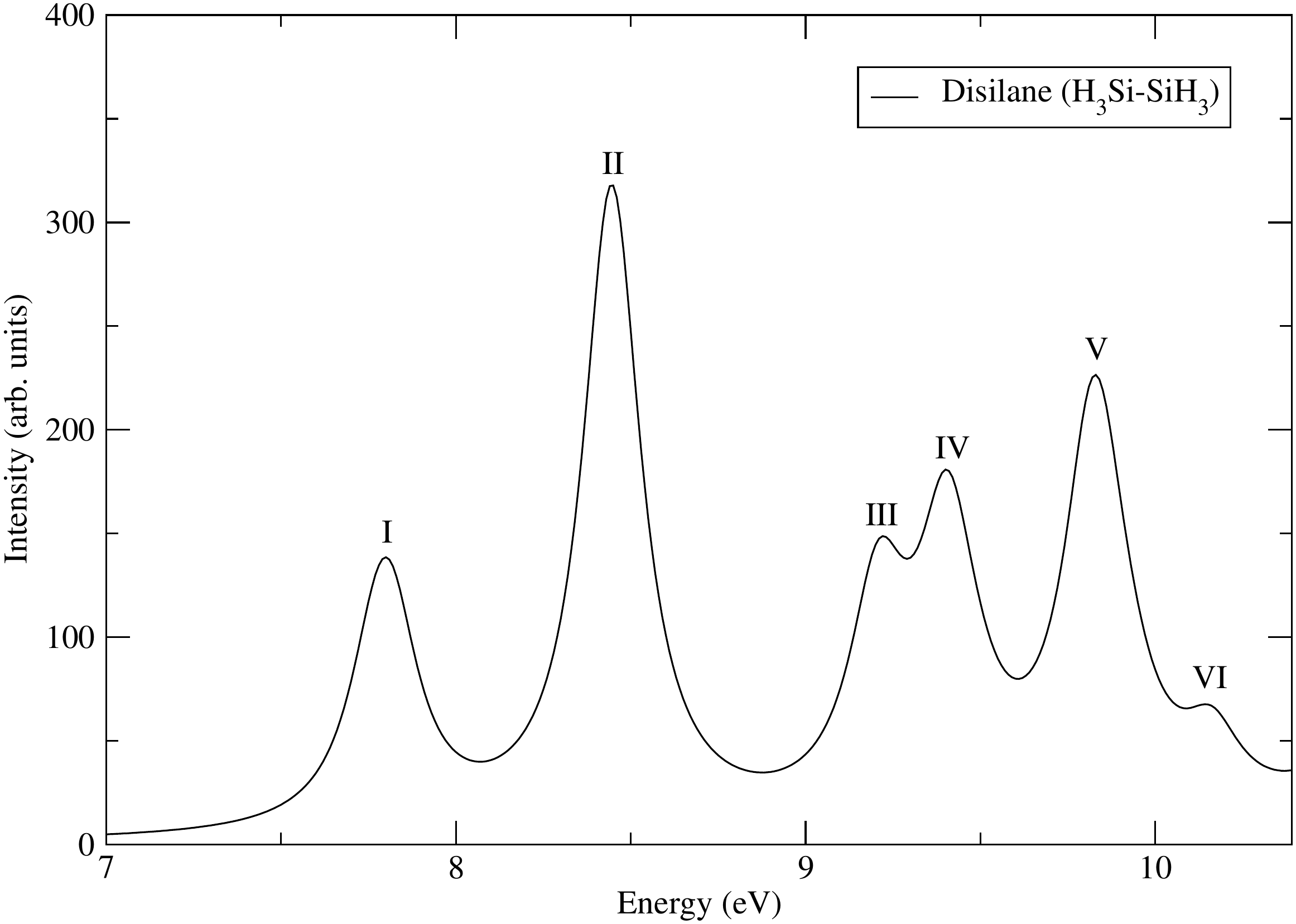}
\par\end{centering}
\caption{{\small{}Optical absorption spectrum of disilane computed using the
MRSDCI method, and the cc-pVTZ basis set. For plotting the spectrum,
we assumed a uniform line-width of 0.1 eV.\label{fig:Optics-disilane}}}
\end{figure}

Our photoabsorption spectrum of disilane is presented in \textcolor{black}{Fig.
\ref{fig:Optics-disilane},} while the optimized geometries of a couple
of excited states are given in Fig. S17 of the Supporting Information.\textcolor{red}{{}
}It is obvious from the absorption spectrum that it consists of six
well-separated peaks of fairly strong intensities, except for the
last one, which is comparatively weaker.

The optical absorption in disilane starts at fairly high energies,
as compared to previously discussed molecules, with a moderately intense
peak near 7.80 eV. This peak is due to two close-lying excited states,
both whose wave function are dominated by the singly-excited configuration
$|H\rightarrow(L+2)\rangle$, where the orbital $L+2$ is doubly degenerate.
The relaxed geometry of this state is twisted with respect to the
ground state about the Si-Si bond axis. Furthermore, Si-Si bond is
a bit longer (2.60 $\text{Å}$) compared to the ground state, whereas
not all the Si-H bond lengths are exactly equal, but still close enough
to their ground state values. The first absorption is followed by
the most intense peak (peak II) of the spectrum located at 8.45 eV,
due to a state whose wave function mainly consists of the single excitation
$\arrowvert H\rightarrow L\rangle$. In the relaxed structure of this
excited state, Si-Si bond is substantially stretched (3.47 $\text{Å}$)
compared to the ground state, but no changes are observed in any of
the Si-H bond lengths. The next peak in the spectrum (peak III) occurs
at 9.21 eV, and is due to a state whose wave function is dominated
by almost equal contributions from the two degenerate excitations
$\arrowvert H-1\rightarrow L+2\rangle$, where orbital $H-1$ is also
doubly degenerate, just like orbital $L+2$. Next we have two peaks
in the spectrum located at 9.41 eV and 9.84 eV, both of which are
due to two closely-lying states each. The first of these peaks (peak
IV) derives its main intensity from a state located at 9.40 eV, with
a small contribution from a state at 9.47 eV. The main contribution
to oscillator strength of peak V comes from a state located at 9.82
eV, with a much smaller intensity derived from a state located 9.92
eV. The wave functions of all the four excited states contributing
to peaks IV and V exhibit strong configuration mixing, and are composed
of various degenerate combinations of singly-excited configurations
$\arrowvert H-1\rightarrow L+2\rangle$, $\arrowvert H-1\rightarrow L\rangle$,
and $\arrowvert H-2\rightarrow L\rangle$ (see Table S8 of Supporting
Information). Final peak of the computed spectrum (peak VI) located
at 10.16 eV is a relatively weaker one, and is due to a state whose
wave function consists mainly of the configuration $|H\rightarrow L+4\rangle$,
with a smaller contribution from the excitation $\arrowvert H-1\rightarrow L+2\rangle$.

In Table \ref{tab:Comparing-optical} we compare our calculated peak
locations to the experimentally measured values of Itoh et al.\cite{Si2H6_optical_expt},
and the Bethe-Salpeter equation based calculations of Rohlfing et
al.,\cite{Si2H6_optical_theory} and find that our results are in
very good agreement with the experiments. This level of agreement
between our calculations and the experiments for the case of disilane,
the largest studied molecule in this work, suggests that our computational
methodology is sound and trustworthy.

\begin{table}[H]
\caption{{\small{}Comparison of the peak locations (in eV) in the calculated
optical absorption spectra of disilane, with the experimental values
reported by Itoh }\emph{\small{}et al}{\small{}.,\cite{Si2H6_optical_expt}}\emph{\small{}
}{\small{}and the theoretical values reported by Louie }\emph{\small{}et
al}{\small{}.,\cite{Si2H6_optical_theory} obtained from their Bethe-Salpeter
equation based calculations.\label{tab:Comparing-optical}}}

\begin{tabular}{ccccc}
\hline 
This work & \hspace{1cm} & Expt. (Ref. \cite{Si2H6_optical_expt}) & \hspace{1cm} & Theory (Ref. \cite{Si2H6_optical_theory})\textsuperscript{}\tabularnewline
 &  &  &  & \tabularnewline
\hline 
\hline 
7.80 &  & 7.6 &  & 7.6\tabularnewline
8.45 &  & 8.4 &  & 9.0\tabularnewline
9.41 &  & 9.5 &  & 9.6\tabularnewline
9.84 &  & 9.9 &  & 9.8\tabularnewline
\hline 
\end{tabular}
\end{table}

\subsection{Si$_{2}$H$_{2}$}

\begin{figure}[H]
\centering{}\subfloat[Dibridged disilyne]{\begin{centering}
\includegraphics[scale=0.16]{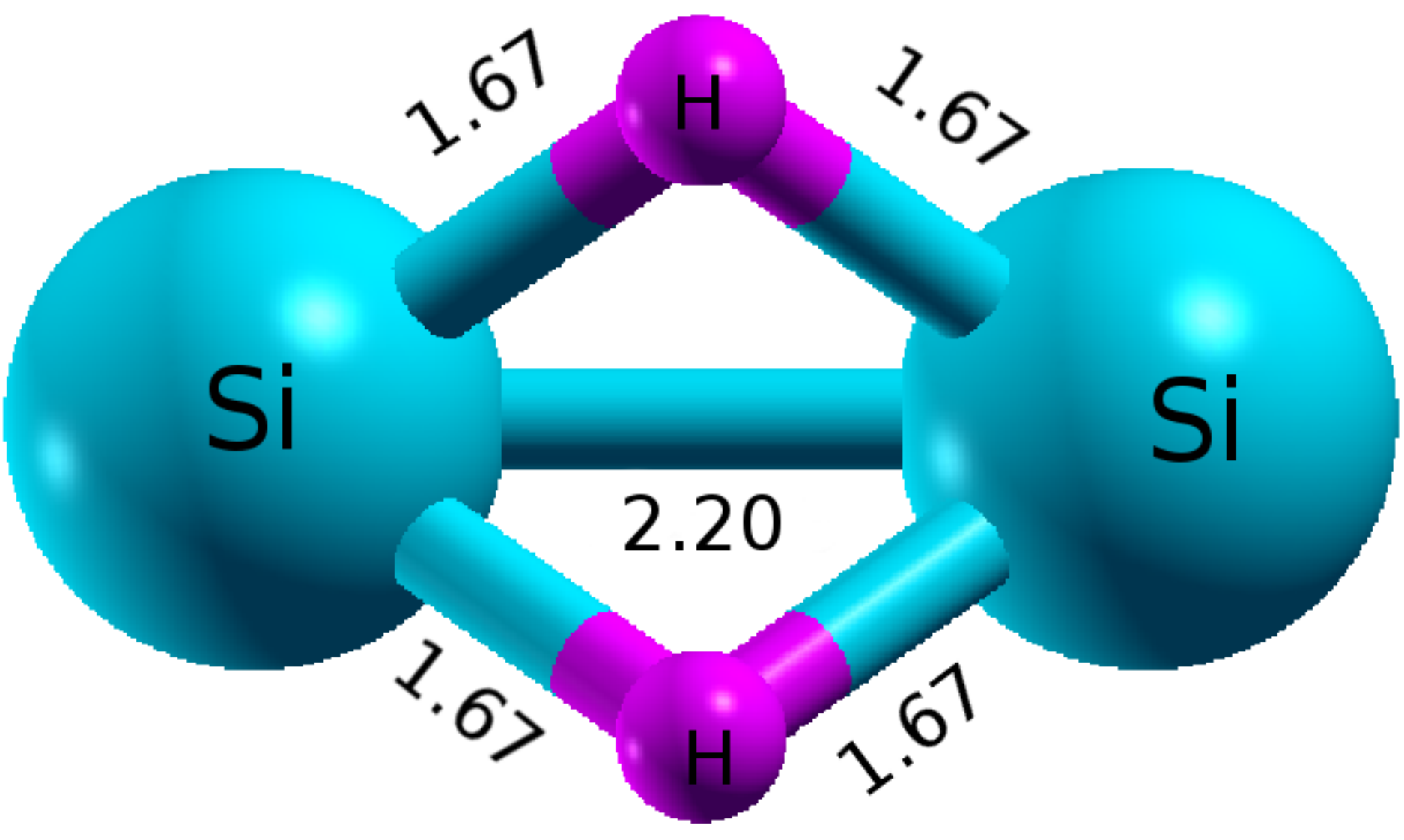}
\par\end{centering}
}~~\subfloat[Mono-bridged]{\begin{centering}
\includegraphics[scale=0.2]{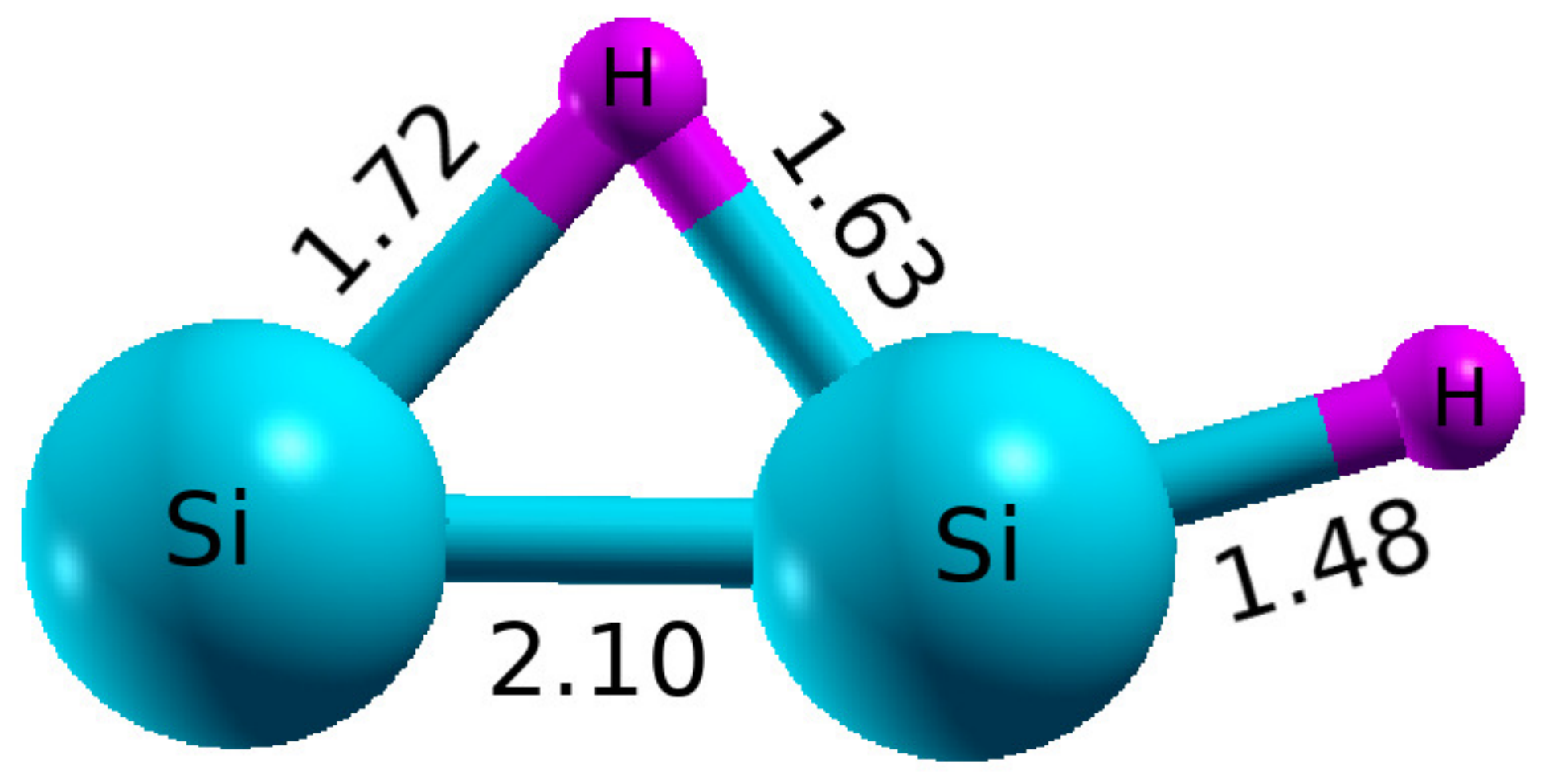}
\par\end{centering}
}~~\subfloat[Disilavinylidene]{\begin{centering}
\includegraphics[scale=0.2]{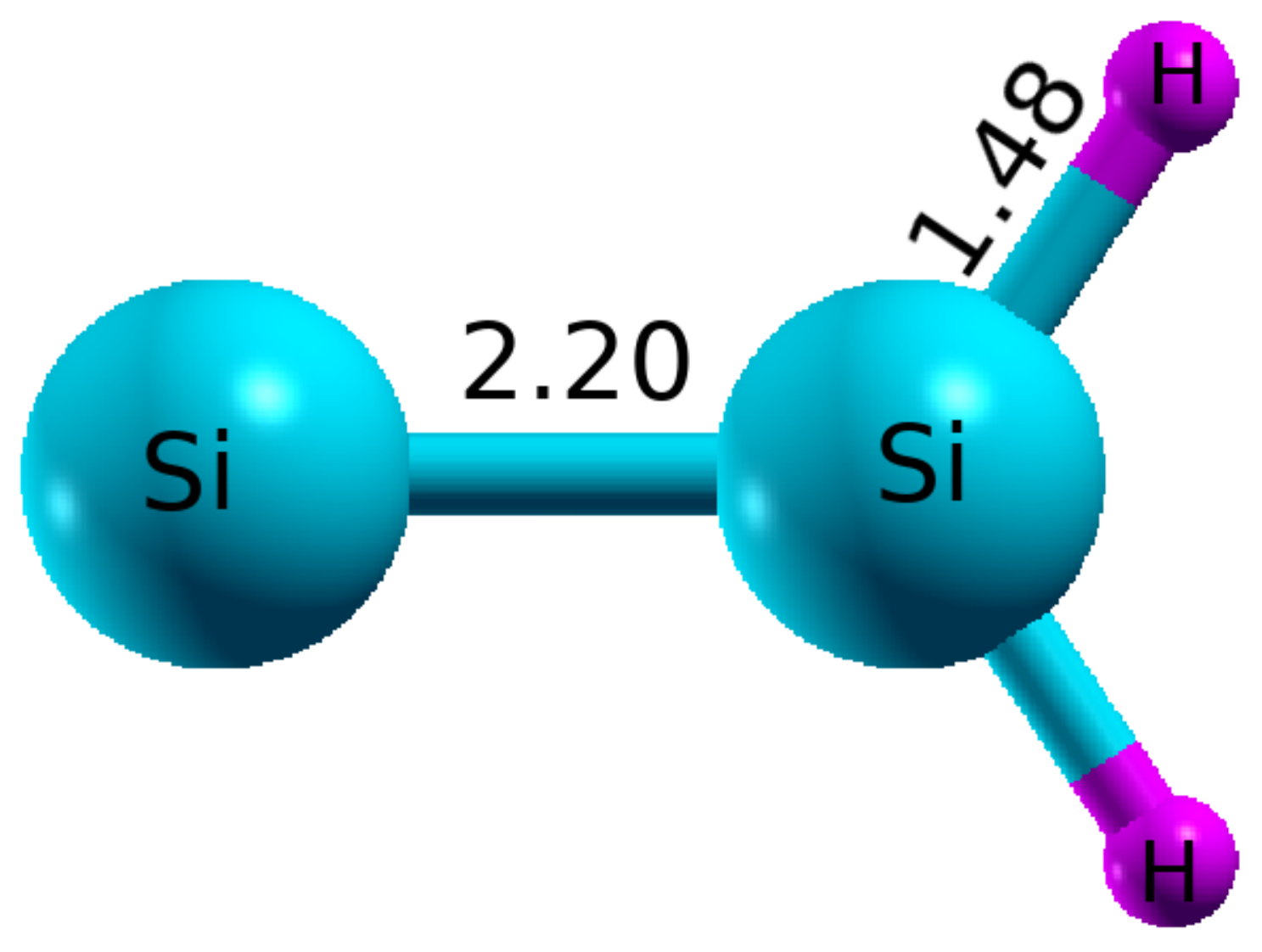}
\par\end{centering}
} ~~\subfloat[Trans-bent]{\begin{centering}
\includegraphics[scale=0.2]{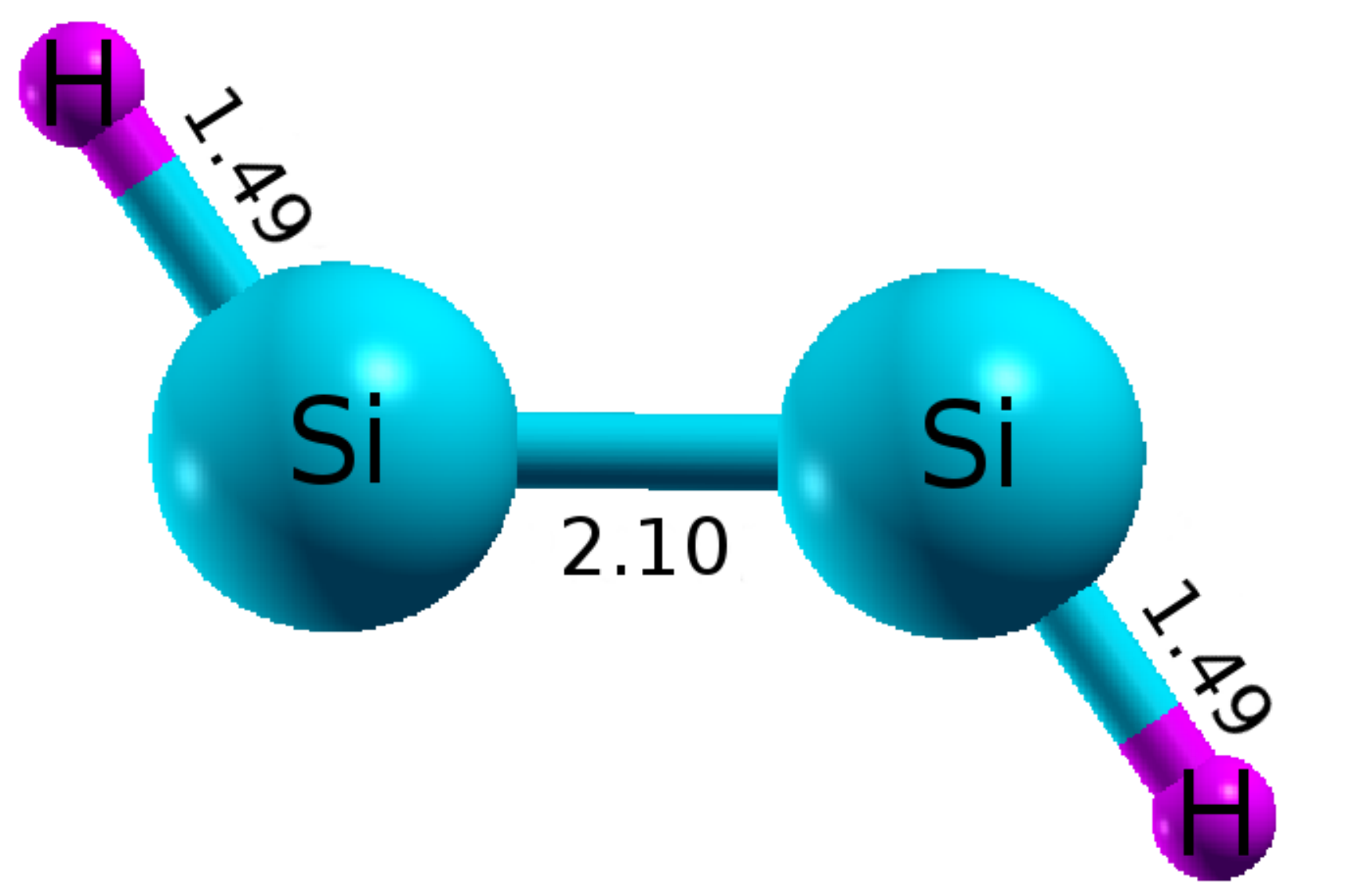}
\par\end{centering}
}\caption{{\small{}Ground state geometries of various isomers of }Si\protect\textsubscript{2}H\protect\textsubscript{2}{\small{}
molecule, optimized using the CCSD(T) method, and the cc-pVTZ basis
set. All the bond lengths are in $\textrm{\AA}$ unit.\label{fig:geometries-si2h2}}}
\end{figure}

Si$_{2}$H$_{2}$ is the smallest member of the Si$_{2}$H$_{2n}$
class of molecules, several of whose conformers have been studied
in the past.\cite{Sannigrahi_Nandi,doi:10.1021/j100377a036,Roger_and_Henry,doi:10.1021/ja00360a016}
The five most studied conformers of Si\textsubscript{2}H\textsubscript{2}
molecule are dibridged disilyne, mono-bridged structure, disilavinylidene,
trans-bent structure, and the planar dibridged disilyne. The stability
analysis of these conformers revealed that all of them are stable,
except for the planar dibridged disilyne structure, which corresponds
to a saddle point, or a transition state, on the potential energy
surface.\cite{Sannigrahi_Nandi,doi:10.1021/j100377a036,Roger_and_Henry,doi:10.1021/ja00360a016}
As a result, we restricted the present study to the four stable structures,
and we optimized their geometries using the the cc-pVTZ basis set,
and the coupled-cluster singles-doubles-perturbative-triples {[}CCSD(T){]}
method, as implemented in the PSI4 computer program.\cite{PSI4} The
optimized geometries are presented in Fig. \ref{fig:geometries-si2h2},
while Table \ref{tab:Ground-state-energies} contains their total
and relative energies. From the table it is obvious that the dibridged
disilyne is the lowest energy isomer, followed by the monobridged
structure which is 0.44 eV higher. Next are two conformers disilavinylidene
and trans-bent structure, which are higher in energy by 0.58 eV, and
0.80 eV, respectively, as compared to the lowest-energy dibridged
disilyne conformer. It is obvious that all the higher energy conformers
are within 1 eV of the lowest-energy structure, and even closer to
each other. Therefore, it will be interesting to see if their optical
absorption spectra, which were computed using the MRSDCI approach
using these geometries, are different enough to facilitate their optical
detection. In the following sections we discuss the optimized geometries
and optical absorption spectra of individual conformers.

\subsubsection{\emph{Dibridged disilyne (Si-H}\protect\textsubscript{\emph{2}}\emph{-Si)}}

Our optimized structure of dibridged disilyne conformer, as shown
in Fig. \ref{fig:geometries-si2h2}(a), consists of two three-center
Si-H-Si bonds with identical Si-H bond lengths of 1.67{\small{} $\textrm{\AA}$},
which are non-planar leading to the $C_{2v}$ point-group symmetry,
instead of $D_{2h}$, had they been planar. Furthermore, we obtained
the optimized Si-Si distance to be 2.20 $\textrm{\AA}$, along with
Si-H-Si, H-Si-H and H-Si-Si bond angles as 82.5$^{o}$, 72.3$^{o}$
and 48.8$^{o}$, respectively, whereas the HSi-SiH dihedral angle
is 103.4$^{\circ}$. Our optimized geometry parameters are in good
agreement with the experimental results of Bogey et al.\cite{Destombes}
obtained using millimeter- and submillimeter-wave spectroscopy, as
also with their ab initio theoretical results. Our results are also
in good agreement with the theoretical calculations reported by Gerv
and Schaefer\cite{Roger_and_Henry}, Adamczyk and Broadbelt\cite{Andrew_and_Linda},
Jursic,\cite{Jursic}\textcolor{red}{{} }and Sannigrahi and Nandi.\cite{Sannigrahi_Nandi}
Atomic coordinates corresponding to the optimized ground state geometry
are presented in Table S9 of the Supporting Information.

\begin{figure}[H]
\begin{centering}
\includegraphics[scale=0.4]{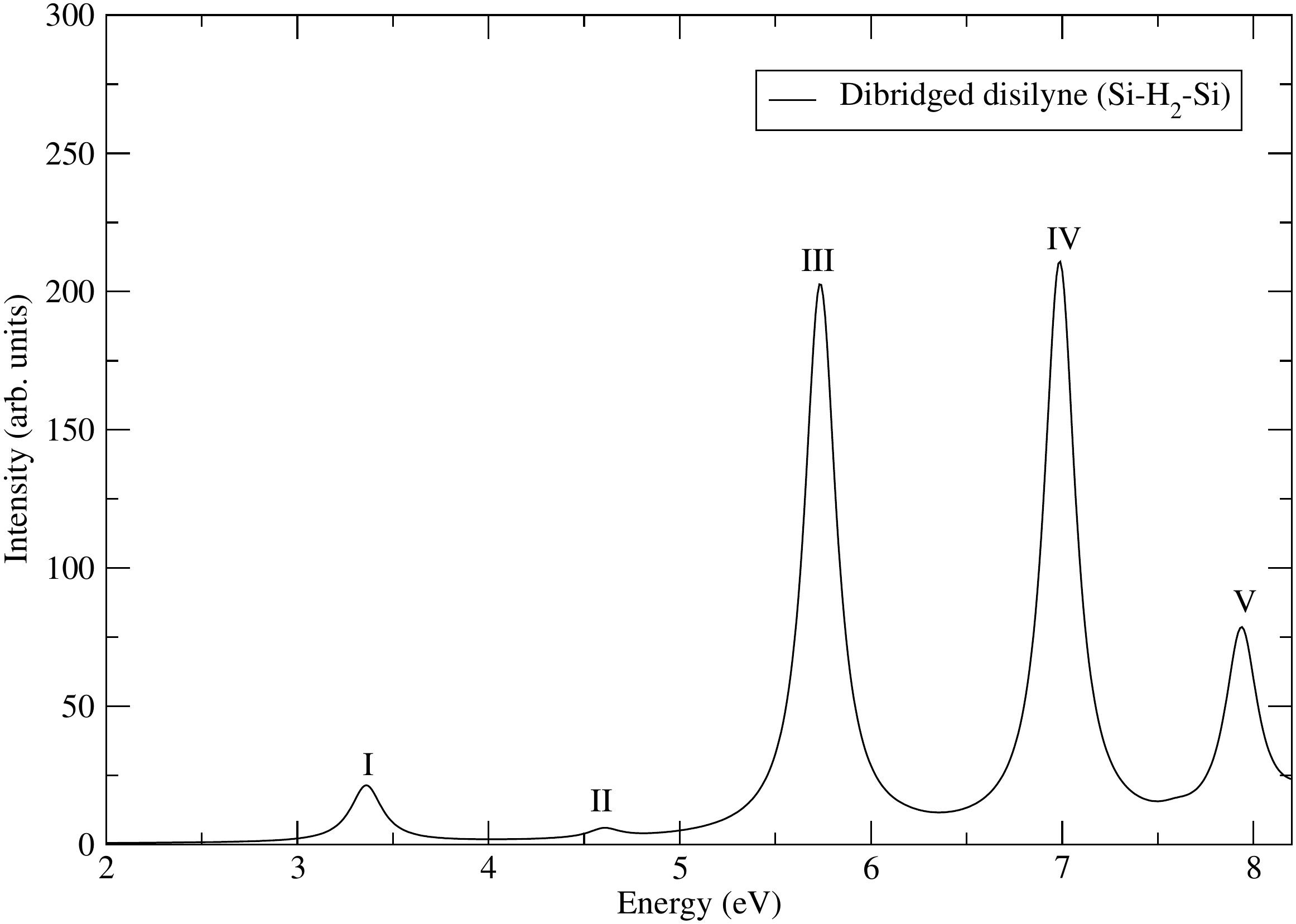}
\par\end{centering}
\caption{{\small{}Optical absorption spectrum of non-planar disilyne (Si-H$_{2}$-Si)
conformer, computed using the MRS- DCI method, and the cc-pVTZ basis
set. For plotting the spectrum, 0.1 eV uniform line-width was used.\label{fig:Optics-dibridged-disilyne}}}
\end{figure}

The calculated photoabsorption spectrum of this conformer is presented
in Fig. \ref{fig:Optics-dibridged-disilyne}, while the relaxed excited-state
structures corresponding to some of the frontier peaks of the spectrum
are given in Fig. S10 of the Supporting Information. The absorption
starts with a relatively weak peak near 3.36 eV, followed by an even
weaker peak at 4.60 eV. The first of these peaks (peak I) is due to
an excited state whose wave function is dominated by the singly-excited
configuration $\arrowvert H\rightarrow L\rangle$, while that corresponding
to the second peak (peak II) consists mainly of the configuration
$\arrowvert H\rightarrow L+2\rangle$.\textcolor{red}{{} }The optimized
geometry of the excited state corresponding to peak I is almost similar
to the ground-state geometry, except for the elongated Si-Si, and
Si-H bond lengths, 2.36 $\text{Å}$, and 1.73 $\text{Å}$, respectively.\textcolor{red}{{}
}These weak peaks are followed by the two of the most intense peaks
(peaks III and IV) of the spectrum, located near 5.74 eV, and 6.99
eV. Peak III is due to two excited states, whose wave functions exhibit
significant configuration mixing. The first state at 5.73 eV contributes
dominantly to the intensity of the peak, and its wave function consists
mainly of configurations $\arrowvert H-1\rightarrow L\rangle$ and
$\arrowvert H\rightarrow L+4\rangle$. The wave function of the second
state around 5.77 eV is dominated by doubly-excited configurations
$|H\rightarrow L+3;\thinspace H\rightarrow L+3\rangle$ and $\arrowvert H\rightarrow L;\thinspace H\rightarrow L\rangle$.
The optimized geometry of the excited state located at 5.73 eV looks
like a monobridged structure, with the Si-H bond distances 1.72 $\text{Å}$,
and 1.68 $\text{Å}$, whereas the other H-atom is connected with a
Si-atom with a bond length 1.51 $\text{Å}$. Furthermore, the Si-Si
bond length is elongated compared to the ground-state, with a bond
distance 2.45 $\text{Å}$.

The wave function of the excited state contributing to peak IV is
dominated by single excitations $\arrowvert H\rightarrow L+4\rangle$
and $\arrowvert H-3\rightarrow L+1\rangle$. The optimized geometry
of this peak has a trans-like structure with Si-Si and Si-H bond lengths,
2.30 $\text{Å}$, and 1.50 $\text{Å}$, respectively. The final peak
of the spectrum (peak V) located close to 7.93 eV, is weaker compared
to these two, and is due to two excited states located at 7.91 eV
and 7.94 eV. The wave function of the first of these is dominated
by the double excitation $\arrowvert H-1\rightarrow L+3;\thinspace H-1\rightarrow L+3\rangle$,
while that of the second one derives main contributions from single
excitations $\arrowvert H-1\rightarrow L+2\rangle$ and $\arrowvert H-2\rightarrow L+1\rangle$.\textcolor{red}{{}
}Detailed information about all the excited states contributing to
the peaks in the spectra is presented in Table \textcolor{black}{S1}
of the Supporting Information, while its Fig. S1 contains the plots
of frontier MOs participating in the optical absorption.

\subsubsection{\emph{Monobridged (Si-H-SiH)}}

The monobridged conformer of Si$_{2}$H$_{2}$ molecule is energetically
next in order as compared to the lowest-energy conformer non-planar
dibridged disilyene discussed in the previous section. This structure,
denoted as Si-H-SiH, contains one tricenter Si-H-Si bond, while the
other hydrogen atom is attached to a Si atom to form a conventional
Si-H single bond, with C\textsubscript{s} point group symmetry of
the conformer, and a closed-shell electronic ground state of symmetry
\textsuperscript{1}A$^{'}$. Our optimized geometry parameters are
(see Fig. \ref{fig:geometries-si2h2}(b)): (a) two Si-H bond distances
in the tricenter Si-H-Si bond are 1.72 $\textrm{\AA}$, and 1.63 $\textrm{\AA}$,
(b) the bond length for the single Si-H bond is 1.48 $\textrm{\AA}$,
and (c) the distance between two silicon atoms is 2.10 $\textrm{\AA}$.
Inside the Si-H-Si triangle, Si-H-Si bond angle is 77.7$^{o}$, while
the two H-Si-Si bond angles are, 49.2$^{o}$ and 53.1$^{o}$. The
external H-Si-H bond angle is computed to be 109$^{o}$, where as
the HSi-SiH dihedral angle is 0.6$^{\circ}$. Our optimized geometry
parameters are in good agreement with the experimental and theoretical
results of Cordonnier et al., \cite{si2h2-monobridged-exp} and with
the theoretical results reported by Colegrove and Schaefer,\cite{doi:10.1021/j100377a036}
Adamczyk and Broadbelt,\cite{Andrew_and_Linda} Grev and Schaefer,\cite{Roger_and_Henry}
Sannigrahi and Nandi,\cite{Sannigrahi_Nandi} Curtiss \emph{et al},\cite{Pople}
and Koseki and Gordon.\cite{Gordon} Atomic coordinates corresponding
to our optimized ground state geometry of this conformer are presented
in Table S10 of the Supporting Information.

\begin{figure}[H]
\begin{centering}
\includegraphics[scale=0.35]{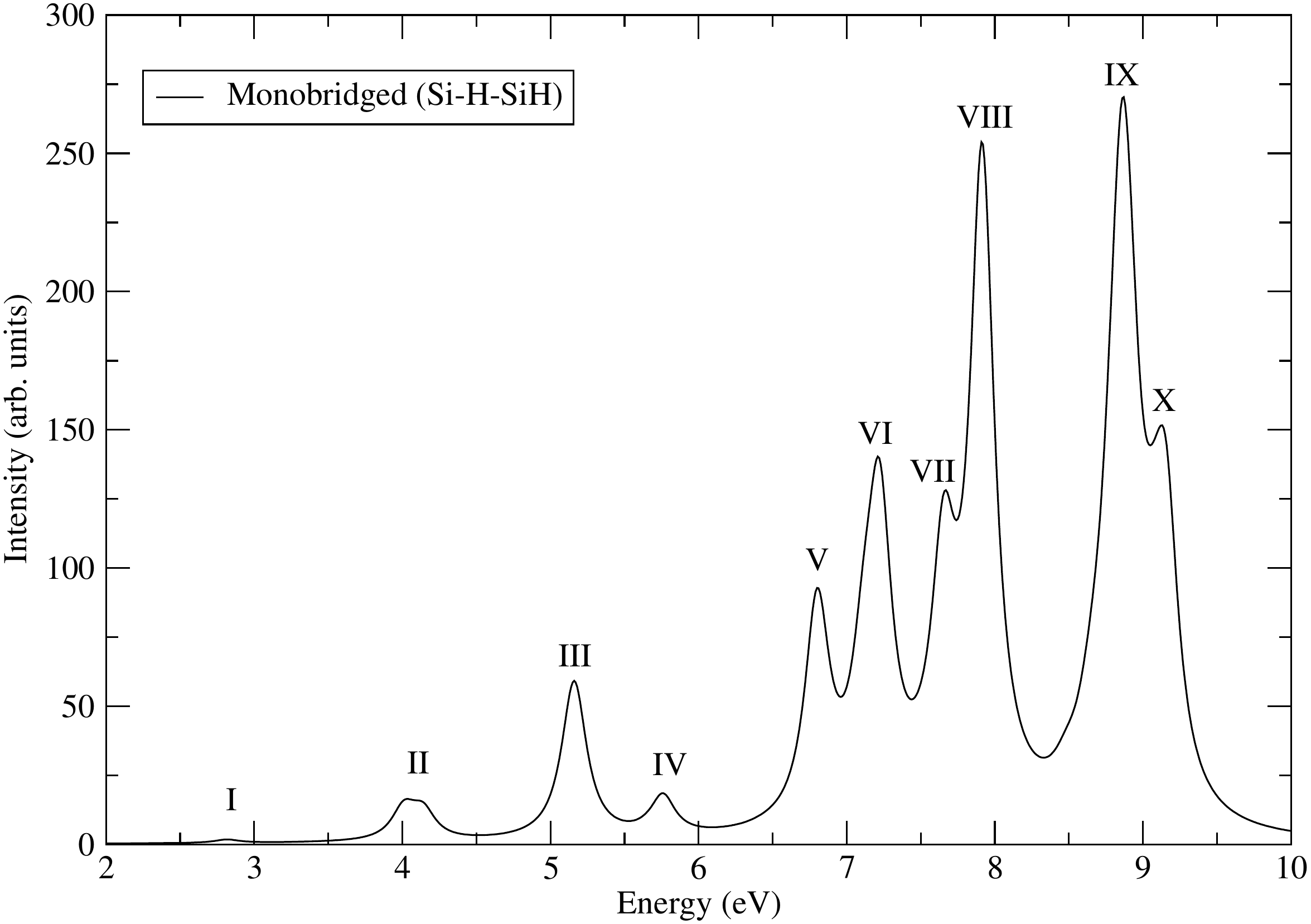}
\par\end{centering}
\caption{{\small{}Optical absorption spectrum of monobridged (Si-H-SiH) conformer
computed using the MRSDCI method, and cc-pVTZ basis set. For plotting
the spectrum, 0.1 eV uniform line-width was assumed.\label{fig:Optics-monobridged-si2h2}}}
\end{figure}

Using these geometry parameters, the calculated photoabsorption spectrum
of the mono-bridged conformer is presented in \textcolor{black}{Fig.
\ref{fig:Optics-monobridged-si2h2}, }while detailed information about
the excited states contributing to the peaks in the spectrum are presented
in Table \textcolor{black}{S2} of the Supporting Information. The
optimized geometries of the select excited states corresponding to
some of the frontier peaks in the spectrum are given in Fig. S11 of
the Supporting Information. The calculated photoabsorption spectrum
of this conformer is spread over a wide energy range starting from
2.8 eV to about 9.5 eV. The absorption begins with a very feeble peak
near 2.81 eV, due to an excited state whose wave function largely
consists of the $\arrowvert H\rightarrow L\rangle$ configuration.\textcolor{red}{{}
}It is followed by three comparatively stronger peaks near 4.08 eV,
5.16 eV, and 5.75 eV, all due to excited states whose wave functions
are dominated by the singly-excited configurations. The first of these
peaks (peak II) derives almost equal intensity from two closely located
states exhibiting significant mixing of configurations $\arrowvert H\rightarrow L+1\rangle$,
$\arrowvert H-1\rightarrow L\rangle$, and $\arrowvert H\rightarrow L+2\rangle$,
$\arrowvert H\rightarrow L+4\rangle$, respectively. The relaxed geometry
of this excited state is a disilavinylidene type of structure with
the Si-Si and equal Si-H bond length 2.26 $\text{Å}$ and 1.48 $\text{Å}$,
respectively. The optimized geometry of the excited state corresponding
to peak III also seems to be a disilavinylidene like structure, with
the Si-Si bond length 2.13 $\text{Å}$, and equal Si-H bond lengths
of 1.47 $\text{Å}$. The most intense peaks of the spectrum are peaks
VIII and IX, located at 7.91 eV and 8.86 eV, respectively. Wave functions
of the excited states giving rise to these peaks exhibit strong mixing
of singly excited configurations. The wave function corresponding
to peak VIII is a mixture of characterized of singly-excited configurations
$\arrowvert H-1\rightarrow L+4\rangle$ and $\arrowvert H-2\rightarrow L+2\rangle$,
while that corresponding to peak IX is largely composed of $\arrowvert H-1\rightarrow L+4\rangle$
and $\arrowvert H-3\rightarrow L\rangle$. The relaxed geometry of
the excited state corresponding to one of the most intense peaks,
i.e., peak VIII, has the elongated Si-Si bond length of almost 2.45
$\text{Å},$ and the H-atom of the single Si-H bond comes out of the
plane with H-Si-Si bond angle 95.5$^{o}$, as compared to the ground
state structure, which is nearly planar. The spectrum terminates with
peak X, which appears to be a shoulder of the preceding peak, and
is due to an excited state located at 9.16 eV, whose wave function
derives dominant contributions from the double-excitations $|H\rightarrow L+1;\thinspace H-1\rightarrow L+2\rangle$
and $|H\rightarrow L+2;\thinspace H\rightarrow L+2\rangle$. When
we compare the absorption spectrum of this conformer to that of the
lowest-energy structure, we find significant differences. Therefore,
it should be possible to distinguish between the two conformers using
absorption spectroscopy.

Fig. S2 of the Supporting Information presents the plots of the Hartree-Fock
MOs which participate in the photobsorption in the monobridged conformer.

\subsubsection{Disilavinylidene (Si-SiH\protect\textsubscript{2})}

This conformer is 0.58 eV higher in energy as compared to the lowest
Si\textsubscript{2}H\textsubscript{2} dibridged disilyne structure,
but just 0.14 eV above the mono-bridged isomer. Its geometrical configuration
is denoted as Si-SiH$_{2}$, and has one Si atom attached to two H
atoms through normal Si-H single bonds, while no H atoms are attached
to the other Si atom, leading to a C\textsubscript{2v} symmetric
structure (see Fig. \ref{fig:geometries-si2h2}(c)), and a ground
state of symmetry \textsuperscript{1}A\textsubscript{1}. Our optimized
geometry parameters with the distance between two Si atoms as 2.20
$\textrm{\AA}$, both the Si-H bonds lengths of 1.48 $\textrm{\AA}$,
and the Si-Si-H and H-Si-H bond angles as 124$^{o}$, and 112$^{o}$,
respectively, are in very good agreement with the theoretically reported
values by several other groups.\cite{Sannigrahi_Nandi,Pople,Gordon,Andrew_and_Linda,Roger_and_Henry,doi:10.1021/j100377a036}
The atomic coordinates corresponding to our optimized ground state
geometry are presented in Table S11 of Supporting information. 

\begin{figure}[H]
\begin{centering}
\includegraphics[scale=0.35]{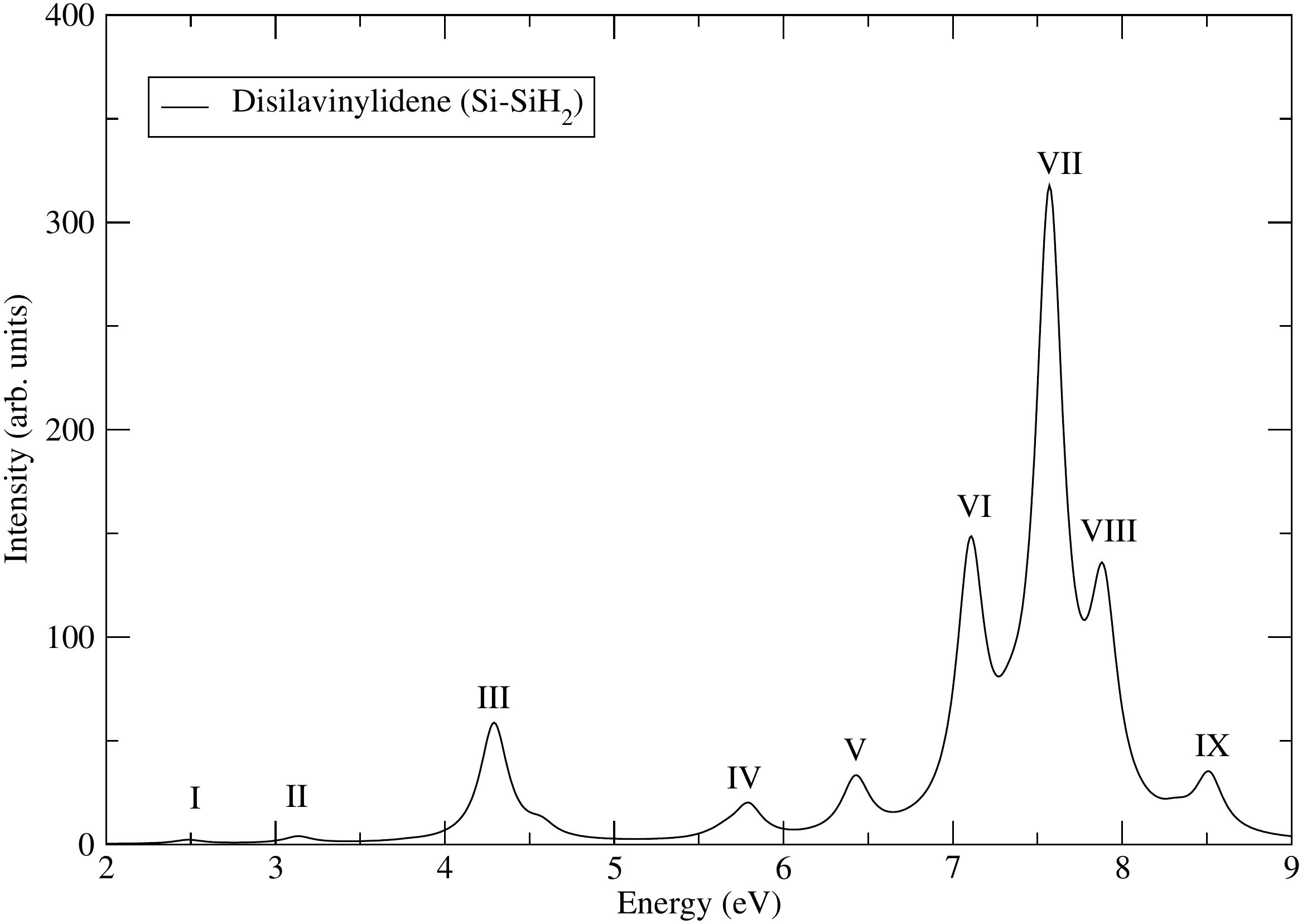}
\par\end{centering}
\caption{{\small{}Optical absorption spectrum of disilavinylidene (Si-SiH$_{2}$)
conformer, calculated using the MRSDCI method, and the cc-pVTZ basis
set. For plotting the spectrum, 0.1 eV uniform line-width was used.\label{fig:Optics-disilavinylidene}}}
\end{figure}

The calculated photoabsorption spectrum of the disilavinylidene conformer,
corresponding to the optimized geometry, is presented in \textcolor{black}{Fig.
\ref{fig:Optics-disilavinylidene}, }while the information related
to its peaks such as the oscillator strengths, dominant many particle
wave functions of the contributing excited states, their excitation
energies etc., are detailed in \textcolor{black}{Table S3} of the
Supporting Information. In Fig. S12 of the Supporting Information
we present the optimized geometries of some of the excited state contributing
to the frontier peaks in the optical absorption spectrum.

The optical absorption spectra of this isomer starts with a tiny peak
at 2.48 eV, followed by four small peaks near 3.13 eV, 4.29 eV, 5.79
eV and 6.43 eV, most of which are dominated by the singly-excited
configurations, with double excitations also making important contributions
in a couple of cases. Wave function of the excited state contributing
to peak I is dominated by single excitation $\arrowvert H-1\rightarrow L\rangle$,
while peak II corresponds to a state dominated by the double excitation
$\arrowvert H\rightarrow L;H\rightarrow L\rangle$. Wave function
of the state corresponding to peak III is dominated by singly-excited
configuration $\arrowvert H\rightarrow L+1\rangle$, while the one
corresponding to peak IV exhibits significant mixing of the single
excitation $\arrowvert H-2\rightarrow L\rangle$ and the double excitation
$\arrowvert H\rightarrow L;H-2\rightarrow L+1\rangle$. The optimized
geometry corresponding to peak III is non-planar, unlike the ground
state structure. It has an elongated Si-Si bond with length 2.70 $\text{Å}$,
and equal Si-H arms, with the bond lengths of 1.52 $\text{Å}$ each.
The most intense peak of the spectrum (peak VII) occurs at 7.57 eV,
and is due to a state whose wave function exhibits strong mixing of
singly-excited configurations $\arrowvert H-1\rightarrow L+2\rangle$
and $\arrowvert H-1\rightarrow L+4\rangle$. The optimized structure
of this excited state is planar, with the Si-Si bond length 2.24 $\text{Å}$,
and equal Si-H bond distances of 1.61 $\text{Å}$. The last peak of
the calculated spectrum (peak IX) is located at 8.51 eV, and is due
to a state whose wave function consists predominantly of the double
excitation $\arrowvert H\rightarrow L;H-2\rightarrow L+1\rangle$.
If we compare the absorption spectrum of disilavinylidene conformer
to those of the two previously discussed structures, including close-lying
monobridged structure, we find significant differences both in terms
of peak locations, and relative intensities, thus making their optical
detection feasible.

The plots of the orbitals contributing to the absorption spectrum
are presented in Fig. S3 of Supporting Information.

\subsubsection{\emph{Trans-bent (HSi-SiH)}}

The trans-bent conformer, which has the highest energy as compared
to rest of the three conformers considered in this work, has the geometrical
configuration denoted as HSi-SiH. This structure is just 0.22 eV higher
in energy as compared to the disilavinylidene conformer, and with
two Si-H single bonds has the point group symmetry $C_{2h}$, as shown
in Fig. \ref{fig:geometries-si2h2}(d). Our optimized geometry parameters
with Si-H and Si-Si bond distances of 1.49 $\textrm{\AA}$, and 2.10
$\textrm{\AA}$, respectively, and the Si-Si-H bond angle of 125.6$^{o}$,
are in good agreement with those reported by Adamczyk and Broadbelt\cite{Andrew_and_Linda}
computed using the G3//B3LYP approach, and by Sannigrahi and Nandi\cite{Sannigrahi_Nandi}
based on \emph{ab} \emph{initio} SCF calculations. Using a DFT based
approach, employing various basis sets, and exchange-correlation functionals,
Jursic obtained the optimized bond lengths of the trans-bent structure,
which are also in good agreement with our results.\cite{Jursic} The
atomic coordinates corresponding to our optimized ground state geomery
of this structure are presented in Table S12 of Supporting Information.

\begin{figure}[H]
\begin{centering}
\includegraphics[scale=0.35]{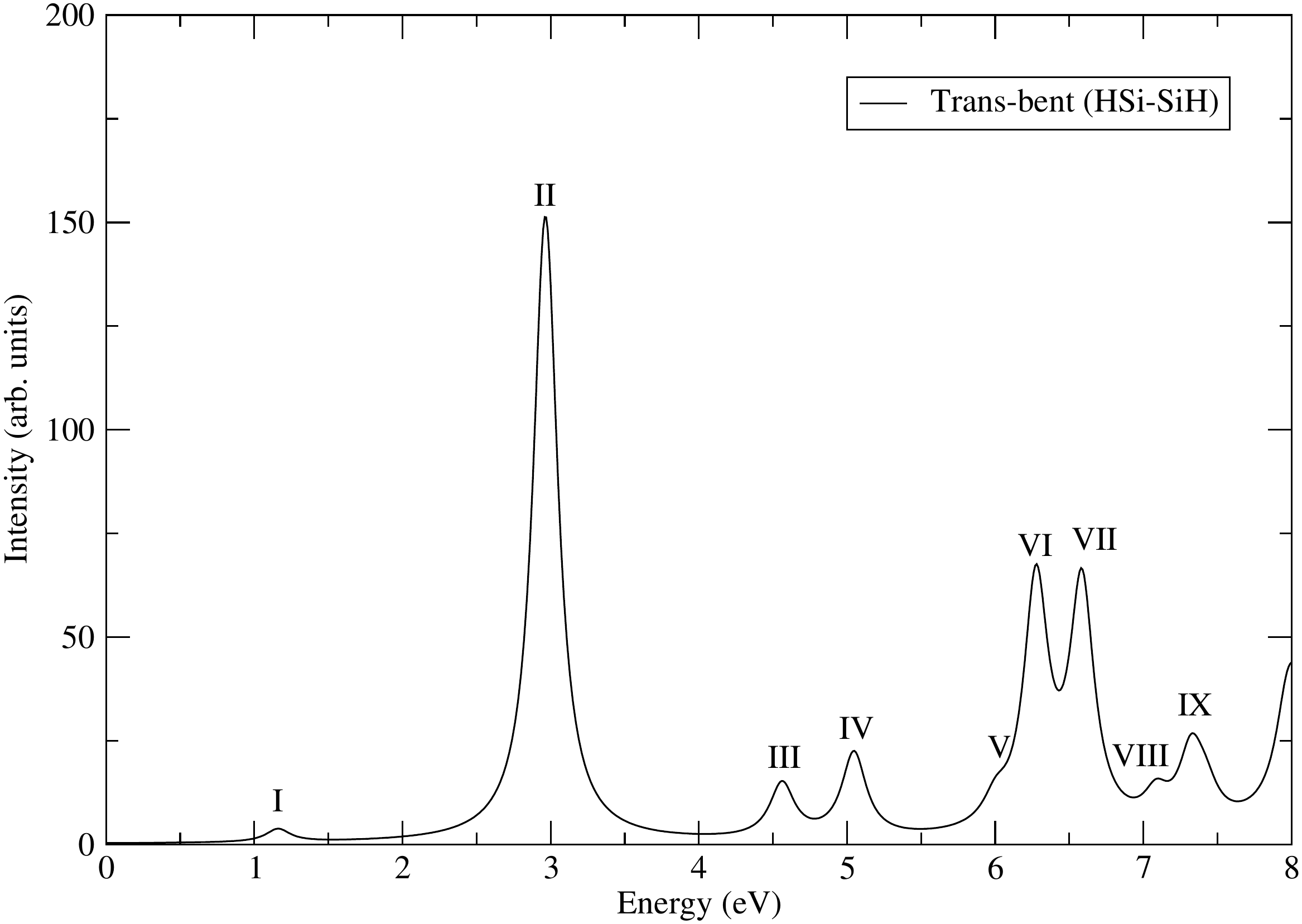}
\par\end{centering}
\caption{{\small{}Optical absorption spectrum of trans-bent (HSi-SiH) conformer
computed using the MRSDCI method and cc-pVTZ basis set. For plotting
the spectrum, a uniform line-width of 0.1 eV was used.\label{fig:Optics-trans-bent}}}
\end{figure}

Using our optimized geometry, the photoabsorption spectrum of the
trans-bent conformer computed using the MRSDCI approach, from its
$^{1}A_{g}$ ground state, is presented in \textcolor{black}{Fig.
\ref{fig:Optics-trans-bent}, }while the detailed information pertaining
to the excited states contributing to various peaks is given in Table
\textcolor{black}{S4} of Supporting Information. The optimized geometries
of a few excited states, corresponding to some important peaks in
absorption spectrum of this isomer, are presented in Fig. S13 of the
Supporting Information. Absorption in the trans-bent conformer starts
with a feeble peak at 1.16 eV, due to a state whose wave function
is dominated by the singly-excited configuration $\arrowvert H\rightarrow L\rangle$,
and the relaxed geometry of this state, unlike the ground state is
neither trans-like, nor planar. The H-atom on one side is out of the
plane, but both the Si-H bond length are still equal (1.49 $\text{Å}$),
whereas the Si-Si bond length is 2.12 $\text{Å}$. This is followed
by the most intense peak of the spectrum at 2.96 eV, due to a state
whose wave function exhibits strong mixing of singly-excited configurations
$\arrowvert H-1\rightarrow L\rangle$ and $\arrowvert H\rightarrow L+1\rangle$.
We note that the location of the most intense peak at a much lower
energy, as compared to the other three conformers, is enough of a
distinguishing feature of this conformer to allow its detection through
optical spectroscopy. However, the relaxed geometry corresponding
to this peak (peak II), is almost similar to the ground state structure,
with the Si-H and Si-Si bond distances of 1.52 $\textrm{\AA}$, and
2.33 $\textrm{\AA}$, respectively. It is followed by two relatively
weaker features III and IV at 4.56 eV, and 5.05 eV, respectively.
The wave functions of the state corresponding to peak III is dominated
by the single excitation $\arrowvert H-1\rightarrow L+1\rangle$,
while that of peak IV exhibits strong mixing of singly-excited configurations
$\arrowvert H\rightarrow L+1\rangle$ and $\arrowvert H-1\rightarrow L\rangle$.
Next three features V, VI, and VII located at 6.00 eV, 6.28 eV, and
6.59 eV, respectively, are due to states whose wave functions derive
strong contributions from doubly-excited configurations. Feature V,
which is a shoulder to peak VI is due to a state whose wave function
is largely composed of the double excitation $\arrowvert H\rightarrow L;\thinspace H\rightarrow L+2\rangle$,
while VI and VII exhibit strong mixing of double and single excitations
$\arrowvert H-1\rightarrow L;\thinspace H\rightarrow L+2\rangle$,
and $\arrowvert H\rightarrow L+3\rangle$. The computed spectrum has
its last peak at 7.35 eV due to two close lying states at 7.32 eV
and 7.42 eV, which exhibit strong mixing of double excitations $|H-2\rightarrow L;\thinspace H\rightarrow L\rangle$,
$\arrowvert H\rightarrow L+2;\thinspace H\rightarrow L+1\rangle$,
and $\arrowvert H-2\rightarrow L;\thinspace H-1\rightarrow L\rangle$,
$\arrowvert H-2\rightarrow L+1;\thinspace H\rightarrow L\rangle$,
respectively.

In Fig. S4 of Supporting Information, the plots of the frontier orbitals
participating in the optical absorption in this conformer are presented.

\subsection{Si$_{2}$H$_{4}$}

\begin{figure}[H]
\centering{}\subfloat[Disilene]{\begin{centering}
\includegraphics[scale=0.32]{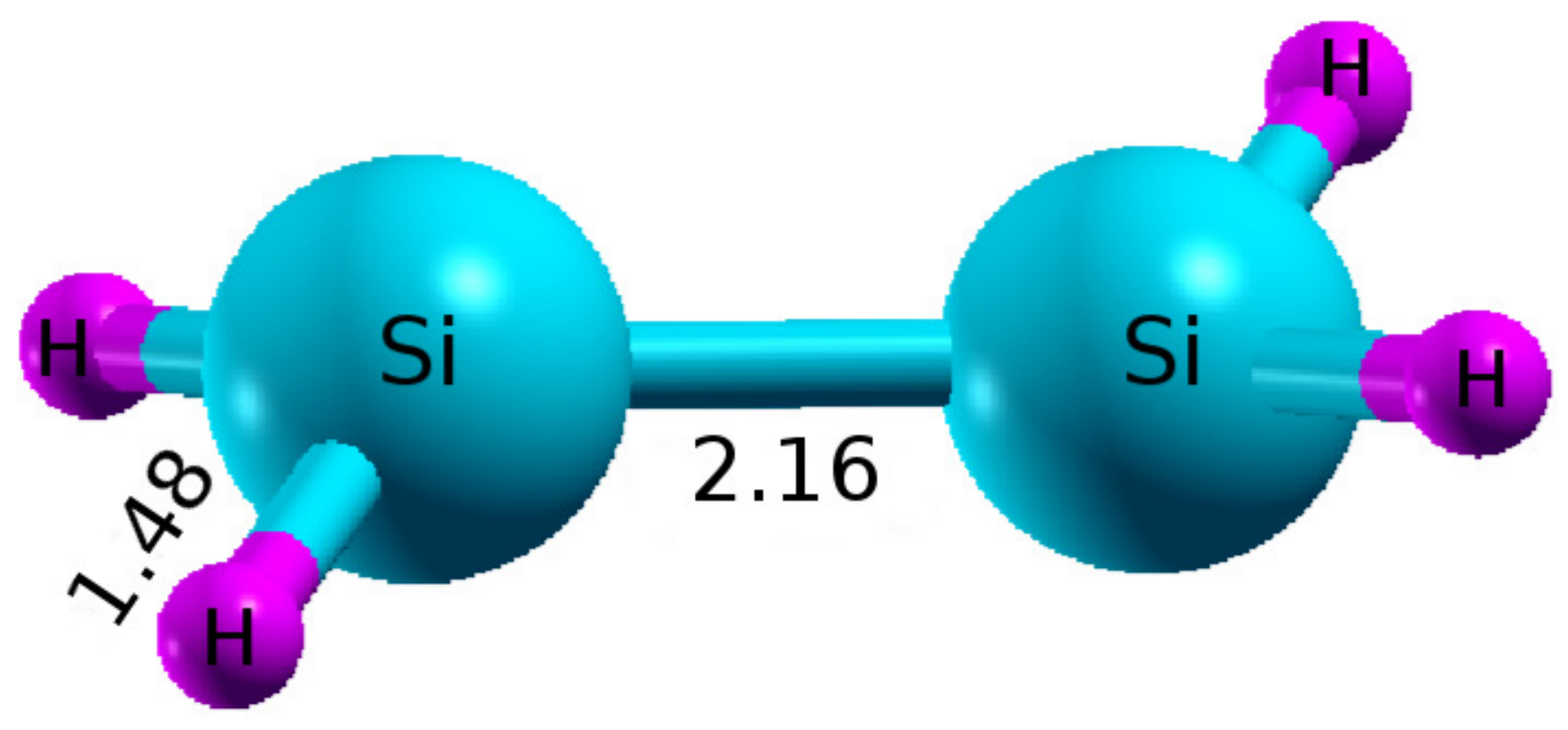}
\par\end{centering}
}\subfloat[Monobridged]{\begin{centering}
\includegraphics[scale=0.32]{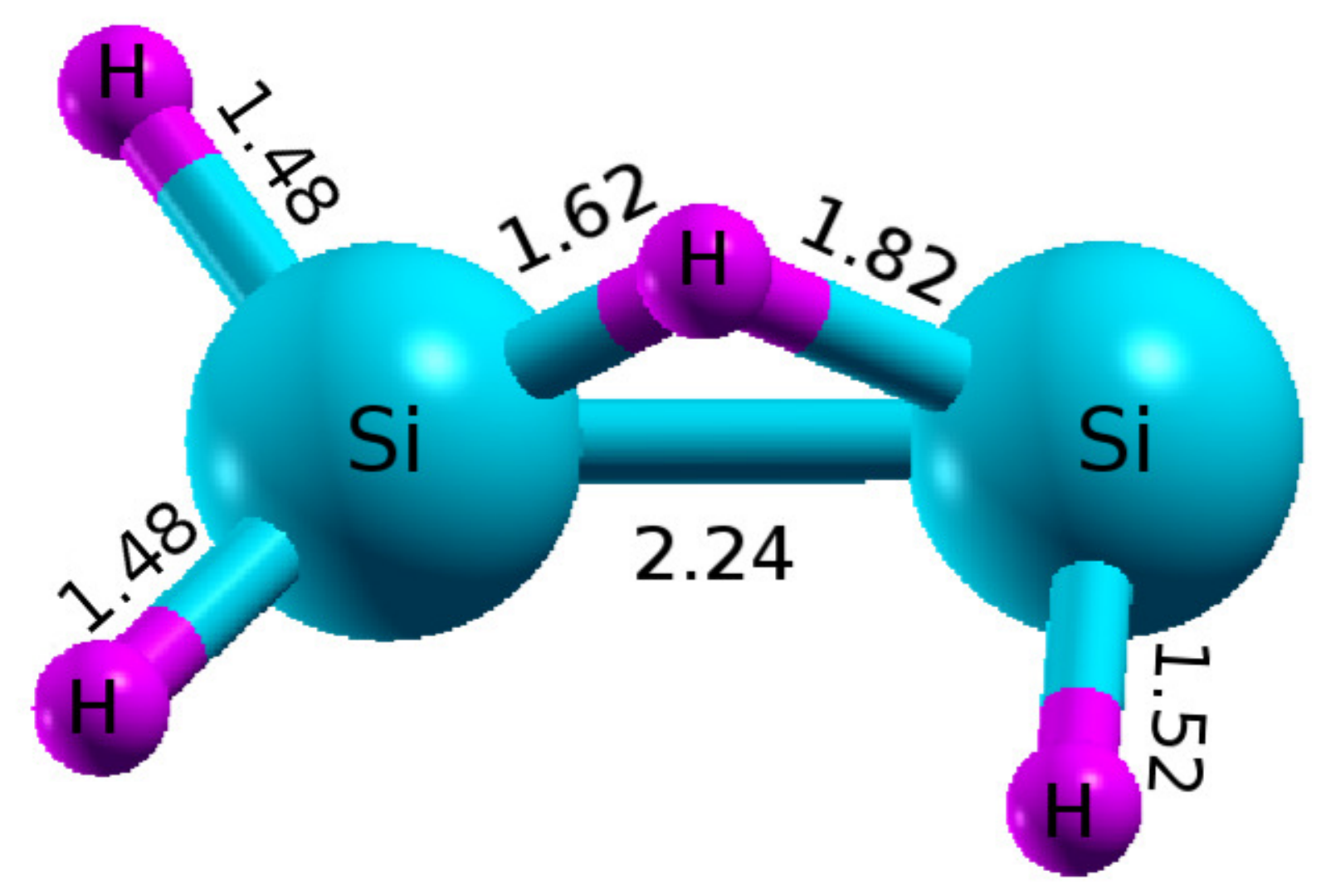}
\par\end{centering}
}\subfloat[Silylsilylene]{\begin{centering}
\includegraphics[scale=0.34]{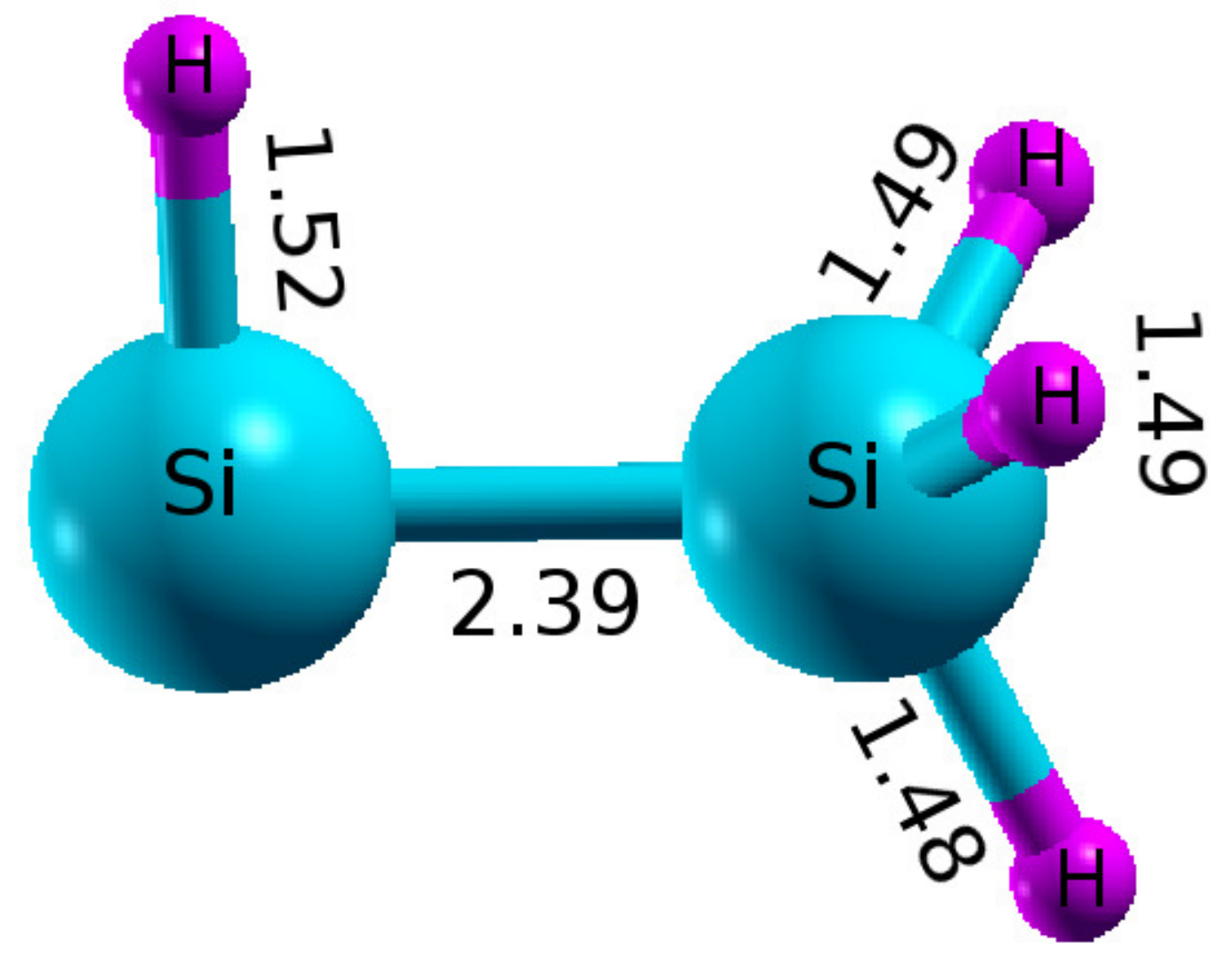}
\par\end{centering}
}\caption{{\small{}Ground state geometries of Si$_{2}$H$_{4}$ isomers, optimized
using the CCSD(T) method, and the cc-pVTZ basis set. All the bond
lengths are in $\textrm{\AA}$ unit.\label{fig:geometry-si2h4}}}
\end{figure}

Geometries and relative stabilities of the isomers of Si$_{2}$H$_{4}$
have been studied extensively over the years using electronic-structure
methods by several authors.{\small{}\cite{doi:10.1021/ja00386a006,LISCHKA1982467,doi:10.1021/j100206a004,Alexander_Sax,Gottfried_Olbrich,Trinquier,doi:10.1021/j100157a052,Pople,Sannigrahi_Nandi,Chaeho_Pak,Thaddeus,Dolgonos}}
Earlier, two main isomers of Si$_{2}$H$_{4}$ were investigated,
namely, disilene containing of a Si-Si double bond, and silylsilylene,
a completely singly-bonded structure.\cite{doi:10.1021/ja00386a006,LISCHKA1982467}
For disilene as well several structures are possible, but two main
structures have been investigated, namely: (a) planar ethylene like
structure consisting of a true Si-Si double bond, with $D_{2h}$ symmetry,
and (b) a trans-bent structure with a nominal Si-Si double bond, with
$C_{2h}$ symmetry.\cite{doi:10.1021/j100206a004,LISCHKA1982467,Alexander_Sax,Trinquier,Sannigrahi_Nandi}
However, it was soon revealed by calculations of vibrational frequencies,
that planar disilene was not stable, and actually corresponded to
a transition state on the potential energy surface.\cite{LISCHKA1982467,doi:10.1021/j100206a004,Trinquier}
Later on a new low-lying isomer with a monobridged structure was discovered
computationally, and found to be stable.\cite{Thaddeus,Monobridged_Si2H4,Dolgonos}
The present-day consensus is that the trans-bent disilene structure
is energetically the lowest, followed by mono-bridged, and silylsilylene
isomers, lying slightly above it.\cite{Gottfried_Olbrich,Thaddeus,Monobridged_Si2H4,Dolgonos}.
We performed our own geometry optimization for these three structures
using the CCSD(T) method, coupled with the cc-pVTZ basis, and the
optimized geometries are shown in Fig. \ref{fig:geometry-si2h4}.
Energetically speaking, we found the ordering trans-bent < mono-bridged
< silylsilylene, with mono-bridged and silylsilylene structures being
almost degenerate and just 0.286 eV and 0.294 eV above the trans-bent
structure (see Table \ref{tab:Ground-state-energies}). Thus, our
results on the geometries and energies of these isomers are in excellent
agreement with the results of recent calculations.\cite{Gottfried_Olbrich,Thaddeus,Monobridged_Si2H4,Dolgonos}
Next, we discuss the geometries and optical absorption spectra of
the individual isomers in the following sections.

\subsubsection{\emph{Disilene (H}\protect\textsubscript{\textbf{\emph{2}}}\emph{Si-SiH}\protect\textsubscript{\textbf{\emph{2}}}\emph{)}}

The trans-bent structure of disilene, with C\textsubscript{2h} point
group symmetry, has the lowest energy among all the Si\textsubscript{2}H\textsubscript{4}
isomers. The optimized geometry of this conformer has equal bond lengths
of 1.48 $\textrm{\AA}$ for the four Si-H bonds. Additionally, Si-Si
bond length was computed to be 2.16 $\textrm{\AA}$, along with the
H-Si-H, and H-Si-Si bond angles of 113$^{o}$, and 119$^{o}$, respectively,
whereas the HSi-SiH dihedral angle is 33$^{\circ}$(see Fig. \ref{fig:geometry-si2h4}(a)).
These values are in good agreement with the geometry parameters reported
in the literature.\cite{Gottfried_Olbrich,Thaddeus,Monobridged_Si2H4,Alexander_Sax,Chaeho_Pak}
The atomic coordinates corresponding to our optimized ground state
geometry of this conformer are presented in Table S13 of Supporting
information.

\begin{figure}[H]
\begin{centering}
\includegraphics[scale=0.35]{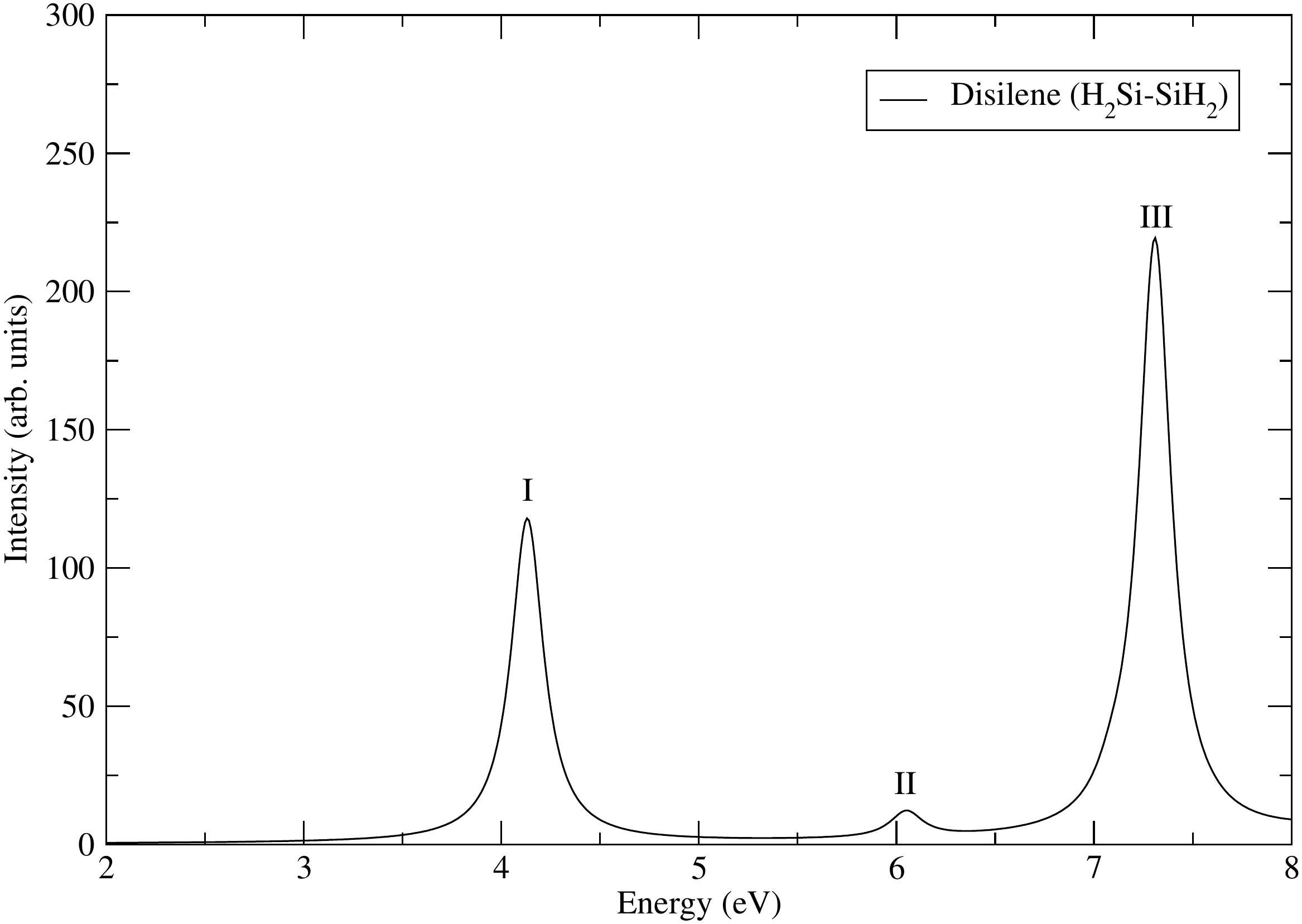}
\par\end{centering}
\caption{{\small{}Optical absorption spectrum of disilene (H$_{2}$Si-SiH$_{2}$)
conformer computed using the MRSDCI method, along with cc-pVTZ basis
set. For plotting the spectrum, a uniform line-width of 0.1 eV was
assumed.\label{fig:Optics-disilene}}}
\end{figure}

The calculated photoabsorption spectrum of the trans-bent disilene
(H\textsubscript{2}Si-SiH\textsubscript{2}) conformer is presented
in \textcolor{black}{Fig. \ref{fig:Optics-disilene},}\textcolor{red}{{}
}and the corresponding dominant many particle wave functions contributing
to the peaks of the optical spectra are given in Table \textcolor{black}{S5}
of the Supporting Information. The relaxed excited state geometries
corresponding to some of the frontier peaks of optical absorption
spectrum of this isomer is given in Fig. S14 of the Supporting Information.

The optical absorption spectrum of this disilene conformer starts
with a intense peak at 4.13 eV, corresponding to an excited state
whose wave function is dominated by singly-excited configuration $\arrowvert H\rightarrow L\rangle$.
As far as the relaxed geometry of this state is concerned, as compared
to the ground state Si-Si bond length becomes longer, around 2.49
$\text{Å}$, whereas the Si-H bond length does not change too much
(1.49 $\text{Å}$). But the dihedral angle (H-Si-Si-H) increases from
33$^{\circ}$ to 57$^{\circ}$ as compared to the ground state structure.
It is followed by a weak peak at 6.05 eV, due to an excited state
with wave function deriving dominant contributions from the singly-excited
configurations $\arrowvert H\rightarrow L+3\rangle$ and $\arrowvert H\rightarrow L+4\rangle$.
The most intense peak, with a very large oscillator strength, is located
at 7.31 eV, with many-particle wave function dominated again by same
set of single excitations $\arrowvert H\rightarrow L+3\rangle$ and
$\arrowvert H\rightarrow L+4\rangle$. In contrast to the ground state,
the optimized geometry of this state (peak III) is completely planar
with the Si-Si bond length 2.16 $\text{Å}$, and equal Si-H bond distances
of 1.47 $\text{Å}$. In Fig. S5 of Supporting Information, the plots
of the frontier orbitals participating in the optical absorption in
disilene are presented.

\subsubsection{\emph{Monobridged (H}\protect\textsubscript{\textbf{\emph{2}}}\emph{Si-H-SiH)}}

The ground state of monobridged conformer has C\textsubscript{1}
symmetry, and is predicted to be only 0.286 eV (see Table \ref{tab:Ground-state-energies})
higher than the ground state of disilene conformer of Si\textsubscript{2}H\textsubscript{4}.
Because of this close energetic proximity of the monobridged isomer
to disilene, their optical absorption spectra become important because
they offer the possibility of optical identification of these close-lying
isomers. Schaefer and co-workers\cite{Thaddeus}, by means of highly-correlated
ab \emph{initio} calculations, were the first ones to predict this
isomer. Later on, the same group\cite{Monobridged_Si2H4} detected
this isomer experimentally using Fourier transform microwave spectroscopy
technique. Our optimized geometry obtained using the CCSD(T) method,
and the ccpVTZ basis set, is shown in Fig. \ref{fig:geometry-si2h4}(b).
The optimized lengths were found to be 2.24 $\textrm{\AA}$ for Si-Si
bond, 1.48 $\textrm{\AA}$ for the two Si-H bonds on the left, and
1.52 $\textrm{\AA}$ for the Si-H bond on the right. On the left side,
the three H-Si-H bond angles are 103.1$^{o}$, 106.8$^{o}$ and 108.3$^{o}$,
while on the right, the H-Si-Si bond angle is 85.4$^{o}$. For the
bridged hydrogen atom, our optimized Si-H bond lengths are 1.62 $\textrm{\AA}$,
and 1.82 $\textrm{\AA}$. Inside the Si-H-Si triangle, the Si-H-Si
bond angle is 81.4$^{o}$, while the two H-Si-Si bond angles are 53.2$^{o}$,
and 45.4$^{o}$, respectively, from left to right. Additionally, H-Si-H
bond angle on the right is optimized to be 84.3$^{o}$. All the HSi-SiH
dihedral angles with the bridged H-atom are near about 85$^{\circ}$.
These results are in good agreement with the values reported by Schaefer
and coworkers.\cite{Thaddeus,Monobridged_Si2H4} The atomic coordinates
corresponding to our optimized ground state geometry of the monobridged
conformer are presented in Table S14 of Supporting information.

\begin{figure}[H]
\begin{centering}
\includegraphics[scale=0.35]{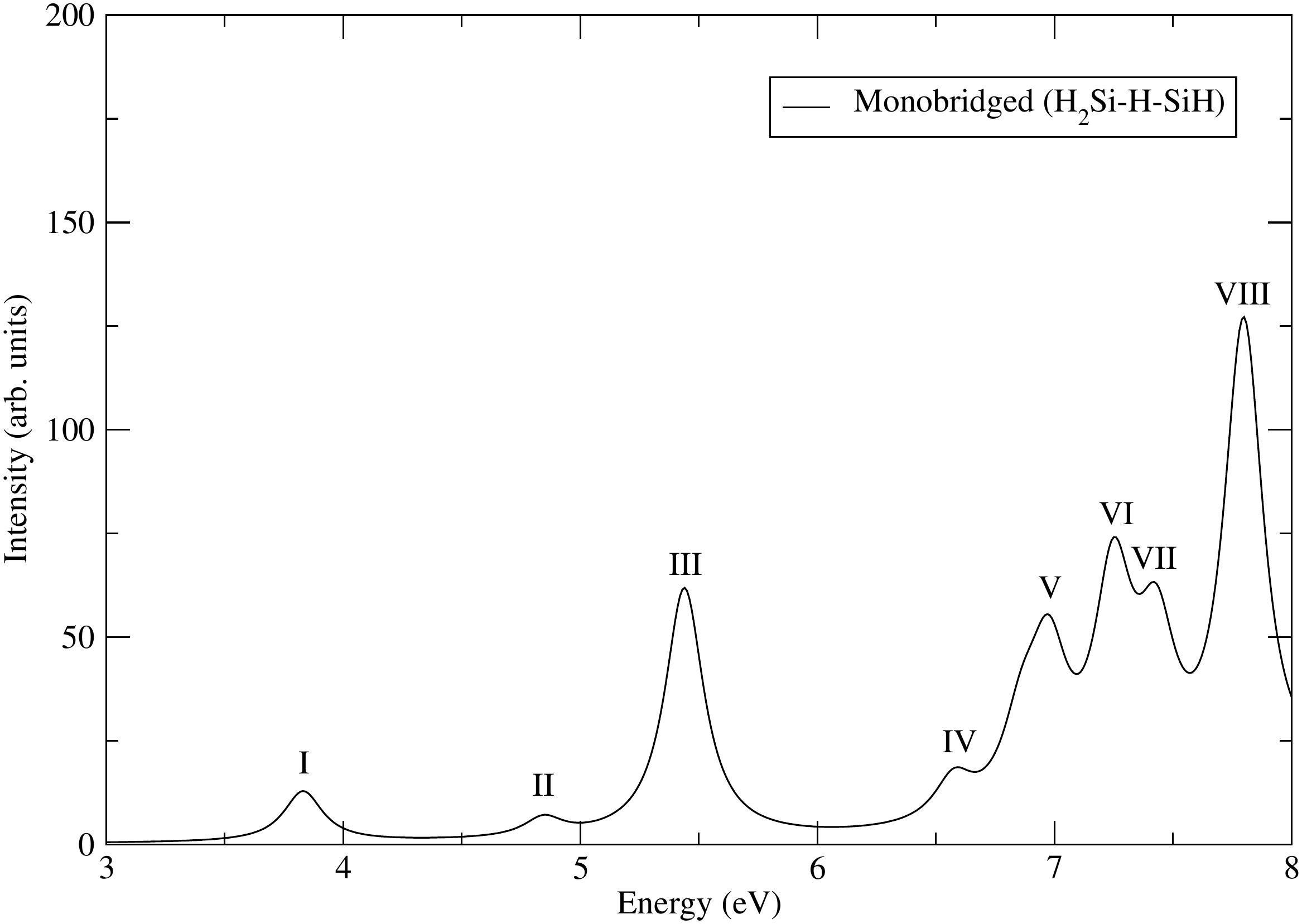}
\par\end{centering}
\caption{{\small{}Optical absorption spectrum of monobridged (H$_{2}$Si-H-SiH)
conformer computed using the MRSDCI method, and cc-pVTZ basis set.
For plotting the spectrum, a uniform line-width of 0.1 eV was assumed.\label{fig:Optics-monobridged}}}
\end{figure}

The calculated photoabsorption spectrum of the monobridged (H\textsubscript{2}Si-H-SiH)
conformer, using our optimized geometry, is presented in \textcolor{black}{Fig.
}\ref{fig:Optics-monobridged}, while the detailed information about
the excited states contributing to various peaks, including their
many particle wave functions are presented in Table\textcolor{black}{{}
S6} of the Supporting Information. The optimized geometries of excited
states contributing to some of the frontier peaks of optical absorption
spectrum are presented in Fig. S15 of the Supporting Information.

As compared to the disilene conformer, optical photoabsorption spectrum
of the monobridged isomer consists of several well-separated peaks
over the same energy range. It starts with a small peaks at 3.83 eV,
due to a state whose many-particle wave function is dominated by the
singly-excited configuration $\arrowvert H\rightarrow L\rangle$.
In the relaxed geometry of this excited state, the Si-Si bond length
(2.36 $\text{Å}$) is elongated a bit more compared to the ground
state. Furthermore, the geometry is qualitatively different as compared
to the ground state, with the bridged H-atom is now connected with
just one Si-atom. Consequently, one Si-atom is connected with three
H atoms, while the other one is connected to just one, like silylsilylene
but with all Si-H bond lengths being close to 1.48 $\text{Å}$. This
is followed by an even weaker peak at 4.84 eV, due to a state whose
wave function derives its main contribution from the configuration
$\arrowvert H-1\rightarrow L\rangle$, and whose optimized geometry
is also similar to a silylsilylene-type structure. The Si-Si bond
length is 2.45 $\text{Å}$, but with equal Si-H bond lengths on the
end where three H atoms are bonded. However, the Si-H bond where the
Si-atom is bonded to only one H-atom, is comparatively longer (1.60
$\text{Å}$).\textcolor{cyan}{{} }A comparatively intense peak appears
at 5.44 eV, with the wave function of the excited state dominated
by the singly-excited configuration $\arrowvert H\rightarrow L+1\rangle$.
The relaxed geometry of this excited state is almost similar to that
of the first one, with same Si-Si and Si-H bond lengths, but with
totally different HSi-SiH dihedral angles. This is followed by a series
of five well-separated peaks ranging from 6.58 eV to the most intense
peak (peak VIII) of the spectrum located at 7.80 eV. Various properties
of the excited states, along with their wave functions for this conformer
are presented in Table S6 of the Supporting Information, from where
it is obvious that the wave functions of all the states are dominated
by singly-excited configurations, except for peak VII whose wave function
is dominated by a configuration consisting of HOMO-LUMO double excitation
$|H\rightarrow L;H\rightarrow L\rangle$. The wave function of the
excited state giving rise to the most intense peak VIII, is dominated
by single excitations $|H\rightarrow L+4\rangle$, and $|H-3\rightarrow L\rangle$.
If we compare the absorption spectrum of the monobridged isomer with
that of disilene, we note that they are sufficiently different both
in terms of the number of peaks, and their locations, such that the
optical spectroscopy can be used for their identification. The frontier
MOs participating in various optical excitations of this conformer
are presented in Fig. S6 of the Supporting Information.

\subsubsection{Silylsilylene (H\protect\textsubscript{\textbf{\textit{3}}}Si-SiH)}

Similar to the case of disilene, over the years extensive research
has been done on the electronic structure and geometry of silylsilylene
(H$_{3}$Si-SiH) conformer of Si$_{2}$H$_{4}$.\cite{Monobridged_Si2H4,Andrew_and_Linda}
Our optimized geometrical parameters of its ground state of C\textsubscript{s}
symmetry, obtained using the CCSD(T) approach, and cc-pVTZ basis set
are: (a) Si-Si bond length 2.39 $\textrm{\AA}$, and (b) Si-H bond
lengths of 1.49 $\textrm{\AA}$, 1.49 $\textrm{\AA}$, and 1.48 $\textrm{\AA}$
on the right side, with H-Si-H bond angles 109.5$^{o}$, 109.5$^{o}$
and 113.9$^{o}$. On the left side optimized Si-H bond length is 1.52
$\textrm{\AA}$, along with the H-Si-Si bond angle 89.2$^{o}$, whereas
the HSi-SiH dihedral angles are 59.2$^{\circ}$, 59.2$^{\circ}$,
and 180$^{\circ}$, respectively (see Fig. \ref{fig:geometry-si2h4}(c)).
These values are in good agreement with those reported by McCarthy
et al,\cite{Monobridged_Si2H4} and Adamczyk et al.\cite{Andrew_and_Linda}
The atomic coordinates corresponding to our optimized ground state
geometry of this conformer are presented in Table S15 of Supporting
information. 

Energetically speaking, silysilylene is just 0.294 eV higher than
disilene, and a negligible 0.008 eV higher than the monobridged isomer.
Again, the question arises, whether their optical absorption spectra
are sufficiently different so as to allow their identification using
this spectroscopy.

\begin{figure}[H]
\begin{centering}
\includegraphics[scale=0.35]{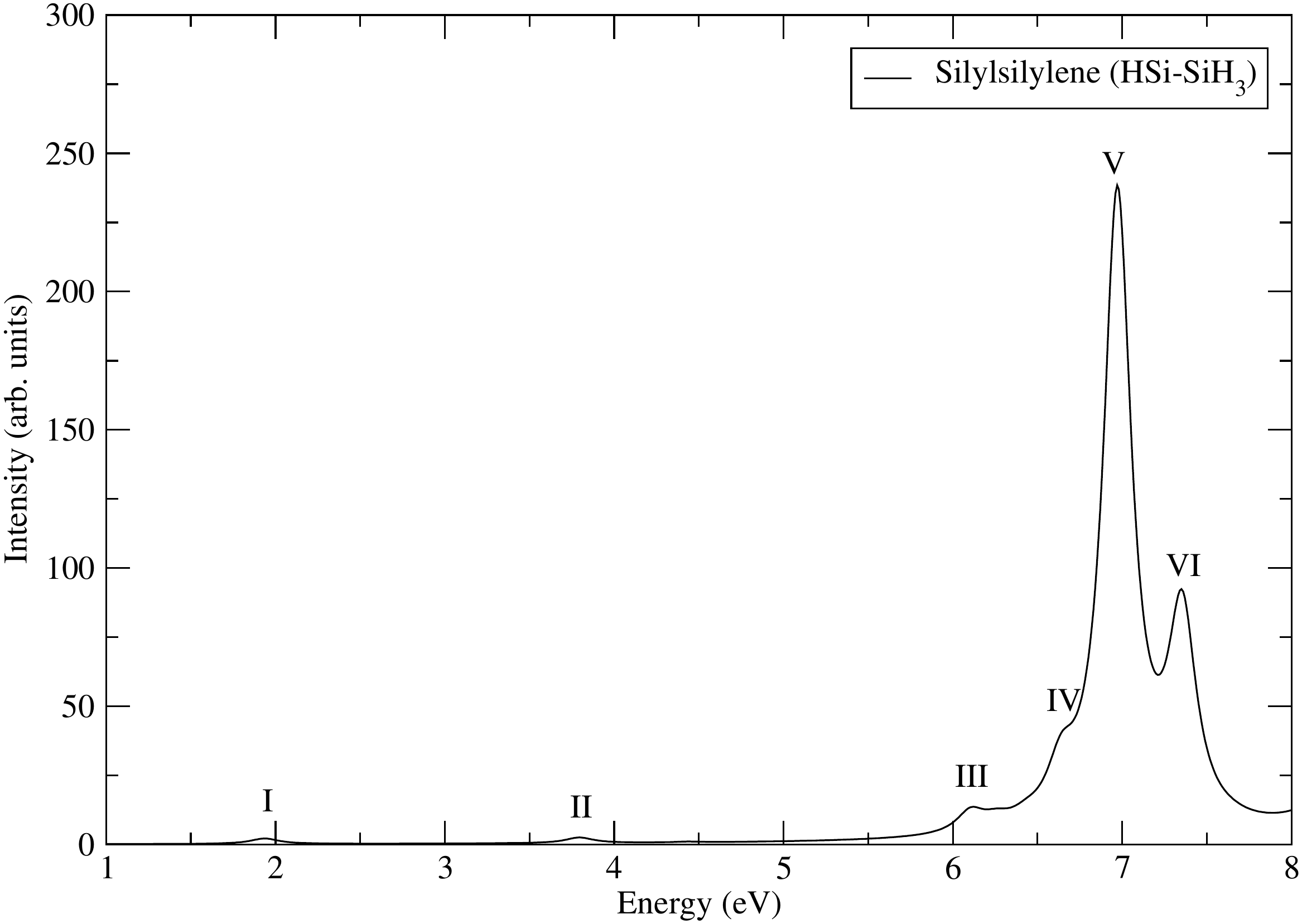}
\par\end{centering}
\caption{{\small{}Optical absorption spectrum of silylsilylene (H$_{3}$Si-SiH)
conformer computed using the MRSDCI method, and cc-pVTZ basis set.
For plotting the spectrum, a uniform line-width of 0.1 eV was used.}\label{fig:Optics-silylsilylene}}
\end{figure}

Using our optimized geometry, we present the calculated photoabsorption
spectrum of silylsilylene in \textcolor{black}{Fig. \ref{fig:Optics-silylsilylene},}
and the optimized excited state geometries corresponding to some of
the peaks of optical absorption spectrum of this isomer are given
in Fig. S16 of the Supporting Information. \textcolor{black}{The most
noteworthy point is that this confomer, except for a couple of very
feeble peaks, does not exhibit any significant absorption till about
6 eV, beyond which the intensity of the absorption rises. This aspect
of its absorption spectrum distinguishes it from those of the two
previous conformers, and can be used as a fingerprint for its optical
detection. }

The absorption spectrum of silylsilene conformer starts with very
weak peaks located at 1.93 eV, and 3.79 eV, due to excited states
whose many-particle wave functions are dominated by the singly-excited
configurations $\arrowvert H\rightarrow L\rangle$, and $\arrowvert H-1\rightarrow L\rangle$,
respectively. The optimized geometry corresponding to peak I has all
Si-H bond lengths are close to 1.48 $\text{Å}$, whereas the Si-Si
bond length is reduced to 2.36 $\text{Å}$. The most intense peak
(peak V) appears at 6.97 eV, due to a state whose wave functions are
dominated by single excitations $\arrowvert H\rightarrow L+1\rangle$,
and $\arrowvert H\rightarrow L+2\rangle$. In the relaxed geometry
of the excited state corresponding to the most intense peak (peak
V), all the Si-H bond lengths of the right side are 1.48 $\text{Å}$,
but the Si-Si bond length, and the Si-H bond distance of the left
side are elongated to 3.06 $\text{Å}$, and 1.56 $\text{Å}$, respectively.
The last peak of our calculated spectrum (peak VI) is located at 7.35
eV, and is due to a state with wave function deriving main contribution
from $\arrowvert H\rightarrow L+3\rangle$ configuration, with a smaller
contribution from the single excitation $|H-1\rightarrow L+1\rangle$.
Detailed information about all the excited states contributing to
various peaks in the absorption spectrum of silylsilylene is presented
in Table S7 of the Supporting Information, whose Fig. S7 contains
plots of MOs participating in the absorption.

\section{Conclusions}

\label{sec:Conclusions}

In this work we presented state of the art quantum chemical calculations,
utilizing large basis sets, aimed at obtaining electronic structure,
geometrical, and optical properties of hydrides of silicon dimer of
the class Si$_{2}$H\textsubscript{$2n$, }$n=1,2,3$. Geometry optimization
was carried out using the CCSD(T) approach, while the optical absorption
spectra was computed by means of large-scale MRSDCI calculations.
For Si$_{2}$H$_{2}$ and Si$_{2}$H$_{4}$ several stable, energetically
low-lying, and very closely spaced, conformers were considered, and
it was discovered that each conformer has a distinct optical absorption
spectrum, suggesting their possible detection and identification using
absorption spectroscopy. Although, no experimental results exist on
the absorption spectra of Si$_{2}$H$_{2}$ and Si$_{2}$H$_{4}$
conformers, but for disilane (Si$_{2}$H$_{6}$) excellent agreement
with experimental measurement was found. This suggests that our computational
methodology based on the MRSDCI method is sound, and, therefore, our
calculated spectra for Si$_{2}$H$_{2}$ and Si$_{2}$H$_{4}$ conformers
must be trustworthy.

Furthermore, we also obtained the relaxed geometries of several important
optically excited states of various clusters, and found that optical
absorption leads to significant geometry changes in these clusters,
as compared to their ground state. We hope that there will be future
experimental efforts to measure the absorption spectra of conformers
of Si$_{2}$H$_{2}$ and Si$_{2}$H$_{4}$, against\textcolor{red}{{}
}which our results could be benchmarked. 

\section*{Supporting Information}

The supporting information consists of a file containing important
information about the excited states contributing to peaks in our
computed absorption spectra of various conformers of Si$_{2}$H$_{2}$,
Si$_{2}$H$_{4}$, and Si$_{2}$H$_{6}$. It consists of plots of
frontier molecular orbitals, excitation energies, oscillator strengths,
dominant terms in their many-body wave functions, polarization directions
of absorbed photons, and the excited state geometries corresponding
to some of the peaks of optical absorption spectra.

\section*{Author Information }

\subsection*{Corresponding Authors}

Alok Shukla:  {*}E-mail: shukla@phy.iitb.ac.in

\subsection*{Notes}

The authors declare no competing financial interests.

\section*{Acknowledgements}

Work of P.B. was supported by a Senior Research Fellowship offered
by University Grants Commission, India.

\section*{TOC Graphic}

\begin{figure}[H]
\includegraphics[width=8.5cm,height=4.75cm]{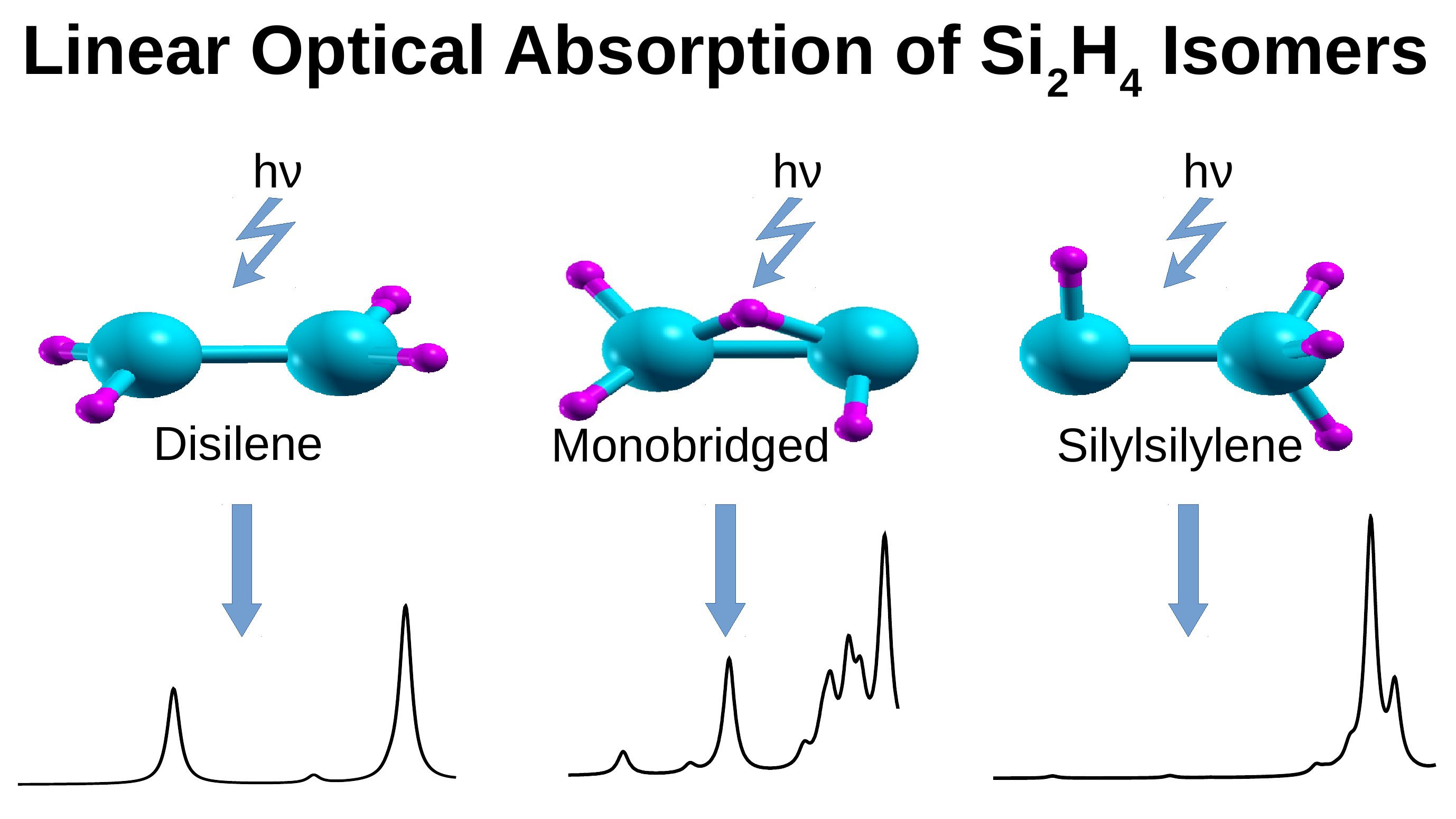}
\end{figure}

\bibliographystyle{achemso}
\addcontentsline{toc}{section}{\refname}\bibliography{sih}

\providecommand{\latin}[1]{#1}
\makeatletter
\providecommand{\doi}
  {\begingroup\let\do\@makeother\dospecials
  \catcode`\{=1 \catcode`\}=2 \doi@aux}
\providecommand{\doi@aux}[1]{\endgroup\texttt{#1}}
\makeatother
\providecommand*\mcitethebibliography{\thebibliography}
\csname @ifundefined\endcsname{endmcitethebibliography}
  {\let\endmcitethebibliography\endthebibliography}{}
\begin{mcitethebibliography}{83}
\providecommand*\natexlab[1]{#1}
\providecommand*\mciteSetBstSublistMode[1]{}
\providecommand*\mciteSetBstMaxWidthForm[2]{}
\providecommand*\mciteBstWouldAddEndPuncttrue
  {\def\EndOfBibitem{\unskip.}}
\providecommand*\mciteBstWouldAddEndPunctfalse
  {\let\EndOfBibitem\relax}
\providecommand*\mciteSetBstMidEndSepPunct[3]{}
\providecommand*\mciteSetBstSublistLabelBeginEnd[3]{}
\providecommand*\EndOfBibitem{}
\mciteSetBstSublistMode{f}
\mciteSetBstMaxWidthForm{subitem}{(\alph{mcitesubitemcount})}
\mciteSetBstSublistLabelBeginEnd
  {\mcitemaxwidthsubitemform\space}
  {\relax}
  {\relax}

\bibitem[Zhao \latin{et~al.}(2016)Zhao, Liu, Yu, Quhe, Zhou, Wang, Liu, Zhong,
  Han, Lu, Yao, and Wu]{Silicene1}
Zhao,~J.; Liu,~H.; Yu,~Z.; Quhe,~R.; Zhou,~S.; Wang,~Y.; Liu,~C.~C.; Zhong,~H.;
  Han,~N.; Lu,~J.; Yao,~Y.; Wu,~K. Rise of silicene: A competitive 2D material.
  \emph{Progress in Materials Science} \textbf{2016}, \emph{83}, 24 --
  151\relax
\mciteBstWouldAddEndPuncttrue
\mciteSetBstMidEndSepPunct{\mcitedefaultmidpunct}
{\mcitedefaultendpunct}{\mcitedefaultseppunct}\relax
\EndOfBibitem
\bibitem[Vogt \latin{et~al.}(2012)Vogt, De~Padova, Quaresima, Avila,
  Frantzeskakis, Asensio, Resta, Ealet, and Le~Lay]{Silicene2}
Vogt,~P.; De~Padova,~P.; Quaresima,~C.; Avila,~J.; Frantzeskakis,~E.;
  Asensio,~M.~C.; Resta,~A.; Ealet,~B.; Le~Lay,~G. Silicene: Compelling
  Experimental Evidence for Graphenelike Two-Dimensional Silicon. \emph{Phys.
  Rev. Lett.} \textbf{2012}, \emph{108}, 155501\relax
\mciteBstWouldAddEndPuncttrue
\mciteSetBstMidEndSepPunct{\mcitedefaultmidpunct}
{\mcitedefaultendpunct}{\mcitedefaultseppunct}\relax
\EndOfBibitem
\bibitem[Ni \latin{et~al.}(2012)Ni, Liu, Tang, Zheng, Zhou, Qin, Gao, Yu, and
  Lu]{Silicene3}
Ni,~Z.; Liu,~Q.; Tang,~K.; Zheng,~J.; Zhou,~J.; Qin,~R.; Gao,~Z.; Yu,~D.;
  Lu,~J. Tunable Bandgap in Silicene and Germanene. \emph{Nano Letters}
  \textbf{2012}, \emph{12}, 113--118, PMID: 22050667\relax
\mciteBstWouldAddEndPuncttrue
\mciteSetBstMidEndSepPunct{\mcitedefaultmidpunct}
{\mcitedefaultendpunct}{\mcitedefaultseppunct}\relax
\EndOfBibitem
\bibitem[Jose and Datta(2014)Jose, and Datta]{Silicene4}
Jose,~D.; Datta,~A. Structures and Chemical Properties of Silicene: Unlike
  Graphene. \emph{Accounts of Chemical Research} \textbf{2014}, \emph{47},
  593--602, PMID: 24215179\relax
\mciteBstWouldAddEndPuncttrue
\mciteSetBstMidEndSepPunct{\mcitedefaultmidpunct}
{\mcitedefaultendpunct}{\mcitedefaultseppunct}\relax
\EndOfBibitem
\bibitem[Kingon \latin{et~al.}(2000)Kingon, Maria, and Streiffer]{Review_1}
Kingon,~A.~I.; Maria,~J.-P.; Streiffer,~S.~K. Alternative dielectrics to
  silicon dioxide for memory and logic devices. \emph{Nature} \textbf{2000},
  \emph{406}, 1032--1038\relax
\mciteBstWouldAddEndPuncttrue
\mciteSetBstMidEndSepPunct{\mcitedefaultmidpunct}
{\mcitedefaultendpunct}{\mcitedefaultseppunct}\relax
\EndOfBibitem
\bibitem[Thomson \latin{et~al.}(2016)Thomson, Zilkie, Bowers, Komljenovic,
  Reed, Vivien, Marris-Morini, Cassan, Virot, Fedeli, Hartmann, Schmid, Xu,
  Boeuf, O'Brien, Mashanovich, and Nedeljkovic]{Review_2}
Thomson,~D. \latin{et~al.}  Roadmap on silicon photonics. \emph{Journal of
  Optics} \textbf{2016}, \emph{18}, 073003\relax
\mciteBstWouldAddEndPuncttrue
\mciteSetBstMidEndSepPunct{\mcitedefaultmidpunct}
{\mcitedefaultendpunct}{\mcitedefaultseppunct}\relax
\EndOfBibitem
\bibitem[Aberle(2000)]{Review_3}
Aberle,~A.~G. Surface passivation of crystalline silicon solar cells: a review.
  \emph{Progress in Photovoltaics: Research and Applications} \textbf{2000},
  \emph{8}, 473--487\relax
\mciteBstWouldAddEndPuncttrue
\mciteSetBstMidEndSepPunct{\mcitedefaultmidpunct}
{\mcitedefaultendpunct}{\mcitedefaultseppunct}\relax
\EndOfBibitem
\bibitem[Stefaan \latin{et~al.}(2012)Stefaan, Antoine, C., and
  Christophe]{Review_4}
Stefaan,~D.~W.; Antoine,~D.; C.,~H.~Z.; Christophe,~B. High-efficiency Silicon
  Heterojunction Solar Cells: A Review. \emph{green} \textbf{2012}, \emph{2},
  7--24\relax
\mciteBstWouldAddEndPuncttrue
\mciteSetBstMidEndSepPunct{\mcitedefaultmidpunct}
{\mcitedefaultendpunct}{\mcitedefaultseppunct}\relax
\EndOfBibitem
\bibitem[Tiedje \latin{et~al.}(1981)Tiedje, Cebulka, Morel, and Abeles]{sih_1}
Tiedje,~T.; Cebulka,~J.~M.; Morel,~D.~L.; Abeles,~B. Evidence for Exponential
  Band Tails in Amorphous Silicon Hydride. \emph{Phys. Rev. Lett.}
  \textbf{1981}, \emph{46}, 1425--1428\relax
\mciteBstWouldAddEndPuncttrue
\mciteSetBstMidEndSepPunct{\mcitedefaultmidpunct}
{\mcitedefaultendpunct}{\mcitedefaultseppunct}\relax
\EndOfBibitem
\bibitem[Tsai \latin{et~al.}(1992)Tsai, Li, Kinosky, Qian, Hsu, Irby, Banerjee,
  Tasch, Campbell, Hance, and White]{sih_2}
Tsai,~C.; Li,~K.; Kinosky,~D.~S.; Qian,~R.; Hsu,~T.; Irby,~J.~T.;
  Banerjee,~S.~K.; Tasch,~A.~F.; Campbell,~J.~C.; Hance,~B.~K.; White,~J.~M.
  Correlation between silicon hydride species and the photoluminescence
  intensity of porous silicon. \emph{Applied Physics Letters} \textbf{1992},
  \emph{60}, 1700--1702\relax
\mciteBstWouldAddEndPuncttrue
\mciteSetBstMidEndSepPunct{\mcitedefaultmidpunct}
{\mcitedefaultendpunct}{\mcitedefaultseppunct}\relax
\EndOfBibitem
\bibitem[Cody \latin{et~al.}(1980)Cody, Wronski, Abeles, Stephens, and
  Brooks]{sih_3}
Cody,~G.; Wronski,~C.; Abeles,~B.; Stephens,~R.; Brooks,~B. Optical
  characterization of amorphous silicon hydride films. \emph{Solar Cells}
  \textbf{1980}, \emph{2}, 227 -- 243\relax
\mciteBstWouldAddEndPuncttrue
\mciteSetBstMidEndSepPunct{\mcitedefaultmidpunct}
{\mcitedefaultendpunct}{\mcitedefaultseppunct}\relax
\EndOfBibitem
\bibitem[Gates \latin{et~al.}(1989)Gates, Kunz, and Greenlief]{sih_4}
Gates,~S.~M.; Kunz,~R.~R.; Greenlief,~C. Silicon hydride etch products from the
  reaction of atomic hydrogen with Si(100). \emph{Surface Science}
  \textbf{1989}, \emph{207}, 364 -- 384\relax
\mciteBstWouldAddEndPuncttrue
\mciteSetBstMidEndSepPunct{\mcitedefaultmidpunct}
{\mcitedefaultendpunct}{\mcitedefaultseppunct}\relax
\EndOfBibitem
\bibitem[Swihart and Girshick(1999)Swihart, and Girshick]{sih_5}
Swihart,~M.~T.; Girshick,~S.~L. Thermochemistry and Kinetics of Silicon Hydride
  Cluster Formation during Thermal Decomposition of Silane. \emph{The Journal
  of Physical Chemistry B} \textbf{1999}, \emph{103}, 64--76\relax
\mciteBstWouldAddEndPuncttrue
\mciteSetBstMidEndSepPunct{\mcitedefaultmidpunct}
{\mcitedefaultendpunct}{\mcitedefaultseppunct}\relax
\EndOfBibitem
\bibitem[Ring and O'Neal(1992)Ring, and O'Neal]{doi:10.1021/j100205a046}
Ring,~M.~A.; O'Neal,~H.~E. Mechanism of the thermally induced gas-phase
  decomposition of silane: a revisitation. \emph{The Journal of Physical
  Chemistry} \textbf{1992}, \emph{96}, 10848--10855\relax
\mciteBstWouldAddEndPuncttrue
\mciteSetBstMidEndSepPunct{\mcitedefaultmidpunct}
{\mcitedefaultendpunct}{\mcitedefaultseppunct}\relax
\EndOfBibitem
\bibitem[Collins and Ferlauto(2002)Collins, and Ferlauto]{si-cvd}
Collins,~R.; Ferlauto,~A. Advances in plasma-enhanced chemical vapor deposition
  of silicon films at low temperatures. \emph{Current Opinion in Solid State
  and Materials Science} \textbf{2002}, \emph{6}, 425 -- 437\relax
\mciteBstWouldAddEndPuncttrue
\mciteSetBstMidEndSepPunct{\mcitedefaultmidpunct}
{\mcitedefaultendpunct}{\mcitedefaultseppunct}\relax
\EndOfBibitem
\bibitem[Kenichi~Tonokura and Koshi(2002)Kenichi~Tonokura, and
  Koshi]{doi:10.1021/jp015523n}
Kenichi~Tonokura,~T.~M.; Koshi,~M. Formation Mechanism of Hydrogenated Silicon
  Clusters during Thermal Decomposition of Disilane. \emph{The Journal of
  Physical Chemistry B} \textbf{2002}, \emph{106}, 555--563\relax
\mciteBstWouldAddEndPuncttrue
\mciteSetBstMidEndSepPunct{\mcitedefaultmidpunct}
{\mcitedefaultendpunct}{\mcitedefaultseppunct}\relax
\EndOfBibitem
\bibitem[Stuckelberger \latin{et~al.}(2017)Stuckelberger, Biron, Wyrsch, Haug,
  and Ballif]{hydrogenated-a-si-solar-cells-review}
Stuckelberger,~M.; Biron,~R.; Wyrsch,~N.; Haug,~F.-J.; Ballif,~C. Review:
  Progress in solar cells from hydrogenated amorphous silicon. \emph{Renewable
  and Sustainable Energy Reviews} \textbf{2017}, \emph{76}, 1497 -- 1523\relax
\mciteBstWouldAddEndPuncttrue
\mciteSetBstMidEndSepPunct{\mcitedefaultmidpunct}
{\mcitedefaultendpunct}{\mcitedefaultseppunct}\relax
\EndOfBibitem
\bibitem[Wei \latin{et~al.}(2017)Wei, Dai, and Huang]{silicene-hydrogenation1}
Wei,~W.; Dai,~Y.; Huang,~B. Hydrogenation of silicene on Ag(111) and formation
  of half-silicane. \emph{J. Mater. Chem. A} \textbf{2017}, \emph{5},
  18128--18137\relax
\mciteBstWouldAddEndPuncttrue
\mciteSetBstMidEndSepPunct{\mcitedefaultmidpunct}
{\mcitedefaultendpunct}{\mcitedefaultseppunct}\relax
\EndOfBibitem
\bibitem[Medina \latin{et~al.}(2017)Medina, Salomon, Lay, and
  Angot]{silicene-hydrogenation2}
Medina,~D.~B.; Salomon,~E.; Lay,~G.~L.; Angot,~T. Hydrogenation of silicene
  films grown on Ag(111). \emph{Journal of Electron Spectroscopy and Related
  Phenomena} \textbf{2017}, \emph{219}, 57 -- 62, SI: The electronic structure
  of 2D and layered materials\relax
\mciteBstWouldAddEndPuncttrue
\mciteSetBstMidEndSepPunct{\mcitedefaultmidpunct}
{\mcitedefaultendpunct}{\mcitedefaultseppunct}\relax
\EndOfBibitem
\bibitem[Mironov and Nepomnina(1961)Mironov, and Nepomnina]{sih_expt_1}
Mironov,~V.~F.; Nepomnina,~V.~V. The synthesis of alkenylsilanes by the
  high-temperature condensation of unsaturated compounds with silicon hydrides.
  \emph{Bulletin of the Academy of Sciences of the USSR, Division of chemical
  science} \textbf{1961}, \emph{10}, 1759--1761\relax
\mciteBstWouldAddEndPuncttrue
\mciteSetBstMidEndSepPunct{\mcitedefaultmidpunct}
{\mcitedefaultendpunct}{\mcitedefaultseppunct}\relax
\EndOfBibitem
\bibitem[Speier \latin{et~al.}(1957)Speier, Webster, and Barnes]{sih_expt_2}
Speier,~J.~L.; Webster,~J.~A.; Barnes,~G.~H. The Addition of Silicon Hydrides
  to Olefinic Double Bonds. Part II. The Use of Group VIII Metal Catalysts.
  \emph{Journal of the American Chemical Society} \textbf{1957}, \emph{79},
  974--979\relax
\mciteBstWouldAddEndPuncttrue
\mciteSetBstMidEndSepPunct{\mcitedefaultmidpunct}
{\mcitedefaultendpunct}{\mcitedefaultseppunct}\relax
\EndOfBibitem
\bibitem[Andrews and Wang(2002)Andrews, and Wang]{Wang}
Andrews,~L.; Wang,~X. Infrared Spectra of the Novel Si2H2 and Si2H4 Species and
  the SiH1,2,3 Intermediates in Solid Neon, Argon, and Deuterium. \emph{The
  Journal of Physical Chemistry A} \textbf{2002}, \emph{106}, 7696--7702\relax
\mciteBstWouldAddEndPuncttrue
\mciteSetBstMidEndSepPunct{\mcitedefaultmidpunct}
{\mcitedefaultendpunct}{\mcitedefaultseppunct}\relax
\EndOfBibitem
\bibitem[B. and J.(1991)B., and
  J.]{:/content/aip/journal/jcp/95/4/10.1063/1.460947}
B.,~R.; J.,~B. Photoionization mass spectrometric studies of the transient
  species Si2Hn (n=2-5). \emph{The Journal of Chemical Physics} \textbf{1991},
  \emph{95}, 2416--2432\relax
\mciteBstWouldAddEndPuncttrue
\mciteSetBstMidEndSepPunct{\mcitedefaultmidpunct}
{\mcitedefaultendpunct}{\mcitedefaultseppunct}\relax
\EndOfBibitem
\bibitem[C. and L.(1991)C., and L.]{PhysRevLett.66.413}
C.,~B. M. B. H.~D.; L.,~D.~J. Nonclassical double-bridged structure in
  silicon-containing molecules: Experimental evidence in
  ${\mathrm{Si}}_{2}$${\mathrm{H}}_{2}$ from its submillimeter-wave spectrum.
  \emph{Phys. Rev. Lett.} \textbf{1991}, \emph{66}, 413--416\relax
\mciteBstWouldAddEndPuncttrue
\mciteSetBstMidEndSepPunct{\mcitedefaultmidpunct}
{\mcitedefaultendpunct}{\mcitedefaultseppunct}\relax
\EndOfBibitem
\bibitem[L. and G.(1994)L., and G.]{Destombes}
L.,~B. M. B. H. C. M. D. C. D.~J.; G.,~C.~A. Millimeter- and submillimeter-wave
  spectroscopy of dibridged Si2H2 isotopomers: Experimental and theoretical
  structure. \emph{The Journal of Chemical Physics} \textbf{1994}, \emph{100},
  8614--8624\relax
\mciteBstWouldAddEndPuncttrue
\mciteSetBstMidEndSepPunct{\mcitedefaultmidpunct}
{\mcitedefaultendpunct}{\mcitedefaultseppunct}\relax
\EndOfBibitem
\bibitem[III and Thaddeus(2003)III, and Thaddeus]{Thaddeus}
III,~L. S. M. C. M. H. F.~S.; Thaddeus,~P. Mono- and Dibridged Isomers of Si2H3
  and Si2H4: the True Ground State Global Minima. Theory and Experiment in
  Concert. \emph{Journal of the American Chemical Society} \textbf{2003},
  \emph{125}, 11409--11417, PMID: 16220964\relax
\mciteBstWouldAddEndPuncttrue
\mciteSetBstMidEndSepPunct{\mcitedefaultmidpunct}
{\mcitedefaultendpunct}{\mcitedefaultseppunct}\relax
\EndOfBibitem
\bibitem[Mohapatra \latin{et~al.}(2016)Mohapatra, Kundu, Paesch, Herbst-Irmer,
  Stalke, Andrada, Frenking, and Roesky]{si2h2-opt-absorp-exp-2016}
Mohapatra,~C.; Kundu,~S.; Paesch,~A.~N.; Herbst-Irmer,~R.; Stalke,~D.;
  Andrada,~D.~M.; Frenking,~G.; Roesky,~H.~W. The Structure of the Carbene
  Stabilized Si2H2 May Be Equally Well Described with Coordinate Bonds as with
  Classical Double Bonds. \emph{Journal of the American Chemical Society}
  \textbf{2016}, \emph{138}, 10429--10432, PMID: 27494691\relax
\mciteBstWouldAddEndPuncttrue
\mciteSetBstMidEndSepPunct{\mcitedefaultmidpunct}
{\mcitedefaultendpunct}{\mcitedefaultseppunct}\relax
\EndOfBibitem
\bibitem[Chaeho~Pak and III(2000)Chaeho~Pak, and III]{Chaeho_Pak}
Chaeho~Pak,~J. C. R.-K.; III,~H. F.~S. Electron Affinities of Silicon Hydrides:
  SiHn (n = 0-4) and Si2Hn (n = 0-6). \emph{The Journal of Physical Chemistry
  A} \textbf{2000}, \emph{104}, 11232--11242\relax
\mciteBstWouldAddEndPuncttrue
\mciteSetBstMidEndSepPunct{\mcitedefaultmidpunct}
{\mcitedefaultendpunct}{\mcitedefaultseppunct}\relax
\EndOfBibitem
\bibitem[Sax and Kalcher(1991)Sax, and Kalcher]{doi:10.1021/j100157a052}
Sax,~A.~F.; Kalcher,~J. Theoretical enthalpies of formation for small silicon
  hydrides. \emph{The Journal of Physical Chemistry} \textbf{1991}, \emph{95},
  1768--1783\relax
\mciteBstWouldAddEndPuncttrue
\mciteSetBstMidEndSepPunct{\mcitedefaultmidpunct}
{\mcitedefaultendpunct}{\mcitedefaultseppunct}\relax
\EndOfBibitem
\bibitem[Sax(1985)]{Alexander_Sax}
Sax,~A.~F. Pseudopotential calculations on Si2H6 and Si2H4. \emph{Journal of
  Computational Chemistry} \textbf{1985}, \emph{6}, 469--477\relax
\mciteBstWouldAddEndPuncttrue
\mciteSetBstMidEndSepPunct{\mcitedefaultmidpunct}
{\mcitedefaultendpunct}{\mcitedefaultseppunct}\relax
\EndOfBibitem
\bibitem[K. \latin{et~al.}(1986)K., D., and C.]{soma-si2h4-1986}
K.,~S.; D.,~A.~R.; C.,~H.~N. Disilene, silylsilylene and their cations.
  \emph{Theoretica Chimica Acta} \textbf{1986}, \emph{70}, 393 -- 406\relax
\mciteBstWouldAddEndPuncttrue
\mciteSetBstMidEndSepPunct{\mcitedefaultmidpunct}
{\mcitedefaultendpunct}{\mcitedefaultseppunct}\relax
\EndOfBibitem
\bibitem[Poirier and Goddard(1981)Poirier, and Goddard]{POIRIER-si2h4-1981}
Poirier,~R.~A.; Goddard,~J.~D. The isomers of Si2H4: disilene and
  silylsilylene. \emph{Chemical Physics Letters} \textbf{1981}, \emph{80}, 37
  -- 41\relax
\mciteBstWouldAddEndPuncttrue
\mciteSetBstMidEndSepPunct{\mcitedefaultmidpunct}
{\mcitedefaultendpunct}{\mcitedefaultseppunct}\relax
\EndOfBibitem
\bibitem[Colegrove and Schaefer(1990)Colegrove, and
  Schaefer]{doi:10.1021/j100377a036}
Colegrove,~B.~T.; Schaefer,~H. F.~I. Disilyne (Si2H2) revisited. \emph{The
  Journal of Physical Chemistry} \textbf{1990}, \emph{94}, 5593--5602\relax
\mciteBstWouldAddEndPuncttrue
\mciteSetBstMidEndSepPunct{\mcitedefaultmidpunct}
{\mcitedefaultendpunct}{\mcitedefaultseppunct}\relax
\EndOfBibitem
\bibitem[Lischka and Kohler(1982)Lischka, and Kohler]{LISCHKA1982467}
Lischka,~H.; Kohler,~H.-J. On the structure and stability of singlet and
  triplet disilene and silylsilylene. \emph{Chemical Physics Letters}
  \textbf{1982}, \emph{85}, 467 -- 471\relax
\mciteBstWouldAddEndPuncttrue
\mciteSetBstMidEndSepPunct{\mcitedefaultmidpunct}
{\mcitedefaultendpunct}{\mcitedefaultseppunct}\relax
\EndOfBibitem
\bibitem[Koehler and Lischka(1982)Koehler, and
  Lischka]{doi:10.1021/ja00386a006}
Koehler,~H.~J.; Lischka,~H. A systematic investigation on the structure and
  stability of the lowest singlet and triplet states of Si2H4 and SiH3SiH and
  the carbon analogous compounds SiH2CH2, SiH3CH, CH3SiH, C2H4, and CH3CH.
  \emph{Journal of the American Chemical Society} \textbf{1982}, \emph{104},
  5884--5889\relax
\mciteBstWouldAddEndPuncttrue
\mciteSetBstMidEndSepPunct{\mcitedefaultmidpunct}
{\mcitedefaultendpunct}{\mcitedefaultseppunct}\relax
\EndOfBibitem
\bibitem[Lischka and Koehler(1983)Lischka, and
  Koehler]{doi:10.1021/ja00360a016}
Lischka,~H.; Koehler,~H.~J. Ab initio investigation on the lowest singlet and
  triplet state of disilyne (Si2H2). \emph{Journal of the American Chemical
  Society} \textbf{1983}, \emph{105}, 6646--6649\relax
\mciteBstWouldAddEndPuncttrue
\mciteSetBstMidEndSepPunct{\mcitedefaultmidpunct}
{\mcitedefaultendpunct}{\mcitedefaultseppunct}\relax
\EndOfBibitem
\bibitem[S. and F.(1992)S., and F.]{Roger_and_Henry}
S.,~G.~R.; F.,~S.~H. The remarkable monobridged structure of Si2H2. \emph{The
  Journal of Chemical Physics} \textbf{1992}, \emph{97}, 7990--7998\relax
\mciteBstWouldAddEndPuncttrue
\mciteSetBstMidEndSepPunct{\mcitedefaultmidpunct}
{\mcitedefaultendpunct}{\mcitedefaultseppunct}\relax
\EndOfBibitem
\bibitem[Sannigrahi and Nandi(1992)Sannigrahi, and Nandi]{Sannigrahi_Nandi}
Sannigrahi,~A.; Nandi,~P. Ab initio SCF study of the nature of bonding in Si2H2
  and Si2H4. \emph{Chemical Physics Letters} \textbf{1992}, \emph{188}, 575 --
  583\relax
\mciteBstWouldAddEndPuncttrue
\mciteSetBstMidEndSepPunct{\mcitedefaultmidpunct}
{\mcitedefaultendpunct}{\mcitedefaultseppunct}\relax
\EndOfBibitem
\bibitem[Jursic(1999)]{Jursic}
Jursic,~B.~S. Density functional theory investigation of the Si2H2 nonclassical
  and tetrahedron distorted structures. \emph{Journal of Molecular Structure:
  \{THEOCHEM\}} \textbf{1999}, \emph{491}, 1 -- 9\relax
\mciteBstWouldAddEndPuncttrue
\mciteSetBstMidEndSepPunct{\mcitedefaultmidpunct}
{\mcitedefaultendpunct}{\mcitedefaultseppunct}\relax
\EndOfBibitem
\bibitem[Olbrich(1986)]{Gottfried_Olbrich}
Olbrich,~G. On the structure and stability of Si2H4. \emph{Chemical Physics
  Letters} \textbf{1986}, \emph{130}, 115 -- 119\relax
\mciteBstWouldAddEndPuncttrue
\mciteSetBstMidEndSepPunct{\mcitedefaultmidpunct}
{\mcitedefaultendpunct}{\mcitedefaultseppunct}\relax
\EndOfBibitem
\bibitem[McCarthy M. C.; Yu~Z. and P.(2006)McCarthy M. C.; Yu~Z., and
  P.]{Monobridged_Si2H4}
McCarthy M. C.; Yu~Z.,~S. L. S. H.~F.; P.,~T. Monobridged Si2H4. \emph{The
  Journal of Chemical Physics} \textbf{2006}, \emph{124}, 074303\relax
\mciteBstWouldAddEndPuncttrue
\mciteSetBstMidEndSepPunct{\mcitedefaultmidpunct}
{\mcitedefaultendpunct}{\mcitedefaultseppunct}\relax
\EndOfBibitem
\bibitem[Cordonnier \latin{et~al.}(1992)Cordonnier, Bogey, Demuynck, and
  Destombes]{si2h2-monobridged-exp}
Cordonnier,~M.; Bogey,~M.; Demuynck,~C.; Destombes,~J.-L. Nonclassical
  structures in silicon-containing molecules: The monobridged isomer of Si2H2.
  \emph{The Journal of Chemical Physics} \textbf{1992}, \emph{97},
  7984--7989\relax
\mciteBstWouldAddEndPuncttrue
\mciteSetBstMidEndSepPunct{\mcitedefaultmidpunct}
{\mcitedefaultendpunct}{\mcitedefaultseppunct}\relax
\EndOfBibitem
\bibitem[Raabe and Michl(1985)Raabe, and Michl]{Michl_2}
Raabe,~G.; Michl,~J. Multiple bonding to silicon. \emph{Chemical Reviews}
  \textbf{1985}, \emph{85}, 419--509\relax
\mciteBstWouldAddEndPuncttrue
\mciteSetBstMidEndSepPunct{\mcitedefaultmidpunct}
{\mcitedefaultendpunct}{\mcitedefaultseppunct}\relax
\EndOfBibitem
\bibitem[WEST \latin{et~al.}(1981)WEST, FINK, and MICHL]{Michl}
WEST,~R.; FINK,~M.~J.; MICHL,~J. Tetramesityldisilene, a Stable Compound
  Containing a Silicon-Silicon Double Bond. \emph{Science} \textbf{1981},
  \emph{214}, 1343--1344\relax
\mciteBstWouldAddEndPuncttrue
\mciteSetBstMidEndSepPunct{\mcitedefaultmidpunct}
{\mcitedefaultendpunct}{\mcitedefaultseppunct}\relax
\EndOfBibitem
\bibitem[Curtiss \latin{et~al.}(1991)Curtiss, Raghavachari, Deutsch, and
  Pople]{Pople}
Curtiss,~L.~A.; Raghavachari,~K.; Deutsch,~P.~W.; Pople,~J.~A. Theoretical
  study of Si2Hn (n=0-6) and Si2H+n (n=0-7): Appearance potentials, ionization
  potentials, and enthalpies of formation. \emph{The Journal of Chemical
  Physics} \textbf{1991}, \emph{95}, 2433--2444\relax
\mciteBstWouldAddEndPuncttrue
\mciteSetBstMidEndSepPunct{\mcitedefaultmidpunct}
{\mcitedefaultendpunct}{\mcitedefaultseppunct}\relax
\EndOfBibitem
\bibitem[Trinquier(1990)]{Trinquier}
Trinquier,~G. Double bonds and bridged structures in the heavier analogs of
  ethylene. \emph{Journal of the American Chemical Society} \textbf{1990},
  \emph{112}, 2130--2137\relax
\mciteBstWouldAddEndPuncttrue
\mciteSetBstMidEndSepPunct{\mcitedefaultmidpunct}
{\mcitedefaultendpunct}{\mcitedefaultseppunct}\relax
\EndOfBibitem
\bibitem[Krogh-Jespersen(1982)]{doi:10.1021/j100206a004}
Krogh-Jespersen,~K. Geometries and relative energies of singlet silylsilylene
  and singlet disilene. \emph{The Journal of Physical Chemistry} \textbf{1982},
  \emph{86}, 1492--1495\relax
\mciteBstWouldAddEndPuncttrue
\mciteSetBstMidEndSepPunct{\mcitedefaultmidpunct}
{\mcitedefaultendpunct}{\mcitedefaultseppunct}\relax
\EndOfBibitem
\bibitem[Dolgonos(2008)]{Dolgonos}
Dolgonos,~G. Relative stability and thermodynamic properties of Si2H4 isomers.
  \emph{Chemical Physics Letters} \textbf{2008}, \emph{466}, 11 -- 15\relax
\mciteBstWouldAddEndPuncttrue
\mciteSetBstMidEndSepPunct{\mcitedefaultmidpunct}
{\mcitedefaultendpunct}{\mcitedefaultseppunct}\relax
\EndOfBibitem
\bibitem[Gutowsky and Stejskal(1954)Gutowsky, and Stejskal]{Gutowsky}
Gutowsky,~H.~S.; Stejskal,~E.~O. The Infrared Spectrum of Disilane. \emph{The
  Journal of Chemical Physics} \textbf{1954}, \emph{22}, 939--943\relax
\mciteBstWouldAddEndPuncttrue
\mciteSetBstMidEndSepPunct{\mcitedefaultmidpunct}
{\mcitedefaultendpunct}{\mcitedefaultseppunct}\relax
\EndOfBibitem
\bibitem[Itoh \latin{et~al.}(1986)Itoh, Toyoshima, Onuki, Washida, and
  Ibuki]{Si2H6_optical_expt}
Itoh,~U.; Toyoshima,~Y.; Onuki,~H.; Washida,~N.; Ibuki,~T. Vacuum ultraviolet
  absorption cross sections of SiH4, GeH4, Si2H6, and Si3H8. \emph{The Journal
  of Chemical Physics} \textbf{1986}, \emph{85}, 4867--4872\relax
\mciteBstWouldAddEndPuncttrue
\mciteSetBstMidEndSepPunct{\mcitedefaultmidpunct}
{\mcitedefaultendpunct}{\mcitedefaultseppunct}\relax
\EndOfBibitem
\bibitem[Bock \latin{et~al.}(1976)Bock, Ensslin, Feher, and
  Freund]{Photoelectron_Si2H6_1}
Bock,~H.; Ensslin,~W.; Feher,~F.; Freund,~R. Photoelectron spectra and
  molecular properties. LI. Ionization potentials of silanes SinH2n+2.
  \emph{Journal of the American Chemical Society} \textbf{1976}, \emph{98},
  668--674\relax
\mciteBstWouldAddEndPuncttrue
\mciteSetBstMidEndSepPunct{\mcitedefaultmidpunct}
{\mcitedefaultendpunct}{\mcitedefaultseppunct}\relax
\EndOfBibitem
\bibitem[EnBlin \latin{et~al.}(1975)EnBlin, Bergmann, and
  Elbel]{Photoelectron_Si2H6_2}
EnBlin,~W.; Bergmann,~H.; Elbel,~S. Photoelectron spectra of polysilanes.
  Conformational analyses of tetra-and penta-silane. \emph{J. Chem. Soc.,
  Faraday Trans. 2} \textbf{1975}, \emph{71}, 913--920\relax
\mciteBstWouldAddEndPuncttrue
\mciteSetBstMidEndSepPunct{\mcitedefaultmidpunct}
{\mcitedefaultendpunct}{\mcitedefaultseppunct}\relax
\EndOfBibitem
\bibitem[Blustin(1976)]{Si2H6_GS_2}
Blustin,~P.~H. A theoretical study of multiple bonding in carbon and silicon.
  \emph{Journal of Organometallic Chemistry} \textbf{1976}, \emph{105}, 161 --
  168\relax
\mciteBstWouldAddEndPuncttrue
\mciteSetBstMidEndSepPunct{\mcitedefaultmidpunct}
{\mcitedefaultendpunct}{\mcitedefaultseppunct}\relax
\EndOfBibitem
\bibitem[Collins \latin{et~al.}(1976)Collins, von R.~Schleyer, Binkley, and
  Pople]{Si2H6_GS_5}
Collins,~J.~B.; von R.~Schleyer,~P.; Binkley,~J.~S.; Pople,~J.~A.
  Self-consistent molecular orbital methods. XVII. Geometries and binding
  energies of second-row molecules. A comparison of three basis sets. \emph{The
  Journal of Chemical Physics} \textbf{1976}, \emph{64}, 5142--5151\relax
\mciteBstWouldAddEndPuncttrue
\mciteSetBstMidEndSepPunct{\mcitedefaultmidpunct}
{\mcitedefaultendpunct}{\mcitedefaultseppunct}\relax
\EndOfBibitem
\bibitem[Nicolas \latin{et~al.}(1976)Nicolas, Barthelat, and
  Durand]{Si2H6_GS_6}
Nicolas,~G.; Barthelat,~J.~C.; Durand,~P. Valence electronic structure and
  internal rotation barrier of the molecules XH3YH3 (X, Y = carbon, silicon,
  germanium) by a pseudopotential method. \emph{Journal of the American
  Chemical Society} \textbf{1976}, \emph{98}, 1346--1350\relax
\mciteBstWouldAddEndPuncttrue
\mciteSetBstMidEndSepPunct{\mcitedefaultmidpunct}
{\mcitedefaultendpunct}{\mcitedefaultseppunct}\relax
\EndOfBibitem
\bibitem[Berkovitch-yellin \latin{et~al.}(1981)Berkovitch-yellin, Ellis, and
  Ratner]{Si2H6_GS_3}
Berkovitch-yellin,~Z.; Ellis,~D.; Ratner,~M.~A. Intramolecular electron
  localization and local-density calculations on silicon-containing molecules:
  Tetramethylsilane and hexamethyldisilane. \emph{Chemical Physics}
  \textbf{1981}, \emph{62}, 21 -- 35\relax
\mciteBstWouldAddEndPuncttrue
\mciteSetBstMidEndSepPunct{\mcitedefaultmidpunct}
{\mcitedefaultendpunct}{\mcitedefaultseppunct}\relax
\EndOfBibitem
\bibitem[Halevi \latin{et~al.}(1985)Halevi, Winkelhofer, Meisl, and
  Janoschek]{Si2H6_GS_4}
Halevi,~E.; Winkelhofer,~G.; Meisl,~M.; Janoschek,~R. Electronic transitions of
  polysilanes and their photochemistry. \emph{Journal of Organometallic
  Chemistry} \textbf{1985}, \emph{294}, 151 -- 161\relax
\mciteBstWouldAddEndPuncttrue
\mciteSetBstMidEndSepPunct{\mcitedefaultmidpunct}
{\mcitedefaultendpunct}{\mcitedefaultseppunct}\relax
\EndOfBibitem
\bibitem[Luke \latin{et~al.}(1986)Luke, Pople, Krogh-Jespersen, Apeloig,
  Chandrasekhar, and Schleyer]{Si2H6_GS_1}
Luke,~B.~T.; Pople,~J.~A.; Krogh-Jespersen,~M.~B.; Apeloig,~Y.;
  Chandrasekhar,~J.; Schleyer,~P. v.~R. A theoretical survey of singly bonded
  silicon compounds. Comparison of the structures and bond energies of silyl
  and methyl derivatives. \emph{Journal of the American Chemical Society}
  \textbf{1986}, \emph{108}, 260--269\relax
\mciteBstWouldAddEndPuncttrue
\mciteSetBstMidEndSepPunct{\mcitedefaultmidpunct}
{\mcitedefaultendpunct}{\mcitedefaultseppunct}\relax
\EndOfBibitem
\bibitem[Ortiz and Mintmire(1988)Ortiz, and Mintmire]{Ortiz_Mintmire}
Ortiz,~J.~V.; Mintmire,~J.~W. Ground states and ionization energies of Si2H6,
  Si3H8, Si4H10, and Si5H12. \emph{Journal of the American Chemical Society}
  \textbf{1988}, \emph{110}, 4522--4527\relax
\mciteBstWouldAddEndPuncttrue
\mciteSetBstMidEndSepPunct{\mcitedefaultmidpunct}
{\mcitedefaultendpunct}{\mcitedefaultseppunct}\relax
\EndOfBibitem
\bibitem[Kawai \latin{et~al.}(1989)Kawai, Kasatani, Kawasaki, Sato, and
  Hirao]{Kawai}
Kawai,~E.; Kasatani,~K.; Kawasaki,~M.; Sato,~H.; Hirao,~K. Vacuum UV
  Photoabsorption Spectra of Silane and Disilane: Molecular Orbital Calculation
  of Electronic States. \emph{Japanese Journal of Applied Physics}
  \textbf{1989}, \emph{28}, 247\relax
\mciteBstWouldAddEndPuncttrue
\mciteSetBstMidEndSepPunct{\mcitedefaultmidpunct}
{\mcitedefaultendpunct}{\mcitedefaultseppunct}\relax
\EndOfBibitem
\bibitem[Rohlfing and Louie(1998)Rohlfing, and Louie]{Si2H6_optical_theory}
Rohlfing,~M.; Louie,~S.~G. Excitonic Effects and the Optical Absorption
  Spectrum of Hydrogenated Si Clusters. \emph{Phys. Rev. Lett.} \textbf{1998},
  \emph{80}, 3320--3323\relax
\mciteBstWouldAddEndPuncttrue
\mciteSetBstMidEndSepPunct{\mcitedefaultmidpunct}
{\mcitedefaultendpunct}{\mcitedefaultseppunct}\relax
\EndOfBibitem
\bibitem[Priya \latin{et~al.}(2017)Priya, Rai, and Shukla]{epjd-pradip}
Priya,~P.~K.; Rai,~D.~K.; Shukla,~A. Photoabsorption in sodium clusters: first
  principles configuration interaction calculations. \emph{Eur. Phys. J. D}
  \textbf{2017}, \emph{71}, 116\relax
\mciteBstWouldAddEndPuncttrue
\mciteSetBstMidEndSepPunct{\mcitedefaultmidpunct}
{\mcitedefaultendpunct}{\mcitedefaultseppunct}\relax
\EndOfBibitem
\bibitem[Shinde and Shukla(2017)Shinde, and Shukla]{epjd-shinde-mg}
Shinde,~R.; Shukla,~A. First principles electron-correlated calculations of
  optical absorption in magnesium clusters. \emph{Eur. Phys. J. D}
  \textbf{2017}, \emph{71}, 301\relax
\mciteBstWouldAddEndPuncttrue
\mciteSetBstMidEndSepPunct{\mcitedefaultmidpunct}
{\mcitedefaultendpunct}{\mcitedefaultseppunct}\relax
\EndOfBibitem
\bibitem[Ravindra and Alok(2012)Ravindra, and Alok]{Shinde_nano_life}
Ravindra,~S.; Alok,~S. Large-scale first principles configuration interaction
  calculations of optical absorption in boron clusters. \emph{Nano LIFE}
  \textbf{2012}, \emph{02}, 1240004\relax
\mciteBstWouldAddEndPuncttrue
\mciteSetBstMidEndSepPunct{\mcitedefaultmidpunct}
{\mcitedefaultendpunct}{\mcitedefaultseppunct}\relax
\EndOfBibitem
\bibitem[Ravindra and Alok(2014)Ravindra, and Alok]{Shinde_PCCP}
Ravindra,~S.; Alok,~S. Large-scale first principles configuration interaction
  calculations of optical absorption in aluminum clusters. \emph{Phys. Chem.
  Chem. Phys.} \textbf{2014}, \emph{16}, 20714--20723\relax
\mciteBstWouldAddEndPuncttrue
\mciteSetBstMidEndSepPunct{\mcitedefaultmidpunct}
{\mcitedefaultendpunct}{\mcitedefaultseppunct}\relax
\EndOfBibitem
\bibitem[Aryanpour \latin{et~al.}(2014)Aryanpour, Shukla, and
  Mazumdar]{Aryanpour_Shukla}
Aryanpour,~K.; Shukla,~A.; Mazumdar,~S. Electron correlations and two-photon
  states in polycyclic aromatic hydrocarbon molecules: A peculiar role of
  geometry. \emph{The Journal of Chemical Physics} \textbf{2014}, \emph{140},
  --\relax
\mciteBstWouldAddEndPuncttrue
\mciteSetBstMidEndSepPunct{\mcitedefaultmidpunct}
{\mcitedefaultendpunct}{\mcitedefaultseppunct}\relax
\EndOfBibitem
\bibitem[Shukla(2002)]{PhysRevB.65.125204Shukla65}
Shukla,~A. Correlated theory of triplet photoinduced absorption in
  phenylene-vinylene chains. \emph{Phys. Rev. B} \textbf{2002}, \emph{65},
  125204\relax
\mciteBstWouldAddEndPuncttrue
\mciteSetBstMidEndSepPunct{\mcitedefaultmidpunct}
{\mcitedefaultendpunct}{\mcitedefaultseppunct}\relax
\EndOfBibitem
\bibitem[Shukla(2004)]{PhysRevB.69.165218Shukla69}
Shukla,~A. Theory of nonlinear optical properties of phenyl-substituted
  polyacetylenes. \emph{Phys. Rev. B} \textbf{2004}, \emph{69}, 165218\relax
\mciteBstWouldAddEndPuncttrue
\mciteSetBstMidEndSepPunct{\mcitedefaultmidpunct}
{\mcitedefaultendpunct}{\mcitedefaultseppunct}\relax
\EndOfBibitem
\bibitem[Chakraborty and Shukla(2013)Chakraborty, and Shukla]{Himanshu}
Chakraborty,~H.; Shukla,~A. Pariser-Parr-Pople Model Based Investigation of
  Ground and Low-Lying Excited States of Long Acenes. \emph{The Journal of
  Physical Chemistry A} \textbf{2013}, \emph{117}, 14220--14229\relax
\mciteBstWouldAddEndPuncttrue
\mciteSetBstMidEndSepPunct{\mcitedefaultmidpunct}
{\mcitedefaultendpunct}{\mcitedefaultseppunct}\relax
\EndOfBibitem
\bibitem[Chakraborty and Shukla(2014)Chakraborty, and Shukla]{himanshu-triplet}
Chakraborty,~H.; Shukla,~A. Theory of triplet optical absorption in
  oligoacenes: From naphthalene to heptacene. \emph{The Journal of Chemical
  Physics} \textbf{2014}, \emph{141}, --\relax
\mciteBstWouldAddEndPuncttrue
\mciteSetBstMidEndSepPunct{\mcitedefaultmidpunct}
{\mcitedefaultendpunct}{\mcitedefaultseppunct}\relax
\EndOfBibitem
\bibitem[Sony and Shukla(2007)Sony, and Shukla]{Priya_Sony}
Sony,~P.; Shukla,~A. Large-scale correlated calculations of linear optical
  absorption and low-lying excited states of polyacenes: Pariser-Parr-Pople
  Hamiltonian. \emph{Phys. Rev. B} \textbf{2007}, \emph{75}, 155208\relax
\mciteBstWouldAddEndPuncttrue
\mciteSetBstMidEndSepPunct{\mcitedefaultmidpunct}
{\mcitedefaultendpunct}{\mcitedefaultseppunct}\relax
\EndOfBibitem
\bibitem[Rai \latin{et~al.}(2018)Rai, Chakraborty, and Shukla]{dkr1}
Rai,~D.~K.; Chakraborty,~H.; Shukla,~A. Tunable Optoelectronic Properties of
  Triply Bonded Carbon Molecules with Linear and Graphyne Substructures.
  \emph{The Journal of Physical Chemistry C} \textbf{2018}, \emph{122},
  1309--1317\relax
\mciteBstWouldAddEndPuncttrue
\mciteSetBstMidEndSepPunct{\mcitedefaultmidpunct}
{\mcitedefaultendpunct}{\mcitedefaultseppunct}\relax
\EndOfBibitem
\bibitem[Basak \latin{et~al.}(2015)Basak, Chakraborty, and Shukla]{Tista1}
Basak,~T.; Chakraborty,~H.; Shukla,~A. Theory of linear optical absorption in
  diamond-shaped graphene quantum dots. \emph{Phys. Rev. B} \textbf{2015},
  \emph{92}, 205404\relax
\mciteBstWouldAddEndPuncttrue
\mciteSetBstMidEndSepPunct{\mcitedefaultmidpunct}
{\mcitedefaultendpunct}{\mcitedefaultseppunct}\relax
\EndOfBibitem
\bibitem[Turney \latin{et~al.}(2012)Turney, Simmonett, Parrish, Hohenstein,
  Evangelista, Fermann, Mintz, Burns, Wilke, Abrams, Russ, Leininger, Janssen,
  Seidl, Allen, Schaefer, King, Valeev, Sherrill, and Crawford]{PSI4}
Turney,~J.~M. \latin{et~al.}  Psi4: An open-source ab initio electronic
  structure program. \emph{WIREs Comput. Mol. Sci.} \textbf{2012}, \emph{2},
  556\relax
\mciteBstWouldAddEndPuncttrue
\mciteSetBstMidEndSepPunct{\mcitedefaultmidpunct}
{\mcitedefaultendpunct}{\mcitedefaultseppunct}\relax
\EndOfBibitem
\bibitem[McMurchie \latin{et~al.}()McMurchie, Elbert, Langhoff, and
  Davidson]{MELD}
McMurchie,~L.~E.; Elbert,~S.~T.; Langhoff,~S.~R.; Davidson,~E.~R. {MELD}
  package from Indiana University. It has been modified by us to handle bigger
  systems.\relax
\mciteBstWouldAddEndPunctfalse
\mciteSetBstMidEndSepPunct{\mcitedefaultmidpunct}
{}{\mcitedefaultseppunct}\relax
\EndOfBibitem
\bibitem[Frisch \latin{et~al.}(2016)Frisch, Trucks, Schlegel, Scuseria, Robb,
  Cheeseman, Scalmani, Barone, Petersson, Nakatsuji, Li, Caricato, Marenich,
  Bloino, Janesko, Gomperts, Mennucci, Hratchian, Ortiz, Izmaylov, Sonnenberg,
  Williams-Young, Ding, Lipparini, Egidi, Goings, Peng, Petrone, Henderson,
  Ranasinghe, Zakrzewski, Gao, Rega, Zheng, Liang, Hada, Ehara, Toyota, Fukuda,
  Hasegawa, Ishida, Nakajima, Honda, Kitao, Nakai, Vreven, Throssell,
  Montgomery, Peralta, Ogliaro, Bearpark, Heyd, Brothers, Kudin, Staroverov,
  Keith, Kobayashi, Normand, Raghavachari, Rendell, Burant, Iyengar, Tomasi,
  Cossi, Millam, Klene, Adamo, Cammi, Ochterski, Martin, Morokuma, Farkas,
  Foresman, and Fox]{g16}
Frisch,~M.~J. \latin{et~al.}  Gaussian 16 {R}evision {B}.01. 2016; Gaussian
  Inc. Wallingford CT\relax
\mciteBstWouldAddEndPuncttrue
\mciteSetBstMidEndSepPunct{\mcitedefaultmidpunct}
{\mcitedefaultendpunct}{\mcitedefaultseppunct}\relax
\EndOfBibitem
\bibitem[Dennington \latin{et~al.}(2016)Dennington, Keith, and Millam]{gv6}
Dennington,~R.; Keith,~T.~A.; Millam,~J.~M. GaussView {V}ersion {6}. 2016;
  Semichem Inc. Shawnee Mission KS\relax
\mciteBstWouldAddEndPuncttrue
\mciteSetBstMidEndSepPunct{\mcitedefaultmidpunct}
{\mcitedefaultendpunct}{\mcitedefaultseppunct}\relax
\EndOfBibitem
\bibitem[Kokalj(1999)]{XCrySDen}
Kokalj,~A. XCrySDen-a new program for displaying crystalline structures and
  electron densities. \emph{Journal of Molecular Graphics and Modelling}
  \textbf{1999}, \emph{17}, 176 -- 179\relax
\mciteBstWouldAddEndPuncttrue
\mciteSetBstMidEndSepPunct{\mcitedefaultmidpunct}
{\mcitedefaultendpunct}{\mcitedefaultseppunct}\relax
\EndOfBibitem
\bibitem[Beagley \latin{et~al.}(1972)Beagley, Conrad, Freeman, Monaghan,
  Norton, and Holywell]{Si2H6_geo_expt_1}
Beagley,~B.; Conrad,~A.; Freeman,~J.; Monaghan,~J.; Norton,~B.; Holywell,~G.
  Electron diffraction studies of the hydrides Si2H6 and P2H4. \emph{Journal of
  Molecular Structure} \textbf{1972}, \emph{11}, 371 -- 380\relax
\mciteBstWouldAddEndPuncttrue
\mciteSetBstMidEndSepPunct{\mcitedefaultmidpunct}
{\mcitedefaultendpunct}{\mcitedefaultseppunct}\relax
\EndOfBibitem
\bibitem[Shotton \latin{et~al.}(1973)Shotton, Lee, and Jones]{Si2H6_geo_expt_2}
Shotton,~K.~C.; Lee,~A.~G.; Jones,~W.~J. The pure-rotational Raman spectrum of
  disilane. \emph{Journal of Raman Spectroscopy} \textbf{1973}, \emph{1},
  243--253\relax
\mciteBstWouldAddEndPuncttrue
\mciteSetBstMidEndSepPunct{\mcitedefaultmidpunct}
{\mcitedefaultendpunct}{\mcitedefaultseppunct}\relax
\EndOfBibitem
\bibitem[Adamczyk and Broadbelt(2011)Adamczyk, and Broadbelt]{Andrew_and_Linda}
Adamczyk,~A.~J.; Broadbelt,~L.~J. The Role of Multifunctional Kinetics during
  Early-Stage Silicon Hydride Pyrolysis: Reactivity of Si2H2 Isomers with SiH4
  and Si2H6. \emph{The Journal of Physical Chemistry A} \textbf{2011},
  \emph{115}, 2409--2422, PMID: 21361329\relax
\mciteBstWouldAddEndPuncttrue
\mciteSetBstMidEndSepPunct{\mcitedefaultmidpunct}
{\mcitedefaultendpunct}{\mcitedefaultseppunct}\relax
\EndOfBibitem
\bibitem[Koseki and Gordon(1989)Koseki, and Gordon]{Gordon}
Koseki,~S.; Gordon,~M.~S. Intrinsic reaction coordinate calculations for very
  flat potential energy surfaces: application to singlet disilenylidene
  isomerization. \emph{The Journal of Physical Chemistry} \textbf{1989},
  \emph{93}, 118--125\relax
\mciteBstWouldAddEndPuncttrue
\mciteSetBstMidEndSepPunct{\mcitedefaultmidpunct}
{\mcitedefaultendpunct}{\mcitedefaultseppunct}\relax
\EndOfBibitem
\end{mcitethebibliography}

\end{document}


\title{Supporting Information\\
Systematic First Principles Configuration-Interaction Calculations
of Linear Optical Absorption Spectra in Silicon Hydrides : Si\textsubscript{$2$}H\textsubscript{$2n$}
($n=1-3$)}
\author{Pritam Bhattacharyya }
\email{pritam.bhattacharyya01@gmail.com }

\author{Deepak Kumar Rai}
\email{dkriitb@gmail.com}

\author{Alok Shukla}
\email{shukla@phy.iitb.ac.in}

\affiliation{Department of Physics, Indian Institute of Technology Bombay, Powai,
Mumbai 400076, INDIA}
\maketitle

\section{Excited state Wave-functions}

In this section we present important information about the excited
states contributing to the peaks in the absorption spectra computed
of various conformers of Si$_{2}$H$_{2}$, Si$_{2}$H$_{4}$, and
Si$_{2}$H$_{6}$, using the MRSDCI approach. It consists of plots
of frontier molecular orbitals, excitation energies, oscillator strengths,
dominant terms in their many-body wave functions, and polarization
directions of absorbed photons.

\begin{figure}[H]
\includegraphics[scale=0.12]{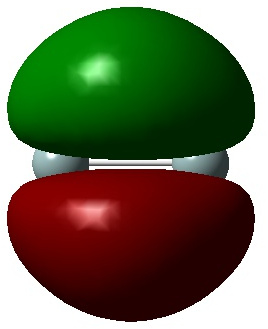}~~~~~\includegraphics[scale=0.12]{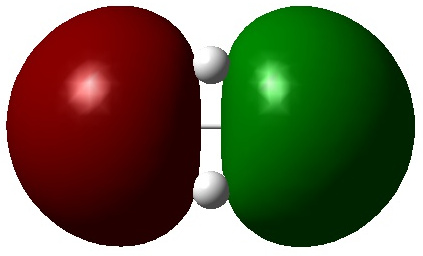}~~~~~\includegraphics[scale=0.12]{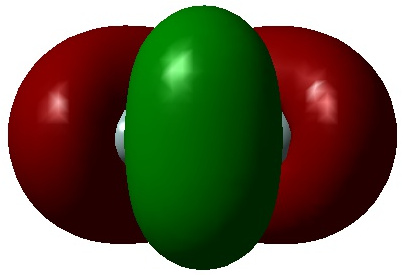}~~~~~\includegraphics[scale=0.12]{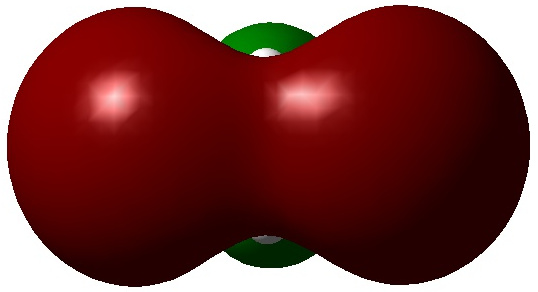}~~~~~\includegraphics[scale=0.12]{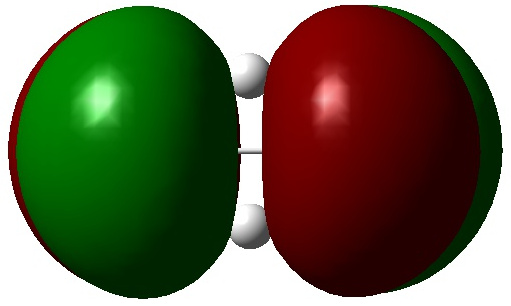}~~~~~\includegraphics[scale=0.1]{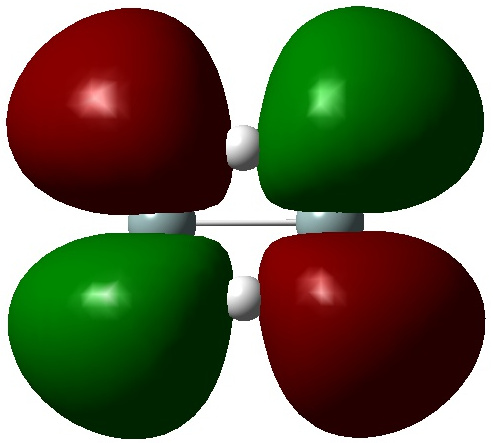}~~~~~\includegraphics[scale=0.1]{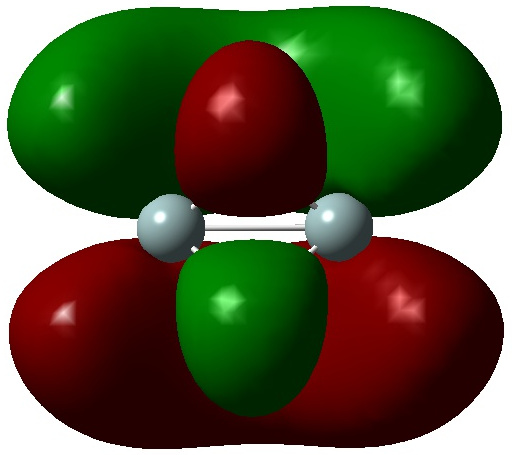}~~~~~\includegraphics[scale=0.1]{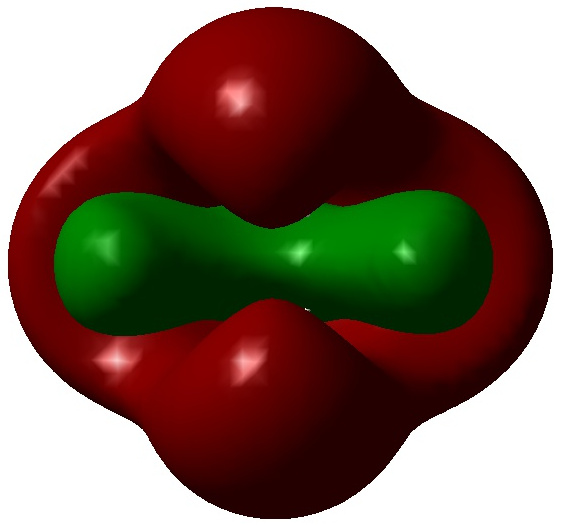}~~~~~\includegraphics[scale=0.1]{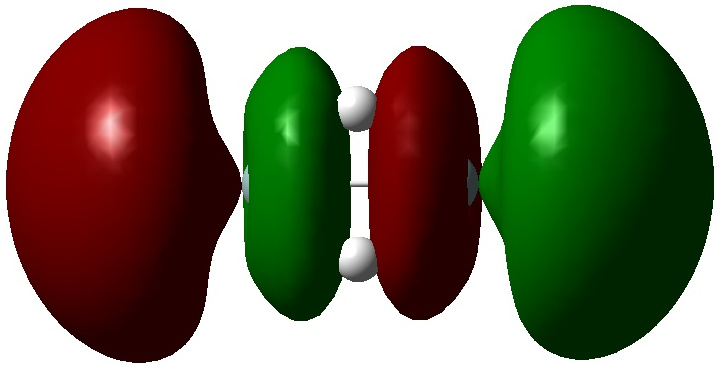}

\textbf{~H-3}~~\textbf{~~~~~~H-2}~~~~~~~~~~~~\textbf{H-1}~~~~~~~~~~~~~~~\textbf{H}~~~~~~~~~~~~~~~\textbf{~~L}~~~~~~~~~~~~~~~\textbf{L+1~~~~~~~~~L+2~~~~~~~~~~L+3~~~~~~~~~~~~~L+4}

\caption{Various frontier molecular orbitals of dibridged disilyne isomer {[}main
article, Fig. 4(a){]} of Si$_{2}$H$_{2}$ molecule, contributing
to optical transitions of comparatively high oscillator strength.
Symbols H and L denote HOMO and LUMO orbitals, while H-n (L+n) represents
n-th orbital below (above) HOMO (LUMO).}
\end{figure}

\begin{table}[H]
\caption{Many-particle wave functions of the excited states contributing to
the peaks in the optical absorption spectrum of dibridged disilyne
(Si-H$_{2}$-Si) isomer {[}main article, Fig. 4(a){]} of Si$_{2}$H$_{2}$
molecule. 'E' corresponds to excitation energy (in eV) of an excited
state, and $f$ (= $\frac{2\times m_{e}}{3\times\hbar^{2}}(E)\sum_{j=x,y,z}|\langle e|O_{j}|R\rangle|^{2}$)
denotes the oscillator strength for a particular electric dipole transition,
where $|e\rangle$,$|R\rangle$ and $O_{j}$ indicate the excited
states of the corresponding peak, reference state, and electric dipole
operator for different Cartesian components, respectively. In the
``|TDM|'' column, we present the magnitudes of the transition dipole
moments (TDMs) to understand the extent of coupling between the relevant
excited state and the ground state. In the ``Polarization'' column,
x, y and z denote the absorption of light polarized along x, y, and
z directions, respectively. 'HF' corresponds to Hartree-Fock configuration.
'$H$' and '$L$' stand for HOMO and LUMO orbitals. In the ``Wave
function'' column, each number inside the parentheses denotes the
coefficient of the corresponding configuration in the CI wave function.
GS indicates the ground states wave function of the isomer, and not
that of an excited state corresponding to a peak. The atomic coordinates
corresponding to the optimized ground state geometry of this isomer
are presented in Table \ref{tab:geo-Dibridged-disilyne-si2h2}.}

\begin{tabular}{ccccccccccc}
\hline 
Peak & \hspace{1cm} & E (eV) & \hspace{1cm} & $f$ & \hspace{1cm} & |TDM| & \hspace{1cm} & Polarization & \hspace{1cm} & Wave function\tabularnewline
\hline 
\hline 
GS &  &  &  &  &  &  &  &  &  & $\arrowvert HF\rangle(0.9190)$\tabularnewline
 &  &  &  &  &  &  &  &  &  & $\arrowvert H\rightarrow L;\thinspace H\rightarrow L\rangle(0.0685)$\tabularnewline
 &  &  &  &  &  &  &  &  &  & \tabularnewline
I &  & 3.360 &  & 1.37  &  & 0.7821 &  & $y$ &  & $\arrowvert H\rightarrow L\rangle(0.8945)$\tabularnewline
 &  &  &  &  &  &  &  &  &  & $\arrowvert H\rightarrow L+12\rangle(0.1101)$\tabularnewline
 &  &  &  &  &  &  &  &  &  & \tabularnewline
II &  & 4.604 &  & 0.23 &  & 0.2738 &  & $x$ &  & $\arrowvert H\rightarrow L+2\rangle(0.8769)$\tabularnewline
 &  &  &  &  &  &  &  &  &  & $\arrowvert H-1\rightarrow L;\thinspace H\rightarrow L+1\rangle(0.1450)$\tabularnewline
 &  &  &  &  &  &  &  &  &  & \tabularnewline
III &  & 5.734 &  & 13.049 &  & 1.8476 &  & $y$ &  & $\arrowvert H-1\rightarrow L\rangle(0.7791)$\tabularnewline
 &  &  &  &  &  &  &  &  &  & $\arrowvert H\rightarrow L+4\rangle(0.3966)$\tabularnewline
 &  & 5.768 &  & 0.352 &  & 0.3024 &  & $z$ &  & $\arrowvert H\rightarrow L+3;\thinspace H\rightarrow L+3\rangle(0.6692)$\tabularnewline
 &  &  &  &  &  &  &  &  &  & $\arrowvert H\rightarrow L;\thinspace H\rightarrow L\rangle(0.5101)$\tabularnewline
 &  &  &  &  &  &  &  &  &  & \tabularnewline
IV &  & 6.986 &  & 13.818 &  & 1.7225 &  & $y$ &  & $\arrowvert H\rightarrow L+4\rangle(0.7267)$\tabularnewline
 &  &  &  &  &  &  &  &  &  & $\arrowvert H-3\rightarrow L+1\rangle(0.4110)$\tabularnewline
 &  &  &  &  &  &  &  &  &  & \tabularnewline
V &  & 7.914 &  & 1.018 &  & 0.4393 &  & $z$ &  & $\arrowvert H-1\rightarrow L+3;\thinspace H-1\rightarrow L+3\rangle(0.7953)$\tabularnewline
 &  &  &  &  &  &  &  &  &  & $\arrowvert H\rightarrow L+5;\thinspace H\rightarrow L+5\rangle(0.2791)$\tabularnewline
 &  & 7.942 &  & 3.719  &  & 0.8381 &  & $x$ &  & $\arrowvert H-1\rightarrow L+2\rangle(0.5350)$\tabularnewline
 &  &  &  &  &  &  &  &  &  & $\arrowvert H-2\rightarrow L+1\rangle(0.4774)$\tabularnewline
\hline 
\end{tabular}
\end{table}

$\pagebreak$

\begin{figure}[H]
\begin{centering}
\includegraphics[scale=0.1]{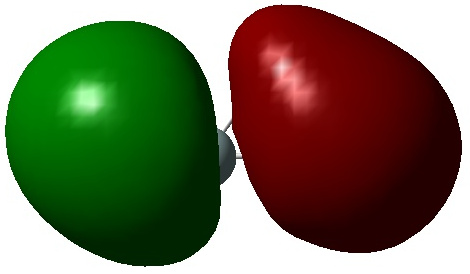}~~~~~~\includegraphics[scale=0.1]{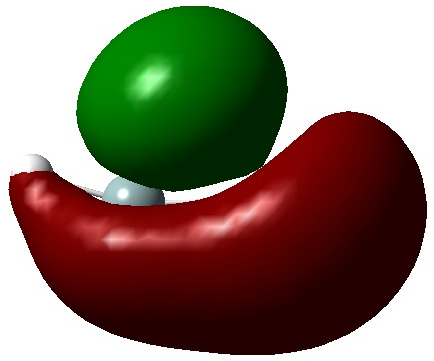}~~~~~~\includegraphics[scale=0.1]{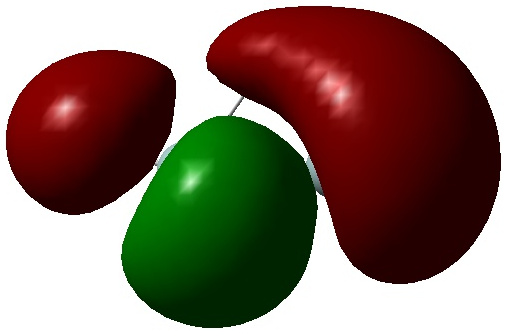}~~~~~~\includegraphics[scale=0.1]{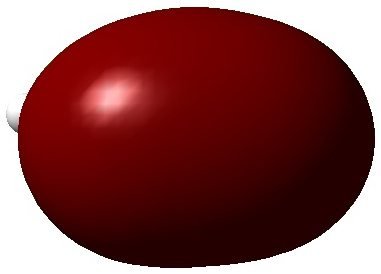}~~~~~~\includegraphics[scale=0.1]{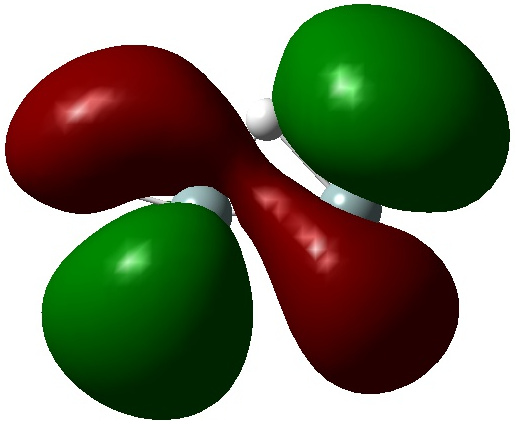}~~~~~~\includegraphics[scale=0.1]{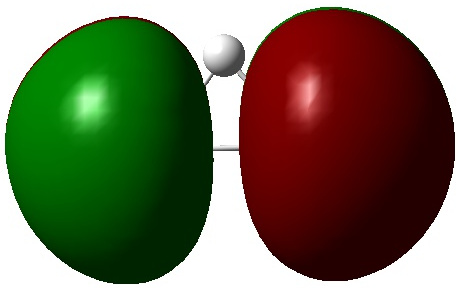}~~~~~~\includegraphics[scale=0.08]{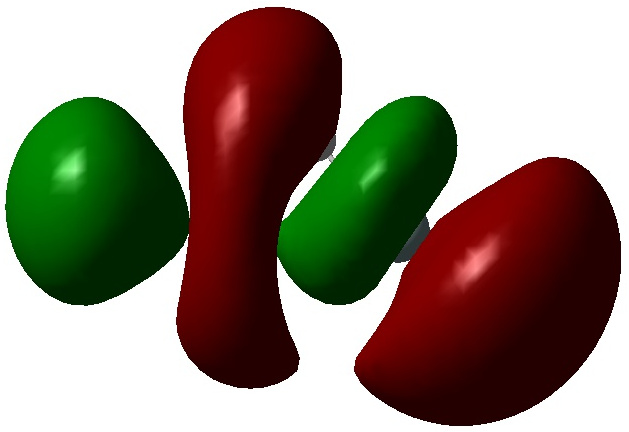}~~~~~~\includegraphics[scale=0.08]{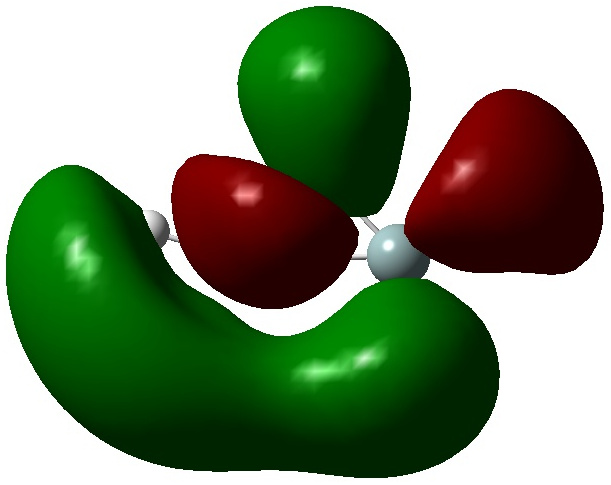}~~~~~~\includegraphics[scale=0.08]{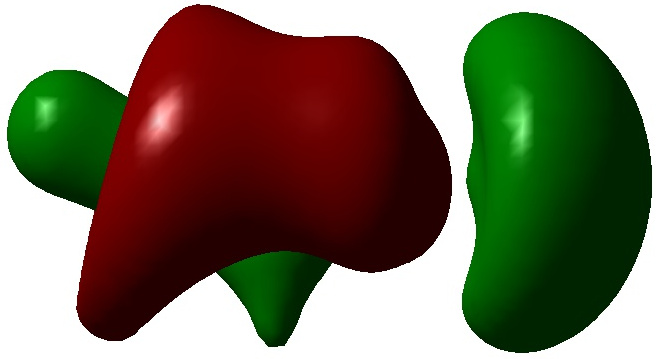}
\par\end{centering}
\begin{centering}
~~\ \textbf{~H-3}~~~\textbf{~~~~~~~H-2}~~~~~~~~~~~~\textbf{~H-1}~~~~~~~~~~~~\textbf{H}~~~~~~~~~~~~~~~\textbf{L}~~~~~~~~~~~~~\textbf{L+1~~~~~~~~~~L+2~~~~~~~~~~L+3~~~~~~~~~~L+4~~~}
\par\end{centering}
{\footnotesize{}\caption{Various frontier molecular orbitals of mono-bridged isomer {[}main
article, Fig. 4(b){]} of Si$_{2}$H$_{2}$ molecule, contributing
to optical transitions of comparatively high oscillator strength.
Symbols H and L denote the HOMO and the LUMO orbitals, while H-n (L+n)
represents n-th orbital below (above) HOMO (LUMO).}
}{\footnotesize\par}
\end{figure}

\begin{table}[H]
\caption{Many-particle wave functions of the excited states contributing to
the peaks in the optical absorption spectrum of mono-bridged (Si-H-SiH)
isomer {[}main article, Fig. 4(b){]} of Si$_{2}$H$_{2}$ molecule.
'E' corresponds to excitation energy (in eV) of an excited state,
whereas $f$, ``TDM'', and ``Polarization'' are defined in the
caption of Table S1. 'HF' corresponds to Hartree-Fock configuration.
'$H$' and '$L$' stand for HOMO and LUMO orbitals. In the ``Wave
function'' column, each number inside the parentheses denotes the
coefficient of the corresponding configuration in the CI wave function.
GS indicates the ground states wave function of the isomer, and not
that of an excited state corresponding to a peak. The atomic coordinates
corresponding to the optimized ground state geometry of this isomer
are presented in Table \ref{tab:geo-Monobridged-si2h2}. }

\begin{tabular}{ccccccccccc}
\hline 
Peak & \hspace{1cm} & E (eV) & \hspace{1cm} & $f$ & \hspace{1cm} & |TDM| & \hspace{1cm} & Polarization & \hspace{1cm} & Wave function\tabularnewline
\hline 
\hline 
GS &  &  &  &  &  &  &  &  &  & $\arrowvert HF\rangle(0.9044)$\tabularnewline
 &  &  &  &  &  &  &  &  &  & $\arrowvert H\rightarrow L+1;\thinspace H\rightarrow L+1\rangle(0.1311)$\tabularnewline
 &  &  &  &  &  &  &  &  &  & \tabularnewline
I &  & 2.81 &  & 0.082 &  & 0.2094 &  & y &  & $\arrowvert H\rightarrow L\rangle(0.8984)$\tabularnewline
 &  &  &  &  &  &  &  &  &  & $\arrowvert H\rightarrow L+2\rangle(0.0859)$\tabularnewline
 &  &  &  &  &  &  &  &  &  & \tabularnewline
II &  & 4.01 &  & 0.755 &  & 0.25/0.47 &  & x/z &  & $\arrowvert H\rightarrow L+1\rangle(0.7516)$\tabularnewline
 &  &  &  &  &  &  &  &  &  & $\arrowvert H-1\rightarrow L\rangle(0.4596)$\tabularnewline
 &  & 4.14 &  & 0.655 &  & 0.4861 &  & y &  & $\arrowvert H\rightarrow L+2\rangle(0.8654)$\tabularnewline
 &  &  &  &  &  &  &  &  &  & $\arrowvert H\rightarrow L+4\rangle(0.1240)$\tabularnewline
 &  &  &  &  &  &  &  &  &  & \tabularnewline
III &  & 5.16 &  & 3.462 &  & 0.97/0.25 &  & x/z &  & $\arrowvert H-1\rightarrow L\rangle(0.6557)$\tabularnewline
 &  &  &  &  &  &  &  &  &  & $\arrowvert H\rightarrow L+1\rangle(0.3348)$\tabularnewline
 &  &  &  &  &  &  &  &  &  & \tabularnewline
IV &  & 5.75 &  & 0.939 &  & 0.39/0.31 &  & x/z &  & $\arrowvert H-1\rightarrow L+2\rangle(0.6559)$\tabularnewline
 &  &  &  &  &  &  &  &  &  & $\arrowvert H-2\rightarrow L\rangle(0.5002)$\tabularnewline
 &  &  &  &  &  &  &  &  &  & \tabularnewline
V &  & 6.80 &  & 4.753 &  & 0.95/0.38 &  & x/z &  & $\arrowvert H-2\rightarrow L\rangle(0.4030)$\tabularnewline
 &  &  &  &  &  &  &  &  &  & $\arrowvert H-1\rightarrow L+2\rangle(0.3875)$\tabularnewline
 &  &  &  &  &  &  &  &  &  & \tabularnewline
VI &  & 7.22 &  & 6.728 &  & 1.1818 &  & y &  & $\arrowvert H\rightarrow L+4\rangle(0.6325)$\tabularnewline
 &  &  &  &  &  &  &  &  &  & $\arrowvert H\rightarrow L+1;\thinspace H\rightarrow L+2\rangle(0.3402)$\tabularnewline
 &  &  &  &  &  &  &  &  &  & \tabularnewline
VII &  & 7.66 &  & 5.660 &  & 0.77/0.72 &  & x/z &  & $\arrowvert H\rightarrow L;\thinspace H-1\rightarrow L+1\rangle(0.4170)$\tabularnewline
 &  &  &  &  &  &  &  &  &  & $\arrowvert H\rightarrow L;\thinspace H\rightarrow L+2\rangle(0.3670)$\tabularnewline
 &  &  &  &  &  &  &  &  &  & \tabularnewline
VIII &  & 7.91 &  & 12.025 &  & 1.09/1.03 &  & x/z &  & $\arrowvert H-1\rightarrow L+4\rangle(0.3523)$\tabularnewline
 &  &  &  &  &  &  &  &  &  & $\arrowvert H-2\rightarrow L+2\rangle(0.3201)$\tabularnewline
 &  &  &  &  &  &  &  &  &  & \tabularnewline
IX &  & 8.86 &  & 11.742 &  & 0.79/1.17 &  & x/z &  & $\arrowvert H-1\rightarrow L+4\rangle(0.4106)$\tabularnewline
 &  &  &  &  &  &  &  &  &  & $\arrowvert H-3\rightarrow L\rangle(0.3441)$\tabularnewline
 &  &  &  &  &  &  &  &  &  & \tabularnewline
X &  & 9.16 &  & 5.864 &  & 0.34/0.92 &  & x/z &  & $\arrowvert H\rightarrow L+1;\thinspace H-1\rightarrow L+2\rangle(0.4375)$\tabularnewline
 &  &  &  &  &  &  &  &  &  & $\arrowvert H\rightarrow L+2;\thinspace H\rightarrow L+2\rangle(0.2899)$\tabularnewline
\hline 
\end{tabular}
\end{table}

\begin{figure}[H]
\includegraphics[scale=0.14]{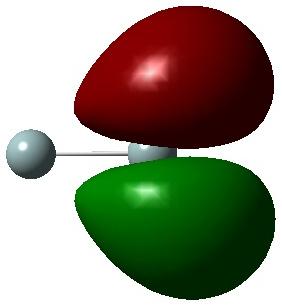}~~~~~~~~\includegraphics[scale=0.12]{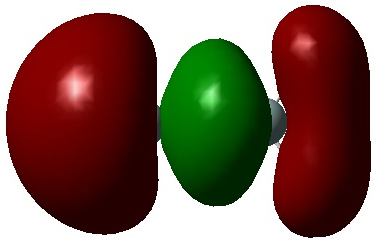}~~~~~~~~\includegraphics[scale=0.14]{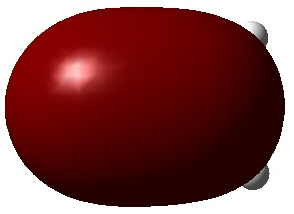}~~~~~~~~\includegraphics[scale=0.12]{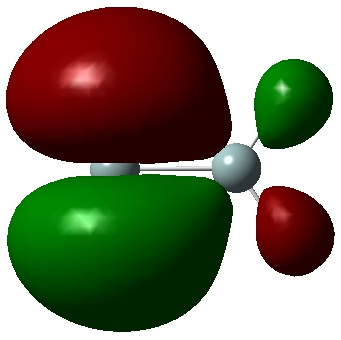}~~~~~~~~\includegraphics[scale=0.14]{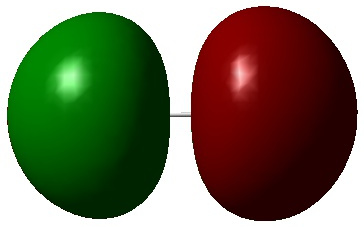}~~~~~~~~\includegraphics[scale=0.12]{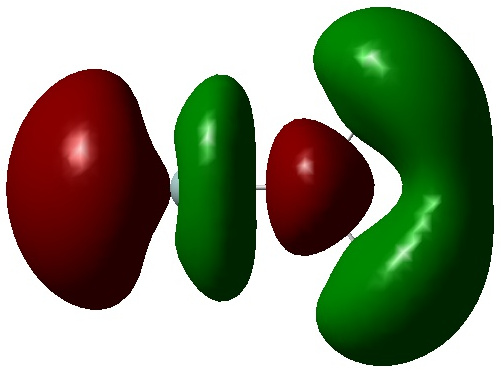}~~~~~~~~\includegraphics[scale=0.12]{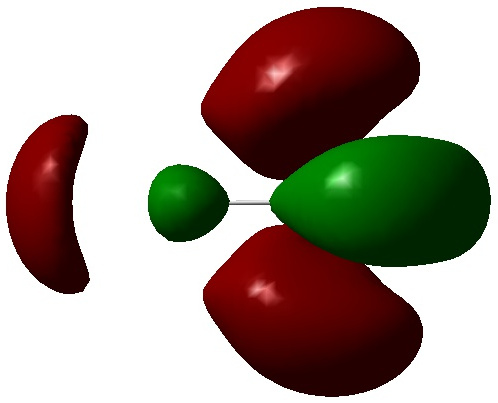}~~~~~~~~\includegraphics[scale=0.12]{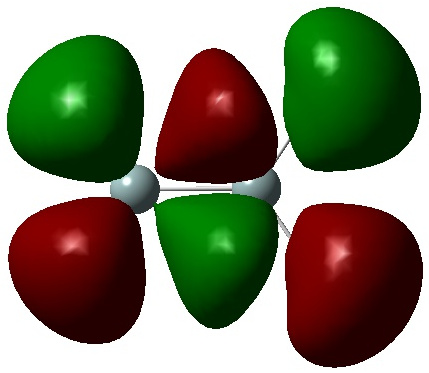}

~~\ \textbf{H-2}~~~\textbf{~~~~~~~~H-1}~~~~~~~~~~~~~\textbf{~H}~~~~~~~~~~~~~~~\textbf{~L}~~~~~~~~~~~~~~\textbf{L+1}~~~~~~~~~~~~~~~\textbf{L+2~~~~~~~~~~~~~~~L+3~~~~~~~~~~~~~L+4}

\caption{Various frontier molecular orbitals of disilavinylidene isomer {[}main
article, Fig. 4(c){]} of Si$_{2}$H$_{2}$ molecule, contributing
to optical transitions of comparatively high oscillator strength.
Symbols H and L denote the HOMO and the LUMO orbitals, while H-n (L+n)
represents n-th orbital below (above) HOMO (LUMO).}
\end{figure}

\begin{table}[H]
\caption{Many-particle wave functions of the excited states contributing to
the peaks in the optical absorption spectrum of disilavinylidene (Si-SiH$_{2}$)
isomer {[}main article, Fig. 4(c){]} of Si$_{2}$H$_{2}$ molecule.
'E' corresponds to excitation energy (in eV) of an excited state,
whereas $f$, ``TDM'', and ``Polarization'' are defined in the
caption of Table S1. 'HF' corresponds to Hartree-Fock configuration.
'$H$' and '$L$' stand for HOMO and LUMO orbitals. In the ``Wave
function'' column, each number inside the parentheses denotes the
coefficient of the corresponding configuration in the CI wave function.
GS indicates the ground states wave function of the isomer, and not
that of an excited state corresponding to a peak. The atomic coordinates
corresponding to the optimized ground state geometry of this isomer
are presented in Table \ref{tab:geo-Disilavinylidene-si2h2}. }

\begin{tabular}{ccccccccccc}
\hline 
Peak & \hspace{0.8cm} & E (eV) & \hspace{0.8cm} & $f$ & \hspace{1cm} & |TDM| & \hspace{1cm} & Polarization & \hspace{1cm} & Wave function\tabularnewline
\hline 
\hline 
GS &  &  &  &  &  &  &  &  &  & $\arrowvert HF\rangle(0.9027)$\tabularnewline
 &  &  &  &  &  &  &  &  &  & $\arrowvert H\rightarrow L+1;\thinspace H\rightarrow L+1\rangle(0.1626)$\tabularnewline
 &  &  &  &  &  &  &  &  &  & \tabularnewline
I &  & 2.48 &  & 0.120 &  & 0.2699 &  & $x$ &  & $\arrowvert H-1\rightarrow L\rangle(0.8898)$\tabularnewline
 &  &  &  &  &  &  &  &  &  & $\arrowvert H-1\rightarrow L;H\rightarrow L+1;H\rightarrow L+1\rangle(0.1306)$\tabularnewline
 &  &  &  &  &  &  &  &  &  & \tabularnewline
II &  & 3.13 &  & 0.213 &  & 0.3196 &  & $y$ &  & $\arrowvert H\rightarrow L;H\rightarrow L\rangle(0.8578)$\tabularnewline
 &  &  &  &  &  &  &  &  &  & $\arrowvert H\rightarrow L+1\rangle(0.1760)$\tabularnewline
 &  &  &  &  &  &  &  &  &  & \tabularnewline
III &  & 4.29 &  & 3.820 &  & 1.1558 &  & $y$ &  & $\arrowvert H\rightarrow L+1\rangle(0.8258)$\tabularnewline
 &  &  &  &  &  &  &  &  &  & $\arrowvert H-1\rightarrow L+2\rangle(0.2004)$\tabularnewline
 &  &  &  &  &  &  &  &  &  & \tabularnewline
IV &  & 5.79 &  & 1.064 &  & 0.5250 &  & $y$ &  & $\arrowvert H-2\rightarrow L\rangle(0.7332)$\tabularnewline
 &  &  &  &  &  &  &  &  &  & $\arrowvert H\rightarrow L;H-2\rightarrow L+1\rangle(0.3659)$\tabularnewline
 &  &  &  &  &  &  &  &  &  & \tabularnewline
V &  & 6.43 &  & 1.783 &  & 0.6450 &  & $z$ &  & $\arrowvert H\rightarrow L+3\rangle(0.6899)$\tabularnewline
 &  &  &  &  &  &  &  &  &  & $\arrowvert H\rightarrow L+4\rangle(0.4886)$\tabularnewline
 &  &  &  &  &  &  &  &  &  & \tabularnewline
VI &  & 7.10 &  & 8.653 &  & 1.3521 &  & $z$ &  & $\arrowvert H\rightarrow L+4\rangle(0.5807)$\tabularnewline
 &  &  &  &  &  &  &  &  &  & $\arrowvert H\rightarrow L+3\rangle(0.5414)$\tabularnewline
 &  &  &  &  &  &  &  &  &  & \tabularnewline
VII &  & 7.57 &  & 19.917 &  & 1.9866 &  & $y$ &  & $\arrowvert H-1\rightarrow L+2\rangle(0.6371)$\tabularnewline
 &  &  &  &  &  &  &  &  &  & $\arrowvert H-1\rightarrow L+4\rangle(0.4699)$\tabularnewline
 &  &  &  &  &  &  &  &  &  & \tabularnewline
VIII &  & 7.89 &  & 6.977 &  & 1.1518 &  & $y$ &  & $\arrowvert H-1\rightarrow L+3\rangle(0.8021)$\tabularnewline
 &  &  &  &  &  &  &  &  &  & $\arrowvert H-1\rightarrow L+2\rangle(0.2299)$\tabularnewline
 &  &  &  &  &  &  &  &  &  & \tabularnewline
IX &  & 8.51 &  & 1.821 &  & 0.5665 &  & $y$ &  & $\arrowvert H\rightarrow L;H-2\rightarrow L+1\rangle(0.7731)$\tabularnewline
 &  &  &  &  &  &  &  &  &  & $\arrowvert H-1\rightarrow L+4\rangle(0.2011)$\tabularnewline
\hline 
\end{tabular}
\end{table}

\begin{figure}[H]
\includegraphics[scale=0.14]{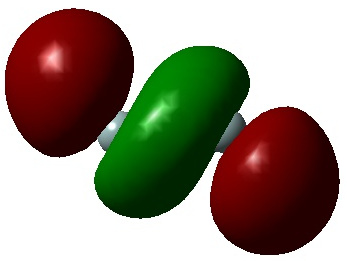}~~~~~~~~\includegraphics[scale=0.14]{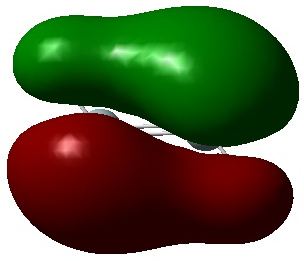}~~~~~~~~\includegraphics[scale=0.14]{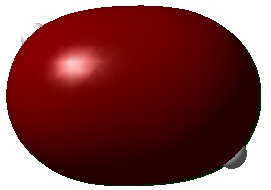}~~~~~~~~\includegraphics[scale=0.14]{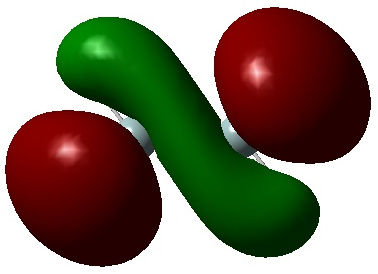}~~~~~~~~\includegraphics[scale=0.14]{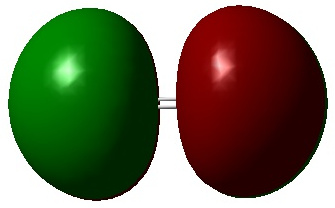}~~~~~~~~\includegraphics[scale=0.14]{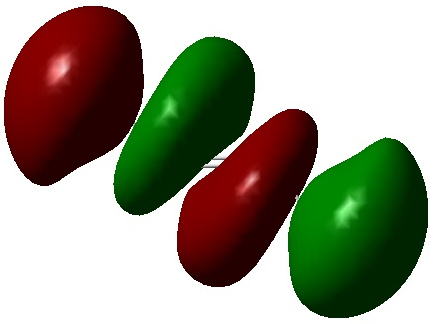}~~~~~~~~\includegraphics[scale=0.14]{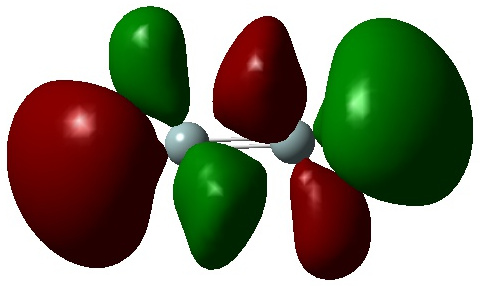}

~~\textbf{H-2}~~~~~~~~~~\textbf{~~~~H-1}~~~~~~~~~~~~~~\textbf{H}~~~~~~~~~~~~~~\textbf{~~~L}~~~~~~~~~~~~~~~\textbf{L+1~~~~~~~~~~~L+2~~~~~~~~~~~~~~~~~L+3}

\caption{{\small{}Various frontier molecular orbitals of trans-bent isomer
{[}main article, Fig. 4(d){]} of Si$_{2}$H$_{2}$ molecule, contributing
to optical transitions of comparatively high oscillator strength.
Symbols H and L denote the HOMO and the LUMO orbitals, while H-n (L+n)
represents n-th orbital below (above) HOMO (LUMO).}}
\end{figure}

\begin{table}[H]
{\small{}\caption{Many-particle wave functions of the excited states contributing to
the peaks in the optical absorption spectrum of trans-bent (HSi-SiH)
isomer {[}main article, Fig. 4(d){]} of Si$_{2}$H$_{2}$ molecule.
'E' corresponds to excitation energy (in eV) of an excited state,
whereas $f$, and ``TDM'' are defined in the caption of Table S1.
In the Palarization column, $\perp$ and $\parallel$ sign indicate
the absorption due to the light polarized perpendicularly, and along
the plane of the isomer, respectively. 'HF' corresponds to Hartree-Fock
configuration. '$H$' and '$L$' stand for HOMO and LUMO orbitals.
In the ``Wave function'' column, each number inside the parentheses
denotes the coefficient of the corresponding configuration in the
CI wave function. GS indicates the ground states wave function of
the isomer, and not that of an excited state corresponding to a peak.
The atomic coordinates corresponding to the optimized ground state
geometry of this isomer are presented in Table \ref{tab:geo-Trans-bent-si2h2}.}
}{\small\par}

{\small{}}%
\begin{tabular}{ccccccccccc}
\hline 
{\small{}Peak} & {\small{}\hspace{1cm}} & {\small{}E (eV)} & {\small{}\hspace{1cm}} & {\small{}$f$} & {\small{}\hspace{1cm}} & {\small{}|TDM|} & {\small{}\hspace{1cm}} & {\small{}Polarization} & {\small{}\hspace{1cm}} & {\small{}Wave function}\tabularnewline
\hline 
\hline 
{\small{}GS} &  &  &  &  &  &  &  &  &  & {\small{}$\arrowvert HF\rangle(0.8889)$}\tabularnewline
 &  &  &  &  &  &  &  &  &  & {\small{}$\arrowvert H-1\rightarrow L+1\rangle(0.1945)$}\tabularnewline
 &  &  &  &  &  &  &  &  &  & \tabularnewline
{\small{}I} &  & {\small{}1.161} &  & {\small{}0.210 } &  & 0.5214 &  & {\small{}$\perp$} &  & {\small{}$\arrowvert H\rightarrow L\rangle(0.9045)$}\tabularnewline
 &  &  &  &  &  &  &  &  &  & {\small{}$\arrowvert H\rightarrow L+11\rangle(0.0871)$}\tabularnewline
 &  &  &  &  &  &  &  &  &  & \tabularnewline
{\small{}II} &  & {\small{}2.965} &  & {\small{}10.079} &  & 2.24/0.29 &  & {\small{}$\parallel$} &  & {\small{}$\arrowvert H-1\rightarrow L\rangle(0.6892)$}\tabularnewline
 &  &  &  &  &  &  &  &  &  & {\small{}$\arrowvert H\rightarrow L+1\rangle(0.5865)$}\tabularnewline
 &  &  &  &  &  &  &  &  &  & \tabularnewline
{\small{}III} &  & {\small{}4.561} &  & {\small{}0.873} &  & 0.5357 &  & {\small{}$\perp$} &  & {\small{}$\arrowvert H-1\rightarrow L+1\rangle(0.8919)$}\tabularnewline
 &  &  &  &  &  &  &  &  &  & {\small{}$\arrowvert H-1\rightarrow L+14\rangle(0.1047)$}\tabularnewline
 &  &  &  &  &  &  &  &  &  & \tabularnewline
{\small{}IV} &  & {\small{}5.047} &  & {\small{}1.363} &  & 0.54/0.34 &  & {\small{}$\parallel$} &  & {\small{}$\arrowvert H\rightarrow L+1\rangle(0.6117)$}\tabularnewline
 &  &  &  &  &  &  &  &  &  & {\small{}$\arrowvert H-1\rightarrow L\rangle(0.5007)$}\tabularnewline
 &  &  &  &  &  &  &  &  &  & \tabularnewline
{\small{}V} &  & {\small{}6.004} &  & {\small{}0.411} &  & 0.19/0.26 &  & {\small{}$\parallel$} &  & {\small{}$\arrowvert H\rightarrow L;\thinspace H\rightarrow L+2\rangle(0.8233)$}\tabularnewline
 &  &  &  &  &  &  &  &  &  & {\small{}$\arrowvert H\rightarrow L;\thinspace H\rightarrow L+4\rangle(0.1640)$}\tabularnewline
 &  &  &  &  &  &  &  &  &  & \tabularnewline
{\small{}VI} &  & {\small{}6.276} &  & {\small{}3.976 } &  & 0.9748 &  & {\small{}$\perp$} &  & {\small{}$\arrowvert H-1\rightarrow L;\thinspace H\rightarrow L+2\rangle(0.5882)$}\tabularnewline
 &  &  &  &  &  &  &  &  &  & {\small{}$\arrowvert H\rightarrow L+3\rangle(0.4964)$}\tabularnewline
 &  &  &  &  &  &  &  &  &  & \tabularnewline
{\small{}VII} &  & {\small{}6.585} &  & {\small{}3.905 } &  & 0.9432 &  & {\small{}$\perp$} &  & {\small{}$\arrowvert H\rightarrow L+3\rangle(0.6696)$}\tabularnewline
 &  &  &  &  &  &  &  &  &  & {\small{}$\arrowvert H-1\rightarrow L;\thinspace H\rightarrow L+2\rangle(0.5031)$}\tabularnewline
 &  &  &  &  &  &  &  &  &  & \tabularnewline
{\small{}VIII} &  & {\small{}7.087} &  & {\small{}0.474} &  & 0.3168 &  & {\small{}$\perp$} &  & {\small{}$\arrowvert H-2\rightarrow L;\thinspace H\rightarrow L\rangle(0.6953)$}\tabularnewline
 &  &  &  &  &  &  &  &  &  & {\small{}$\arrowvert H\rightarrow L+2;\thinspace H\rightarrow L+1\rangle(0.3095)$}\tabularnewline
 &  &  &  &  &  &  &  &  &  & \tabularnewline
{\small{}IX} &  & {\small{}7.323} &  & {\small{}1.140} &  & 0.4833 &  & {\small{}$\perp$} &  & {\small{}$\arrowvert H-2\rightarrow L;\thinspace H\rightarrow L\rangle(0.4417)$}\tabularnewline
 &  &  &  &  &  &  &  &  &  & {\small{}$\arrowvert H\rightarrow L+2;\thinspace H\rightarrow L+1\rangle(0.4411)$}\tabularnewline
 &  & {\small{}7.419} &  & {\small{}0.466} &  & 0.20/0.23 &  & {\small{}$\parallel$} &  & {\small{}$\arrowvert H-2\rightarrow L;\thinspace H-1\rightarrow L\rangle(0.6161)$}\tabularnewline
 &  &  &  &  &  &  &  &  &  & {\small{}$\arrowvert H-2\rightarrow L+1;\thinspace H\rightarrow L\rangle(0.4922)$}\tabularnewline
\hline 
\end{tabular}{\small\par}
\end{table}

\begin{figure}[H]
\includegraphics[scale=0.2]{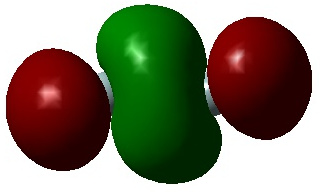}~~~~~~~~\includegraphics[scale=0.15]{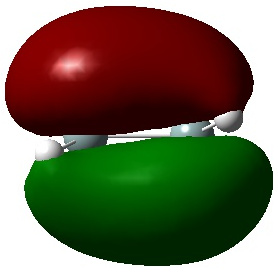}~~~~~~~~\includegraphics[scale=0.15]{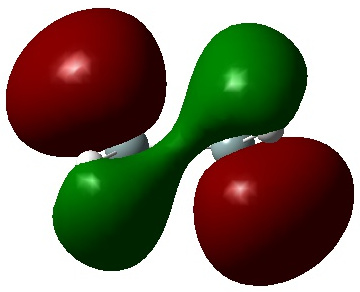}~~~~~~~~\includegraphics[scale=0.15]{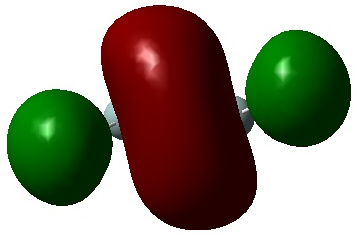}~~~~~~~~\includegraphics[scale=0.15]{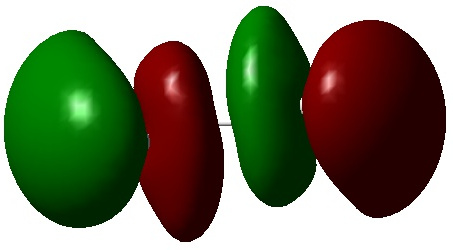}~~~~~~~~\includegraphics[scale=0.15]{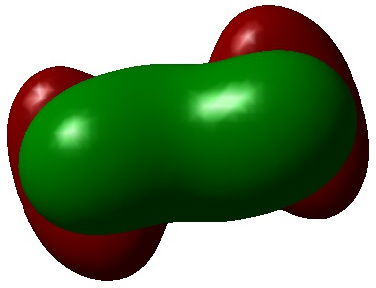}~~~~~~~~\includegraphics[scale=0.15]{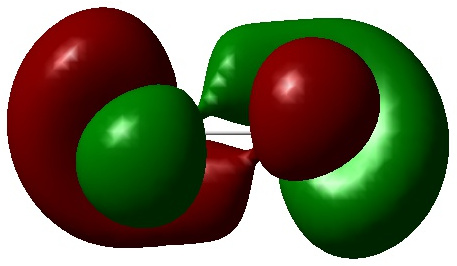}

~~\ ~\textbf{~~H-1}~~~~~~~~~~~~~~~\textbf{~H}~~~~~~~~~~~~~~~~~\textbf{L}~~~~~~~~~~~~~~~~~\textbf{L+1~~~~~~~~~~~~~~L+2~~~~~~~~~~~~~~~L+3~~~~~~~~~~~~~~L+4}

\caption{Various frontier molecular orbitals of disilene isomer {[}main article,
Fig. 9(a){]} of Si$_{2}$H$_{4}$ molecule, contributing to optical
transitions of comparatively high oscillator strength. Symbols H and
L denote the HOMO and the LUMO orbitals, while H-n (L+n) represents
n-th orbital below (above) HOMO (LUMO).}
\end{figure}

\begin{table}[H]
\caption{Many-particle wave functions of the excited states contributing to
the peaks in the optical absorption spectrum of disilene (H$_{2}$Si-SiH$_{2}$)
isomer {[}main article, Fig. 9(a){]} of Si$_{2}$H$_{4}$ molecule.
'E' corresponds to excitation energy (in eV) of an excited state,
whereas $f$, ``TDM'', and ``Polarization'' are defined in the
caption of Table S1. 'HF' corresponds to Hartree-Fock configuration.
'$H$' and '$L$' stand for HOMO and LUMO orbitals. In the ``Wave
function'' column, each number inside the parentheses denotes the
coefficient of the corresponding configuration in the CI wave function.
GS indicates the ground states wave function of the isomer, and not
that of an excited state corresponding to a peak. The atomic coordinates
corresponding to the optimized ground state geometry of this isomer
are presented in Table \ref{tab:geo-Disilene-si2h4}.}

\begin{tabular}{ccccccccccc}
\hline 
Peak & \hspace{1cm} & E (eV) & \hspace{1cm} & $f$ & \hspace{1cm} & |TDM| & \hspace{1cm} & Polarization & \hspace{1cm} & Wave function\tabularnewline
\hline 
\hline 
GS &  &  &  &  &  &  &  &  &  & $\arrowvert HF\rangle(0.9179)$\tabularnewline
 &  &  &  &  &  &  &  &  &  & $\arrowvert H\rightarrow L;\thinspace H\rightarrow L\rangle(0.1926)$\tabularnewline
 &  &  &  &  &  &  &  &  &  & \tabularnewline
I &  & 4.132 &  & 7.816 &  & 0.32/1.65 &  & x/y &  & $\arrowvert H\rightarrow L\rangle(0.8879)$\tabularnewline
 &  &  &  &  &  &  &  &  &  & $\arrowvert H-1\rightarrow L+2\rangle(0.1258)$\tabularnewline
 &  &  &  &  &  &  &  &  &  & \tabularnewline
II &  & 6.05 &  & 0.635 &  & 0.34/0.21 &  & x/y &  & $\arrowvert H\rightarrow L+3\rangle(0.7787)$\tabularnewline
 &  &  &  &  &  &  &  &  &  & $\arrowvert H\rightarrow L+4\rangle(0.4370)$\tabularnewline
 &  &  &  &  &  &  &  &  &  & \tabularnewline
III &  & 7.091 &  & 0.511 &  & 0.3287 &  & z &  & $\arrowvert H\rightarrow L;\thinspace H\rightarrow L+1\rangle(0.6929)$\tabularnewline
 &  &  &  &  &  &  &  &  &  & $\arrowvert H\rightarrow L+6\rangle(0.3992)$\tabularnewline
 &  & 7.309 &  & 14.375 &  & 1.72/0.04 &  & x/y &  & $\arrowvert H\rightarrow L+4\rangle(0.7321)$\tabularnewline
 &  &  &  &  &  &  &  &  &  & $\arrowvert H\rightarrow L+3\rangle(0.4620)$\tabularnewline
\hline 
\end{tabular}
\end{table}

$\pagebreak$

\begin{figure}[H]
\includegraphics[scale=0.12]{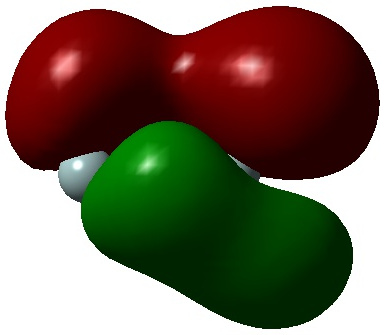}~~~~~~\includegraphics[scale=0.12]{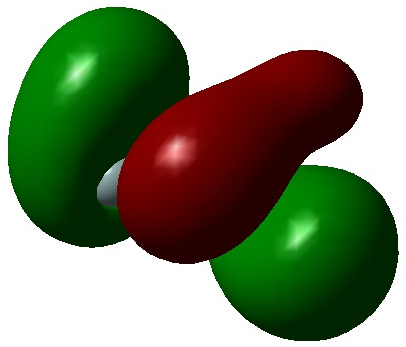}~~~~~~\includegraphics[scale=0.12]{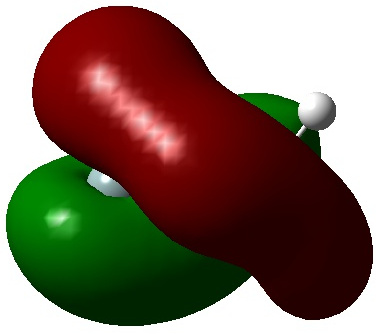}~~~~~~\includegraphics[scale=0.12]{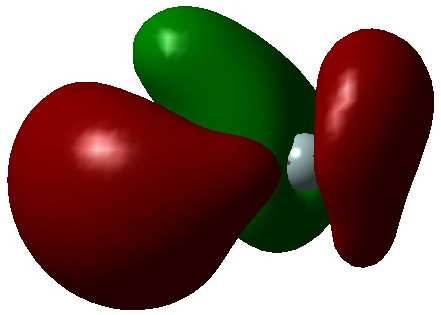}~~~~~~\includegraphics[scale=0.14]{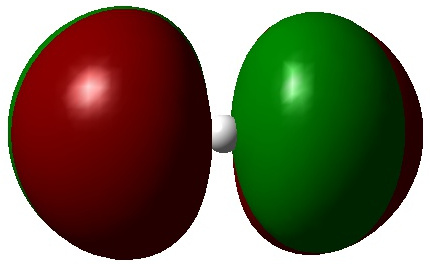}~~~~~~\includegraphics[scale=0.1]{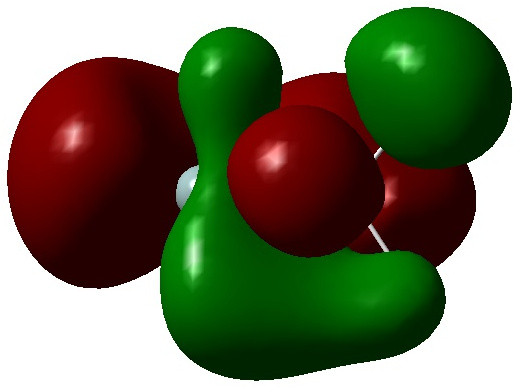}~~~~~~\includegraphics[scale=0.1]{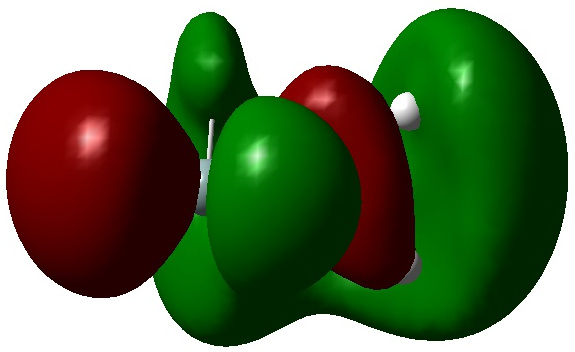}~~~~~~\includegraphics[scale=0.1]{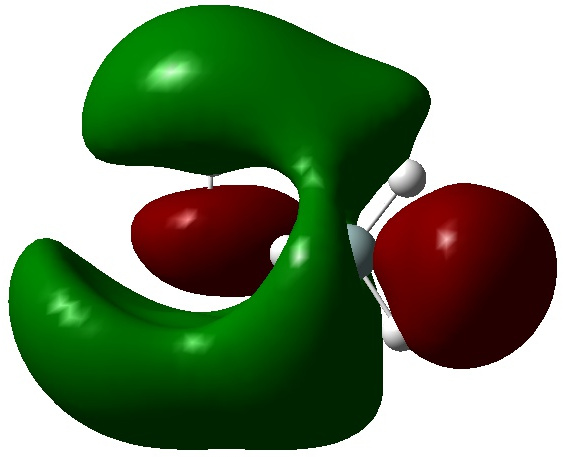}~~~~~~\includegraphics[scale=0.1]{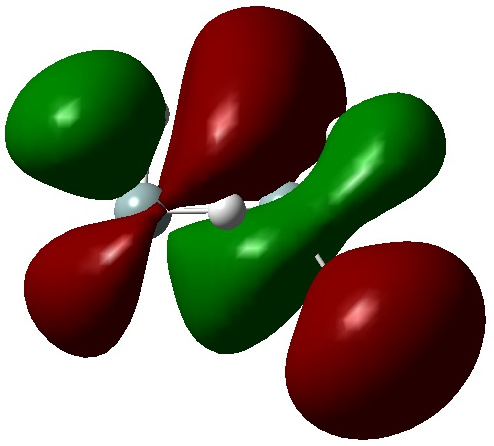}

~~\ \textbf{H-3}~~~~~~\textbf{~~~~H-2}~~~~~~~~~~~~~\textbf{H-1}~~~~~~~~~~~~~~~\textbf{H}~~~~~~~~~~~~~~~~~\textbf{L}~~~~~~~~~~~~~~~\textbf{L+1~~~~~~~~~~~~L+2~~~~~~~~~~~L+3~~~~~~~~~~L+4}

\caption{Various frontier molecular orbitals of mono-bridged isomer {[}main
article, Fig. 9(b){]} of Si$_{2}$H$_{4}$ molecule, contributing
to optical transitions of comparatively high oscillator strength.
Symbols H and L denote the HOMO and the LUMO orbitals, while H-n (L+n)
represents n-th orbital below (above) HOMO (LUMO).}
\end{figure}

\begin{table}[H]
\caption{Many-particle wave functions of the excited states contributing to
the peaks in the optical absorption spectrum of mono-bridged (H$_{2}$Si-H-SiH)
isomer {[}main article, Fig. 9(b){]} of Si$_{2}$H$_{4}$ molecule.
'E' corresponds to excitation energy (in eV) of an excited state,
whereas $f$, ``TDM'', and ``Polarization'' are defined in the
caption of Table S1. 'HF' corresponds to Hartree-Fock configuration.
'$H$' and '$L$' stand for HOMO and LUMO orbitals. In the ``Wave
function'' column, each number inside the parentheses denotes the
coefficient of the corresponding configuration in the CI wave function.
GS indicates the ground states wave function of the isomer, and not
that of an excited state corresponding to a peak. The atomic coordinates
corresponding to the optimized ground state geometry of this isomer
are presented in Table \ref{tab:geo-Mono-bridged-si2h4}. }

\begin{tabular}{ccccccccccc}
\hline 
Peak & \hspace{1cm} & E (eV) & \hspace{1cm} & $f$ & \hspace{1cm} & |TDM| & \hspace{1cm} & Polarization & \hspace{1cm} & Wave function\tabularnewline
\hline 
\hline 
GS &  &  &  &  &  &  &  &  &  & $\arrowvert HF\rangle(0.9248)$\tabularnewline
 &  &  &  &  &  &  &  &  &  & $\arrowvert H\rightarrow L;\thinspace H\rightarrow L\rangle(0.0647)$\tabularnewline
 &  &  &  &  &  &  &  &  &  & \tabularnewline
I &  & 3.831 &  & 0.815  &  & 0.34/0.45 &  & x/z &  & $\arrowvert H\rightarrow L\rangle(0.8789)$\tabularnewline
 &  &  &  &  &  &  &  &  &  & $\arrowvert H\rightarrow L+1\rangle(0.1414)$\tabularnewline
 &  &  &  &  &  &  &  &  &  & \tabularnewline
II &  & 4.845 &  & 0.309  &  & 0.24/0.20 &  & x/z &  & $\arrowvert H-1\rightarrow L\rangle(0.7965)$\tabularnewline
 &  &  &  &  &  &  &  &  &  & $\arrowvert H\rightarrow L+1\rangle(0.3411)$\tabularnewline
 &  &  &  &  &  &  &  &  &  & \tabularnewline
III &  & 5.439 &  & 4.036  &  & 1.0500 &  & x &  & $\arrowvert H\rightarrow L+1\rangle(0.7770)$\tabularnewline
 &  &  &  &  &  &  &  &  &  & $\arrowvert H-1\rightarrow L\rangle(0.3132)$\tabularnewline
 &  &  &  &  &  &  &  &  &  & \tabularnewline
IV &  & 6.576 &  & 0.696  &  & 0.24/0.30 &  & y/z &  & $\arrowvert H-1\rightarrow L+1\rangle(0.6786)$\tabularnewline
 &  &  &  &  &  &  &  &  &  & $\arrowvert H\rightarrow L+2\rangle(0.4134)$\tabularnewline
 &  &  &  &  &  &  &  &  &  & \tabularnewline
V &  & 6.862 &  & 1.260  &  & 0.22/0.47 &  & y/z &  & $\arrowvert H\rightarrow L+2\rangle(0.5174)$\tabularnewline
 &  &  &  &  &  &  &  &  &  & $\arrowvert H-1\rightarrow L+1\rangle(0.4870)$\tabularnewline
 &  & 6.976 &  & 2.346  &  & 0.7070 &  & x &  & $\arrowvert H\rightarrow L+3\rangle(0.5177)$\tabularnewline
 &  &  &  &  &  &  &  &  &  & $\arrowvert H-2\rightarrow L\rangle(0.5144)$\tabularnewline
 &  &  &  &  &  &  &  &  &  & \tabularnewline
VI &  & 7.249 &  & 3.596  &  & 0.84/0.20 &  & x/z &  & $\arrowvert H-2\rightarrow L\rangle(0.6016)$\tabularnewline
 &  &  &  &  &  &  &  &  &  & $\arrowvert H\rightarrow L+4\rangle(0.3915)$\tabularnewline
 &  &  &  &  &  &  &  &  &  & \tabularnewline
VII &  & 7.430 &  & 2.481 &  & 0.47/0.45/0.27 &  & x/y/z &  & $\arrowvert H\rightarrow L;\thinspace H\rightarrow L\rangle(0.5338)$\tabularnewline
 &  &  &  &  &  &  &  &  &  & $\arrowvert H-3\rightarrow L\rangle(0.3569)$\tabularnewline
 &  &  &  &  &  &  &  &  &  & \tabularnewline
VIII &  & 7.801 &  & 6.823  &  & 1.02/0.50/0.18 &  & x/y/z &  & $\arrowvert H\rightarrow L+4\rangle(0.5845)$\tabularnewline
 &  &  &  &  &  &  &  &  &  & $\arrowvert H-3\rightarrow L\rangle(0.4178)$\tabularnewline
\hline 
\end{tabular}
\end{table}

\begin{figure}[H]
\includegraphics[scale=0.15]{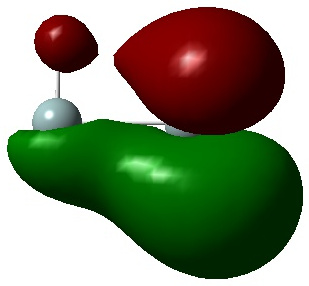}~~~~~~~~\includegraphics[scale=0.15]{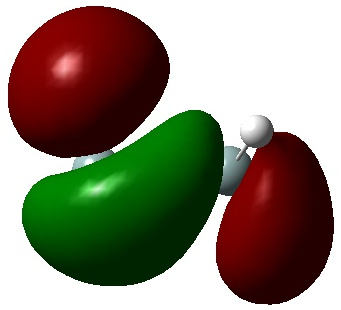}~~~~~~~~\includegraphics[scale=0.15]{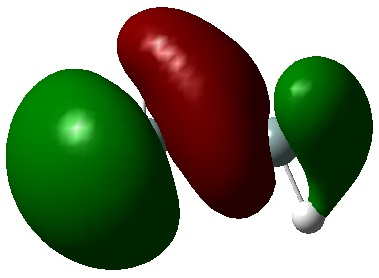}~~~~~~~~\includegraphics[scale=0.15]{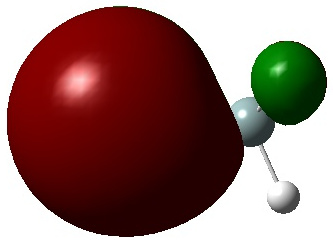}~~~~~~~~\includegraphics[scale=0.12]{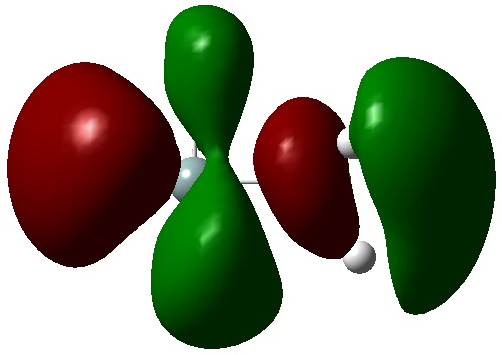}~~~~~~~~\includegraphics[scale=0.12]{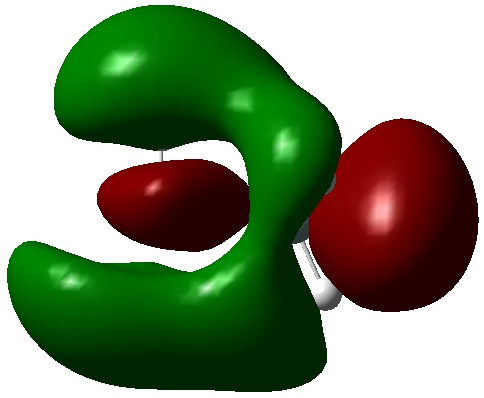}~~~~~~~~\includegraphics[scale=0.12]{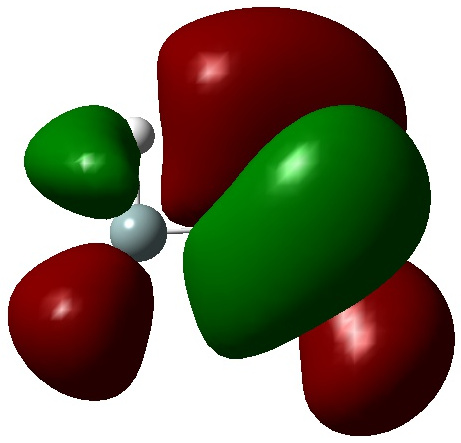}

~~\ \textbf{H-2}~~~~~~~~~~~~~~\textbf{H-1}~~~~~~~~~~~~~~~~~~~\textbf{H}~~~~~~~~~~~~~~~~~\textbf{L}~~~~~~~~~~~~~~~~~\textbf{L+1}~~~~~~~~~~~~~~~~\textbf{L+2~~~~~~~~~~~~~~L+3}

\caption{Various frontier molecular orbitals of silylsilylene isomer {[}main
article, Fig. 9(c){]} of Si$_{2}$H$_{4}$ molecule, contributing
to optical transitions of comparatively high oscillator strength.
Symbols H and L denote the HOMO and the LUMO orbitals, while H-n (L+n)
represents n-th orbital below (above) HOMO (LUMO).}
\end{figure}

\begin{table}[H]
\caption{Many-particle wave functions of the excited states contributing to
the peaks in the optical absorption spectrum of silylsilylene (H$_{3}$Si-SiH)
isomer {[}main article, Fig. 9(c){]} of Si$_{2}$H$_{4}$ molecule.
'E' corresponds to excitation energy (in eV) of an excited state,
whereas $f$, ``TDM'', and ``Polarization'' are defined in the
caption of Table S1. 'HF' corresponds to Hartree-Fock configuration.
'$H$' and '$L$' stand for HOMO and LUMO orbitals. In the ``Wave
function'' column, each number inside the parentheses denotes the
coefficient of the corresponding configuration in the CI wave function.
GS indicates the ground states wave function of the isomer, and not
that of an excited state corresponding to a peak. The atomic coordinates
corresponding to the optimized ground state geometry of this isomer
are presented in Table \ref{tab:geo-Silylsilylene-si2h4}.}

\begin{tabular}{ccccccccccc}
\hline 
Peak & \hspace{1cm} & E (eV) & \hspace{1cm} & $f$ & \hspace{1cm} & |TDM| & \hspace{1cm} & Polarization & \hspace{1cm} & Wave function\tabularnewline
\hline 
\hline 
GS &  &  &  &  &  &  &  &  &  & $\arrowvert HF\rangle(0.9250)$\tabularnewline
 &  &  &  &  &  &  &  &  &  & $\arrowvert H\rightarrow L;\thinspace H\rightarrow L\rangle(0.0978)$\tabularnewline
 &  &  &  &  &  &  &  &  &  & \tabularnewline
I &  & 1.935 &  & 0.131  &  & 0.3187 &  & z &  & $\arrowvert H\rightarrow L\rangle(0.9163)$\tabularnewline
 &  &  &  &  &  &  &  &  &  & $\arrowvert H\rightarrow L+15\rangle(0.0960)$\tabularnewline
 &  &  &  &  &  &  &  &  &  & \tabularnewline
II &  & 3.795 &  & 0.136  &  & 0.2316 &  & z &  & $\arrowvert H-1\rightarrow L\rangle(0.8985)$\tabularnewline
 &  &  &  &  &  &  &  &  &  & $\arrowvert H-1\rightarrow L+15\rangle(0.1136)$\tabularnewline
 &  &  &  &  &  &  &  &  &  & \tabularnewline
III &  & 6.105 &  & 0.505  &  & 0.3522 &  & x &  & $\arrowvert H-1\rightarrow L;\thinspace H\rightarrow L\rangle(0.8374)$\tabularnewline
 &  &  &  &  &  &  &  &  &  & $\arrowvert H\rightarrow L+1\rangle(0.2040)$\tabularnewline
 &  & 6.257 &  & 0.216 &  & 0.2267 &  & z &  & $\arrowvert H-2\rightarrow L\rangle(0.8484)$\tabularnewline
 &  &  &  &  &  &  &  &  &  & $\arrowvert H-2\rightarrow L;\thinspace H\rightarrow L+1\rangle(0.1526)$\tabularnewline
 &  &  &  &  &  &  &  &  &  & \tabularnewline
IV &  & 6.639 &  & 1.186  &  & 0.20/0.48 &  & x/y &  & $\arrowvert H\rightarrow L+2\rangle(0.6048)$\tabularnewline
 &  &  &  &  &  &  &  &  &  & $\arrowvert H\rightarrow L+1\rangle(0.5465)$\tabularnewline
 &  &  &  &  &  &  &  &  &  & \tabularnewline
V &  & 6.971 &  & 15.375  &  & 1.8178 &  & x &  & $\arrowvert H\rightarrow L+1\rangle(0.5637)$\tabularnewline
 &  &  &  &  &  &  &  &  &  & $\arrowvert H\rightarrow L+2\rangle(0.5528)$\tabularnewline
 &  &  &  &  &  &  &  &  &  & \tabularnewline
VI &  & 7.354 &  & 4.986  &  & 0.85/0.54 &  & x/y &  & $\arrowvert H\rightarrow L+3\rangle(0.7725)$\tabularnewline
 &  &  &  &  &  &  &  &  &  & $\arrowvert H-1\rightarrow L+1\rangle(0.2561)$\tabularnewline
\hline 
\end{tabular}
\end{table}

$\pagebreak$

\begin{figure}[H]
~~~~~~~~~~~~~~~~~~~~~~~~~\includegraphics[scale=0.15]{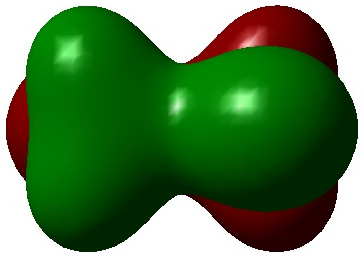}~~~~~~~~~\includegraphics[scale=0.15]{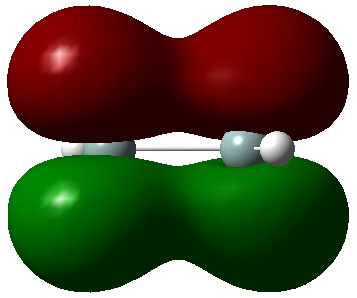}~~~~~~~~~\includegraphics[scale=0.15]{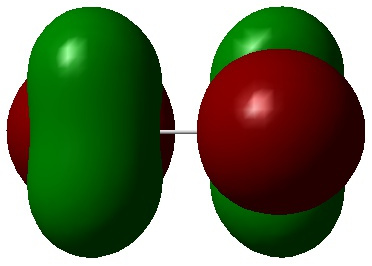}~~~~~~~~~\includegraphics[scale=0.15]{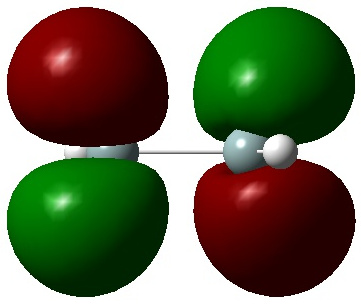}~~~~~~~~~\includegraphics[scale=0.15]{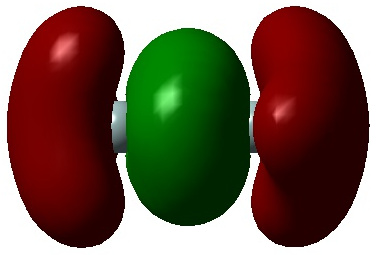}

~~~~~~~~~~~~~~~~~~~~~\textbf{~~~~~(H-2)}\textsubscript{\textbf{2}}~~~~~~~~~~~~\textbf{~(H-2)}\textsubscript{\textbf{1}}~~~~~~~~~\textbf{~~~(H-1)}\textsubscript{\textbf{2}}~~~~~~~~~~~~~\textbf{(H-1)}\textsubscript{\textbf{1}}~~~~~~~~~~~\textbf{~~~~H}

$\newline$

~~~~~~~~~~~~~~~~~~~~~~~~~\includegraphics[scale=0.12]{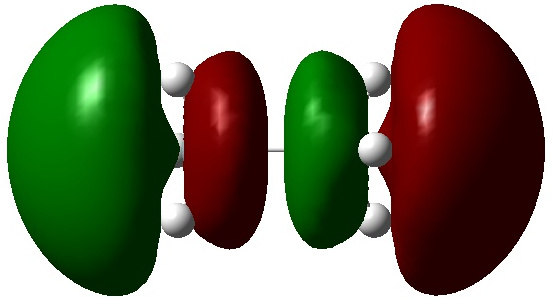}~~~~~~~~~\includegraphics[scale=0.12]{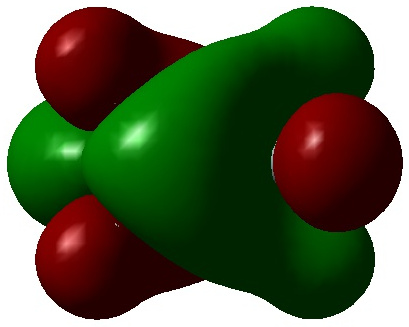}~~~~~~~~~\includegraphics[scale=0.12]{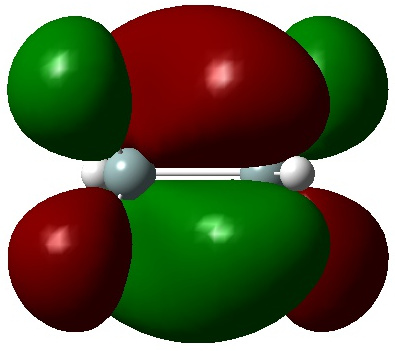}~~~~~~~~~\includegraphics[scale=0.12]{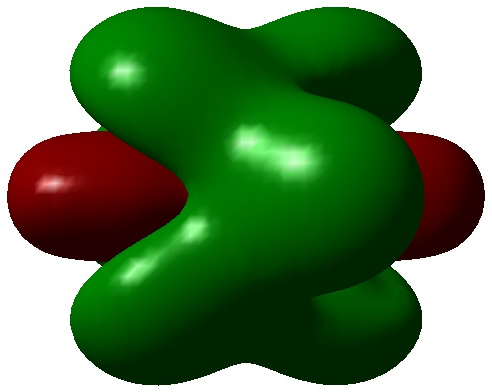}~~~~~~~~~\includegraphics[scale=0.12]{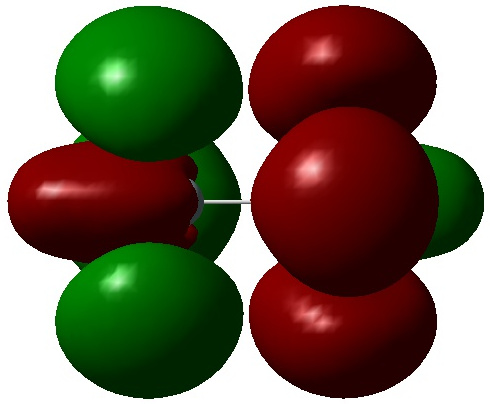}

~~~~~~~~~~~~~~~~~~~~~~~~~\textbf{~~~~~~L}~~~~~~~~~~~~~~~~~~~\textbf{L+1}~~~~~~~~~~~\textbf{~(L+2)}\textsubscript{\textbf{1}}~~~~~~~~~~~~\textbf{(L+2)}\textsubscript{\textbf{2}}~~~~~~~~~~~~~~\textbf{L+4}

\caption{{\small{}Various frontier molecular orbitals of disilane isomer {[}main
article, Fig. 2{]} of Si$_{2}$H$_{6}$ molecule, contributing to
optical transitions of comparatively high oscillator strength. Symbols
H and L denote the HOMO and the LUMO orbitals, while H-n (L+n) represents
n-th orbital below (above) HOMO (LUMO). Subscripts are used to index
the degenerate orbitals.}}
\end{figure}

\begin{table}[H]
{\small{}\caption{Many-particle wave functions of the excited states contributing to
the peaks in the optical absorption spectrum of disilane (H$_{3}$Si-SiH$_{3}$)
isomer {[}main article, Fig. 2{]} of Si$_{2}$H$_{6}$ molecule. 'E'
corresponds to excitation energy (in eV) of an excited state, whereas
$f$, ``TDM'', and ``Polarization'' are defined in the caption
of Table S1. 'HF' corresponds to Hartree-Fock configuration. '$H$'
and '$L$' stand for HOMO and LUMO orbitals. In the ``Wave function''
column, each number inside the parentheses denotes the coefficient
of the corresponding configuration in the CI wave function. GS indicates
the ground states wave function of the isomer, and not that of an
excited state corresponding to a peak. The atomic coordinates corresponding
to the optimized ground state geometry of this isomer are presented
in Table \ref{tab:geo-Disilane-si2h6}.}
}{\small\par}

{\small{}}%
\begin{tabular}{ccccccccccc}
\hline 
{\small{}Peak} & {\small{}\hspace{1cm}} & {\small{}E (eV)} & {\small{}\hspace{1cm}} & {\small{}$f$} & {\small{}\hspace{1cm}} & |TDM| & {\small{}\hspace{1cm}} & {\small{}Polarization} & {\small{}\hspace{1cm}} & {\small{}Wave function}\tabularnewline
\hline 
\hline 
{\small{}GS} &  &  &  &  &  &  &  &  &  & {\small{}$\arrowvert HF\rangle(0.9409)$}\tabularnewline
 &  &  &  &  &  &  &  &  &  & \tabularnewline
{\small{}I} &  & {\small{}7.767} &  & {\small{}2.976} &  & 0.74/0.18 &  & {\small{}x/y} &  & {\small{}$\arrowvert H\rightarrow(L+2)_{2}\rangle(0.8791)$}\tabularnewline
 &  &  &  &  &  &  &  &  &  & {\small{}$\arrowvert(H-1)_{1}\rightarrow L\rangle(0.1246)$}\tabularnewline
 &  & {\small{}7.812} &  & {\small{}5.984} &  & 1.0719 &  & {\small{}z} &  & {\small{}$\arrowvert H\rightarrow(L+2)_{1}\rangle(0.8809)$}\tabularnewline
 &  &  &  &  &  &  &  &  &  & {\small{}$\arrowvert(H-2)_{1}\rightarrow L\rangle(0.1257)$}\tabularnewline
 &  &  &  &  &  &  &  &  &  & \tabularnewline
{\small{}II} &  & {\small{}8.445} &  & {\small{}20.638} &  & 0.31/1.89 &  & {\small{}x/y} &  & {\small{}$\arrowvert H\rightarrow L\rangle(0.8486)$}\tabularnewline
 &  &  &  &  &  &  &  &  &  & {\small{}$\arrowvert(H-1)_{1}\rightarrow(L+2)_{2}\rangle(0.1901)$}\tabularnewline
 &  &  &  &  &  &  &  &  &  & \tabularnewline
{\small{}III} &  & {\small{}9.208} &  & {\small{}6.959} &  & 1.0647 &  & {\small{}z} &  & {\small{}$\arrowvert(H-1)_{1}\rightarrow(L+2)_{1}\rangle(0.6114)$}\tabularnewline
 &  &  &  &  &  &  &  &  &  & {\small{}$\arrowvert(H-1)_{2}\rightarrow(L+2)_{2}\rangle(0.5962)$}\tabularnewline
 &  &  &  &  &  &  &  &  &  & \tabularnewline
{\small{}IV} &  & {\small{}9.404} &  & {\small{}8.843} &  & 0.31/1.15 &  & {\small{}x/y} &  & {\small{}$\arrowvert(H-1)_{1}\rightarrow L\rangle(0.5695)$}\tabularnewline
 &  &  &  &  &  &  &  &  &  & {\small{}$\arrowvert(H-1)_{1}\rightarrow(L+2)_{2}\rangle(0.4532)$}\tabularnewline
 &  & {\small{}9.471} &  & {\small{}0.800} &  & 0.3559 &  & {\small{}z} &  & {\small{}$\arrowvert(H-2)_{1}\rightarrow L\rangle(0.5852)$}\tabularnewline
 &  &  &  &  &  &  &  &  &  & {\small{}$\arrowvert(H-1)_{2}\rightarrow(L+2)_{2}\rangle(0.4249)$}\tabularnewline
 &  &  &  &  &  &  &  &  &  & \tabularnewline
{\small{}V} &  & {\small{}9.816} &  & {\small{}8.668} &  & 0.62/0.97 &  & {\small{}x/y} &  & {\small{}$\arrowvert(H-1)_{1}\rightarrow(L+2)_{2}\rangle(0.6397)$}\tabularnewline
 &  &  &  &  &  &  &  &  &  & {\small{}$\arrowvert(H-1)_{1}\rightarrow L\rangle(0.3825)$}\tabularnewline
 &  & {\small{}9.924} &  & {\small{}1.031} &  & 0.3947 &  & {\small{}z} &  & {\small{}$\arrowvert(H-2)_{1}\rightarrow L\rangle(0.6167)$}\tabularnewline
 &  &  &  &  &  &  &  &  &  & {\small{}$\arrowvert(H-1)_{2}\rightarrow(L+2)_{2}\rangle(0.4176)$}\tabularnewline
 &  &  &  &  &  &  &  &  &  & \tabularnewline
{\small{}VI} &  & {\small{}10.161} &  & {\small{}2.259} &  & 0.53/0.22 &  & {\small{}x/y} &  & {\small{}$\arrowvert H\rightarrow L+4\rangle(0.7547)$}\tabularnewline
 &  &  &  &  &  &  &  &  &  & {\small{}$\arrowvert(H-1)_{2}\rightarrow(L+2)_{1}\rangle(0.3177)$}\tabularnewline
\hline 
\end{tabular}{\small\par}
\end{table}

\section{Influence of Basis Sets on the Optical Absorption Spectrum}

We computed the optical absorption spectra of trans-bent (HSi-SiH)
molecule using both, cc-pVTZ and aug-cc-pVTZ basis sets and MRSDCI
methodology to verify whether the computed spectrum has converged
with respect to the basis set. The main difference between the spectra
computed using two basis sets is the intensity of the main peak. Computed
peak positions are in very good agreement.

\begin{figure}[H]
\begin{centering}
\includegraphics[scale=0.35]{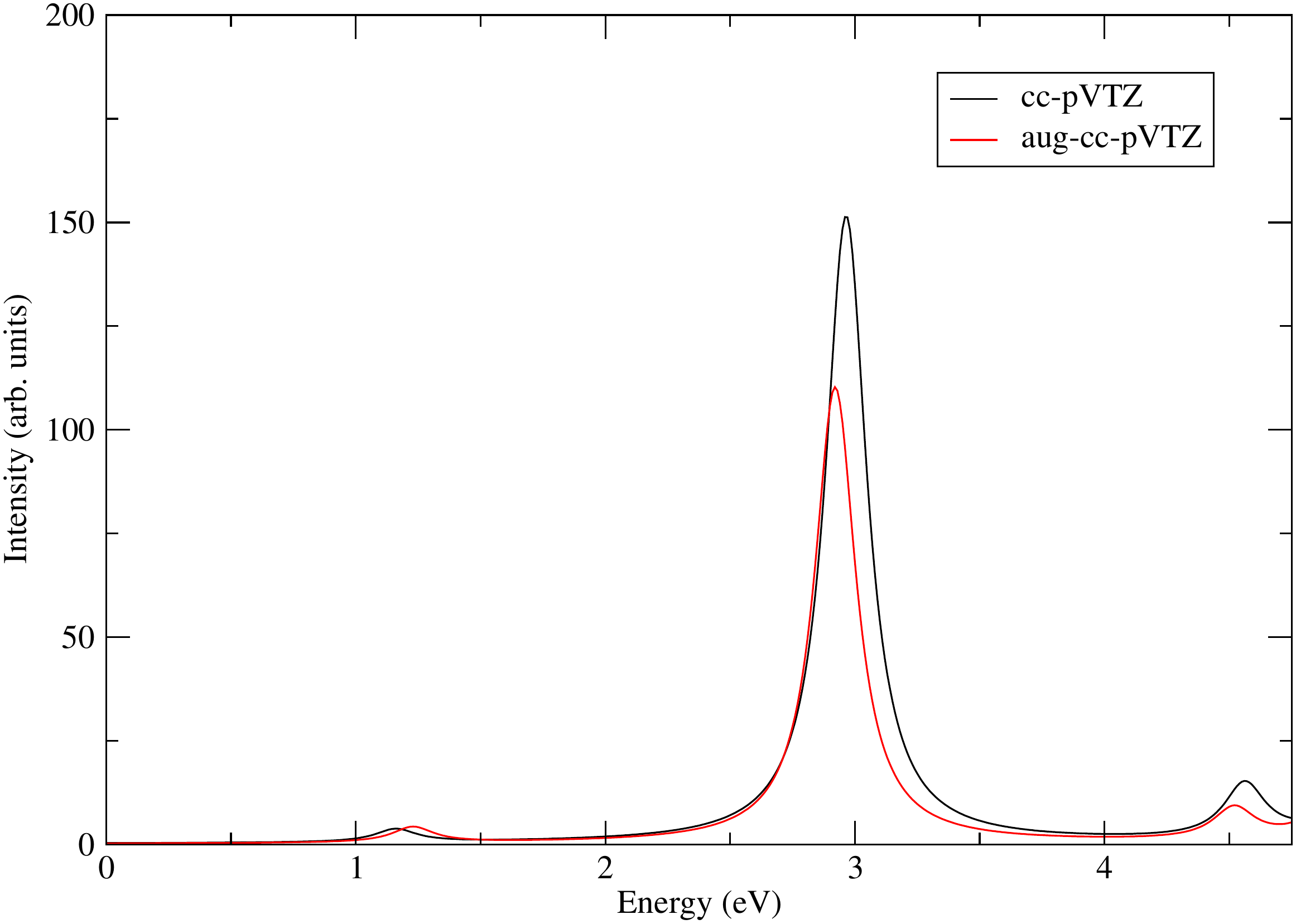}
\par\end{centering}
\caption{{\small{}Optical absorption spectrum of trans-bent (HSi-SiH) conformer,
calculated using the MRSDCI method, and the cc-pVTZ and aug-cc-pVTZ
basis sets. For plotting the spectrum, a uniform line-width of 0.1
eV was used.}}
\end{figure}

\section{Optimized Excited state geometries}

In this section we present the optimized geometries of various excited
states in the optical absorption spectra of various clusters. For
details refer to section 2 of the main text.

\subsection{Dibridged disilyne (Si-H\protect\textsubscript{2}-Si) isomer}

\begin{figure}[H]
\begin{centering}
\subfloat[Peak-I]{\includegraphics[scale=0.35]{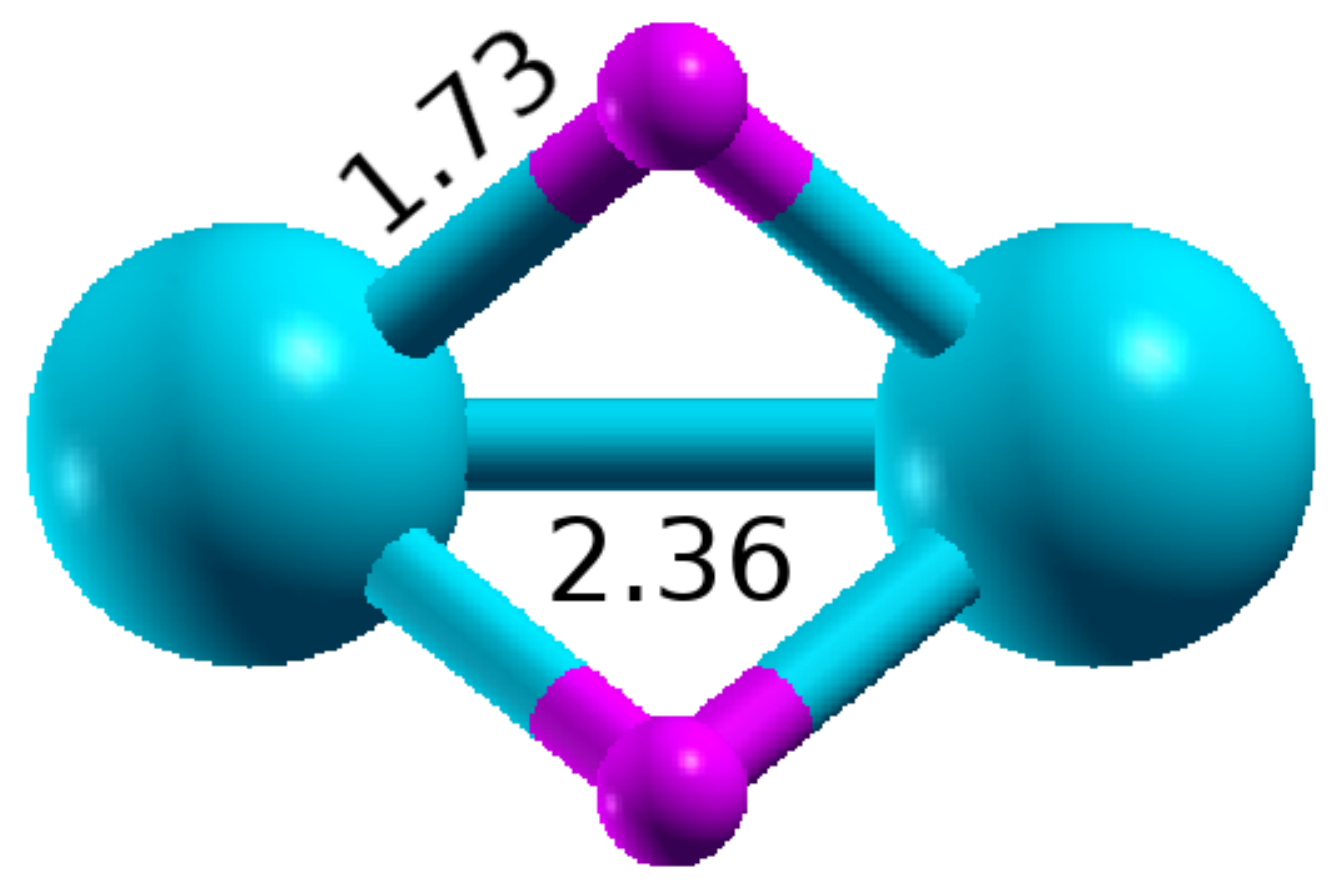}

}\subfloat[Peak-III]{\includegraphics[scale=0.3]{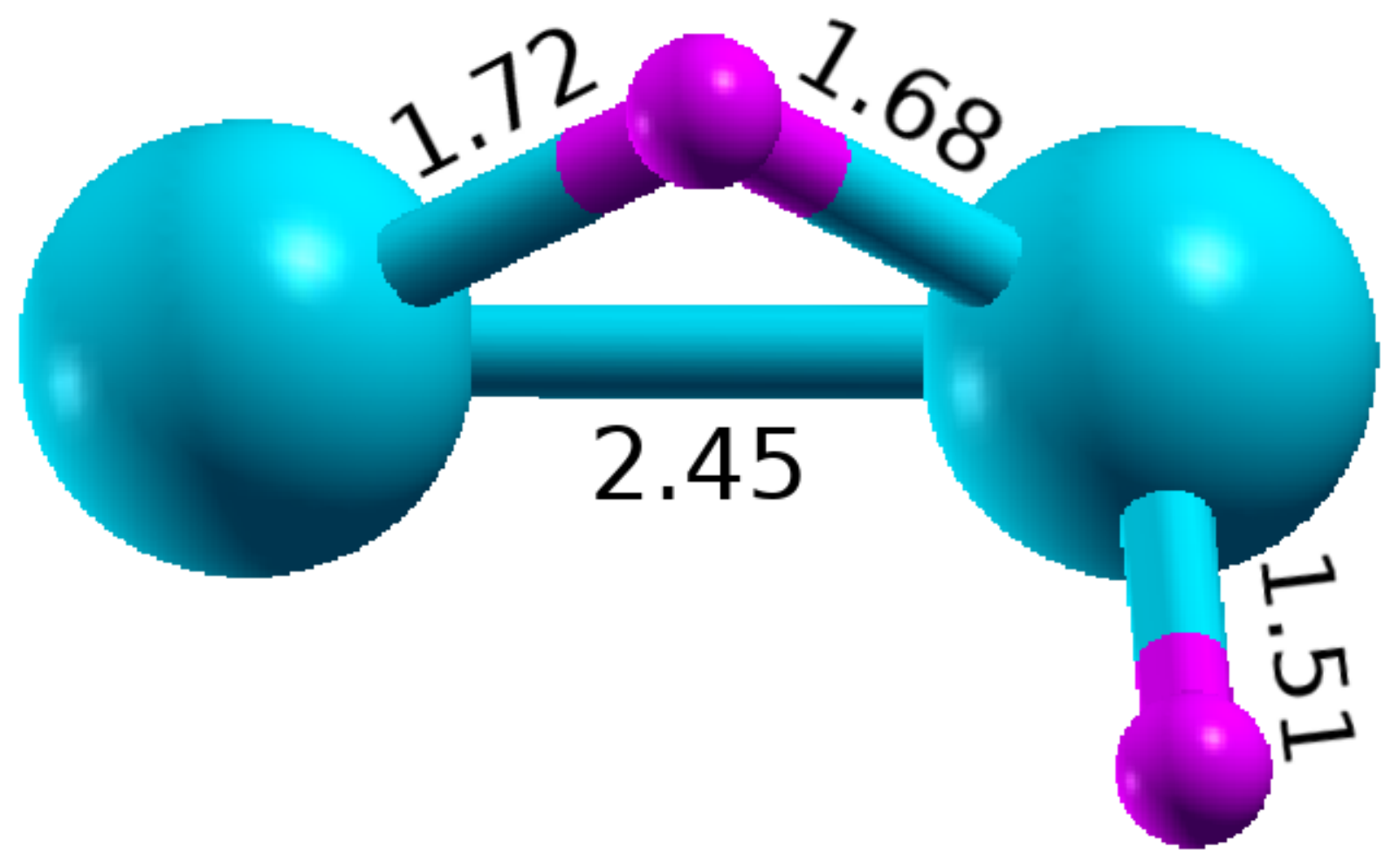}

}\subfloat[Peak-IV]{\includegraphics[scale=0.3]{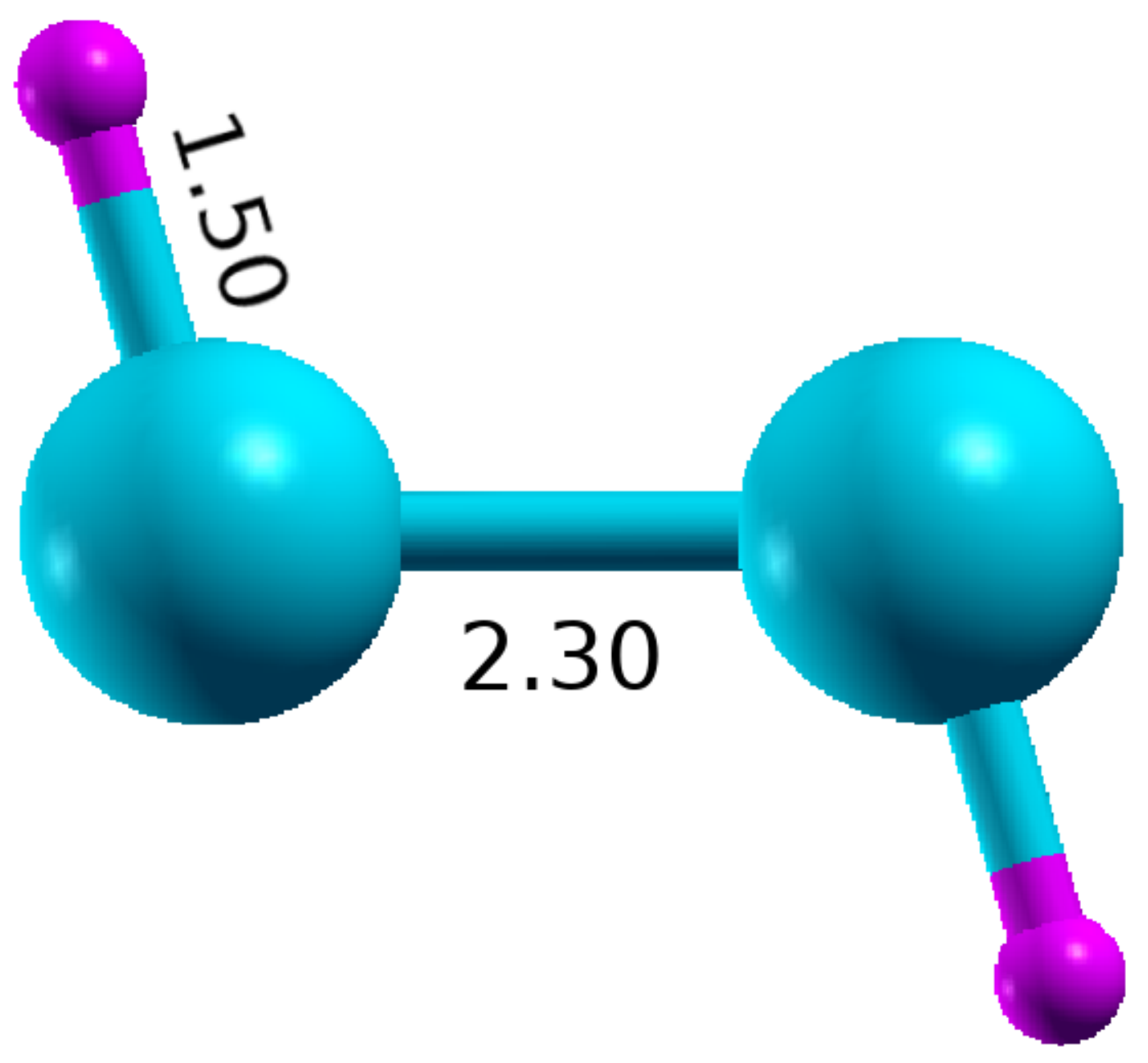}

}
\par\end{centering}
\caption{Optimized geometries of the excited states corresponding to various
peaks of the absorption spectrum of dibridged disilyne (Si-H\protect\textsubscript{2}-Si)
isomer {[}main article, Fig. 4(a){]}. All the bond lengths are in
$\text{Å}$ unit.}
\end{figure}

\subsection{Monobridged (Si-H-SiH) isomer}

\begin{figure}[H]
\begin{centering}
\subfloat[Peak-II]{\includegraphics[scale=0.35]{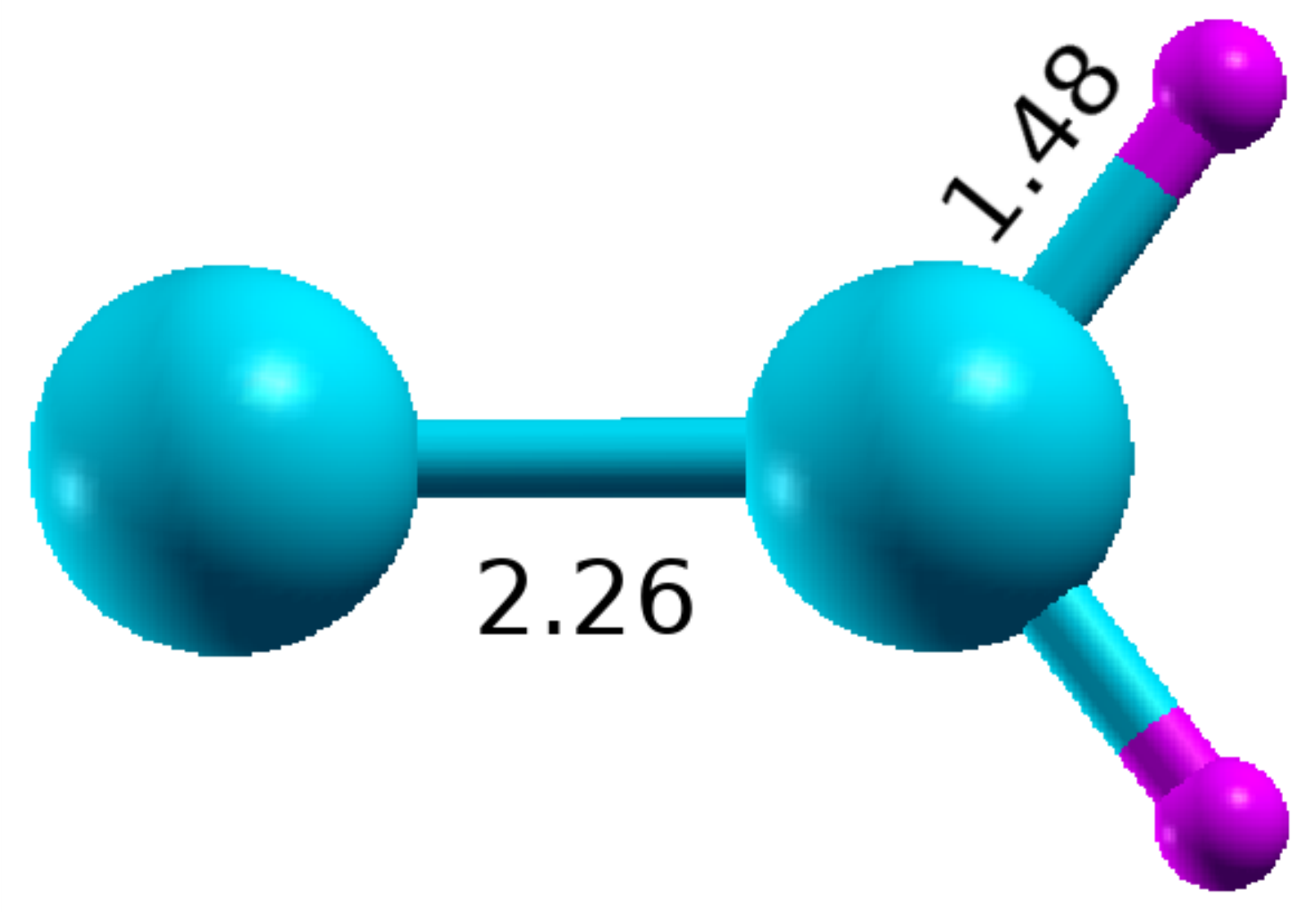}

}\subfloat[Peak-III]{\includegraphics[scale=0.35]{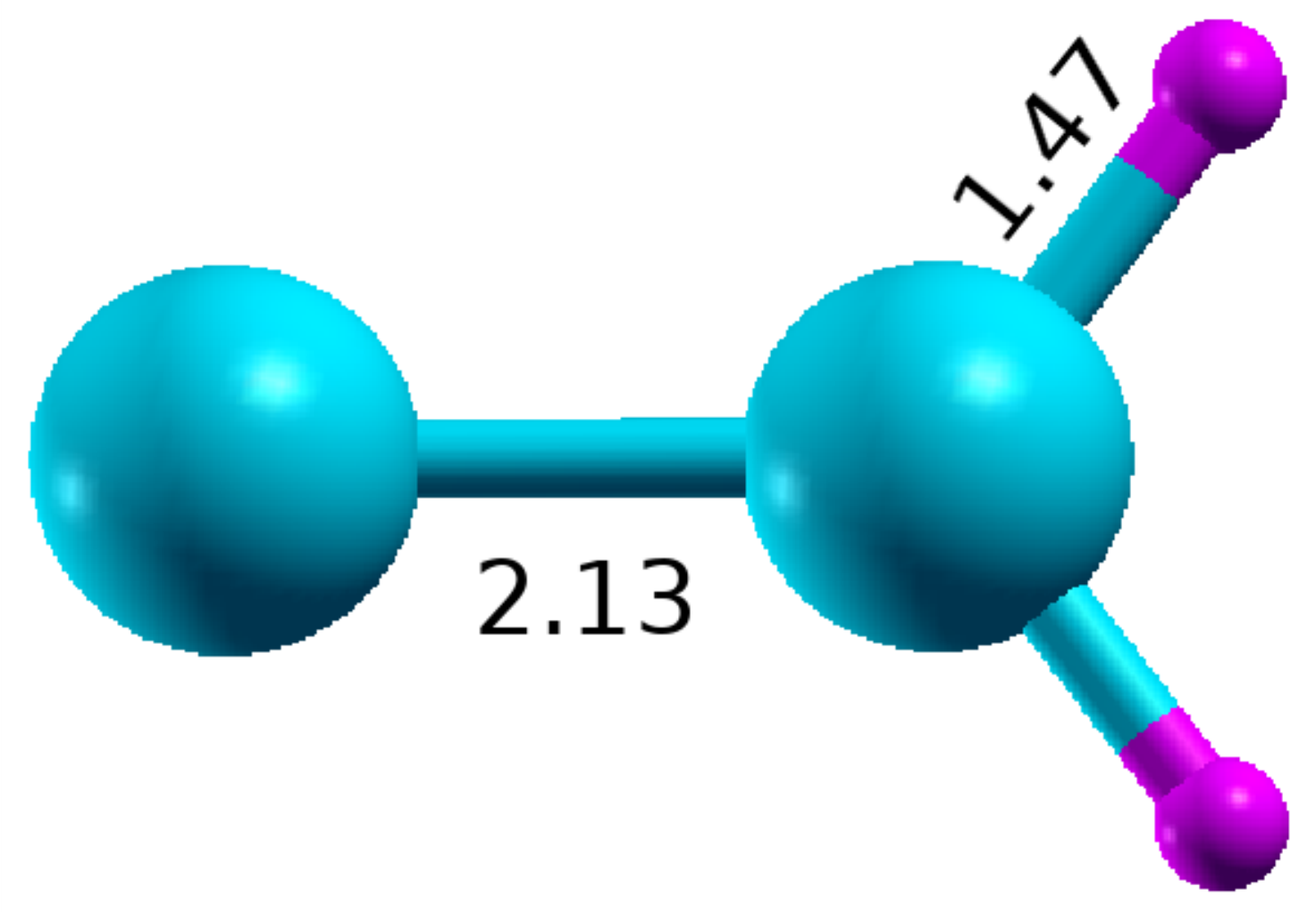}

}\subfloat[Peak-VIII]{\includegraphics[scale=0.3]{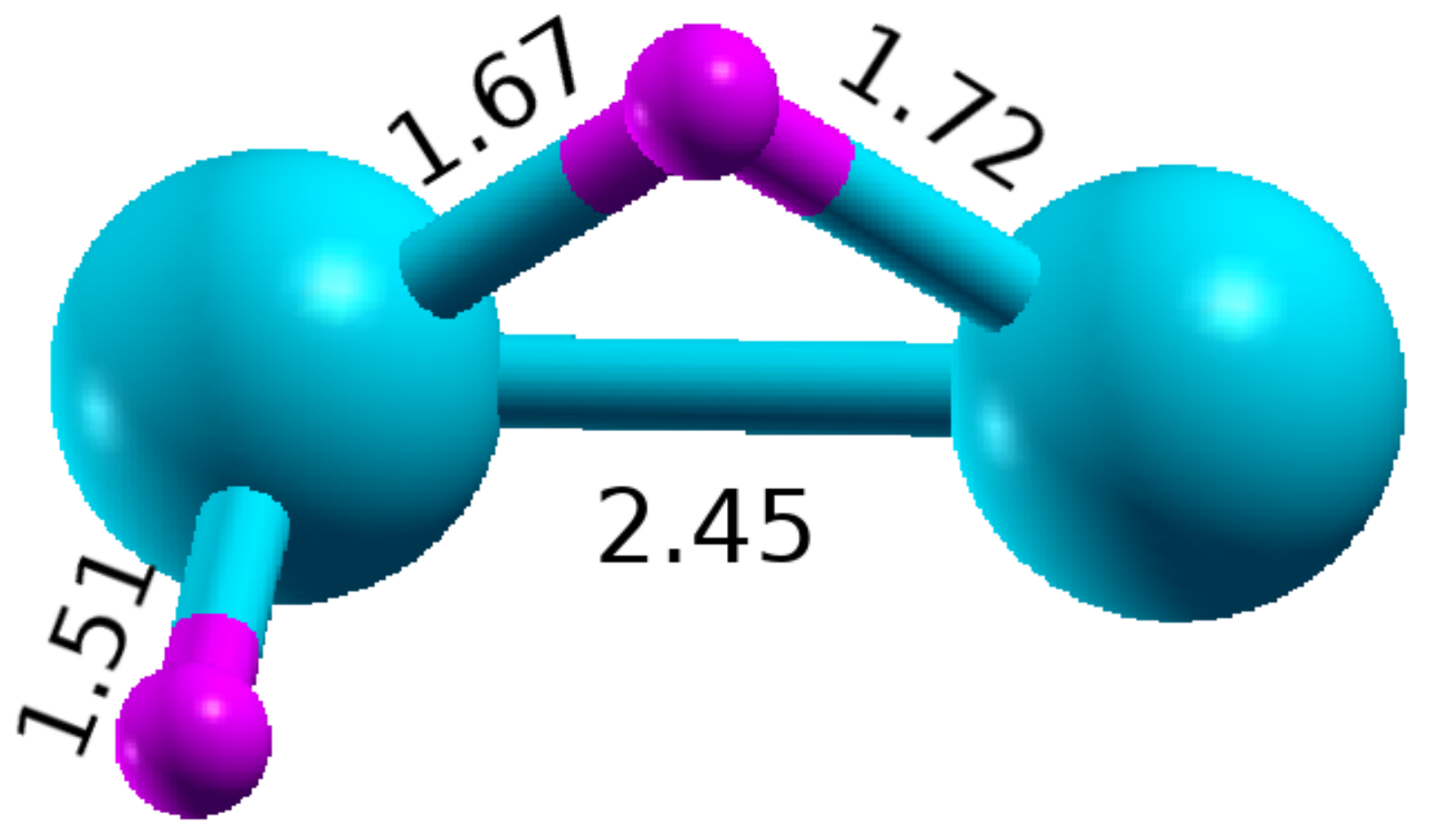}

}
\par\end{centering}
\caption{Optimized excited state geometries corresponding to various peaks
of the absorption spectrum of monobridged (Si-H-SiH) isomer {[}main
article, Fig. 4(b){]}. All the bond lengths are in $\text{Å}$ unit.}
\end{figure}

\subsection{Disilavinylidene (Si-SiH\protect\textsubscript{2}) isomer}

\begin{figure}[H]
\begin{centering}
\subfloat[Peak-III]{\includegraphics[scale=0.35]{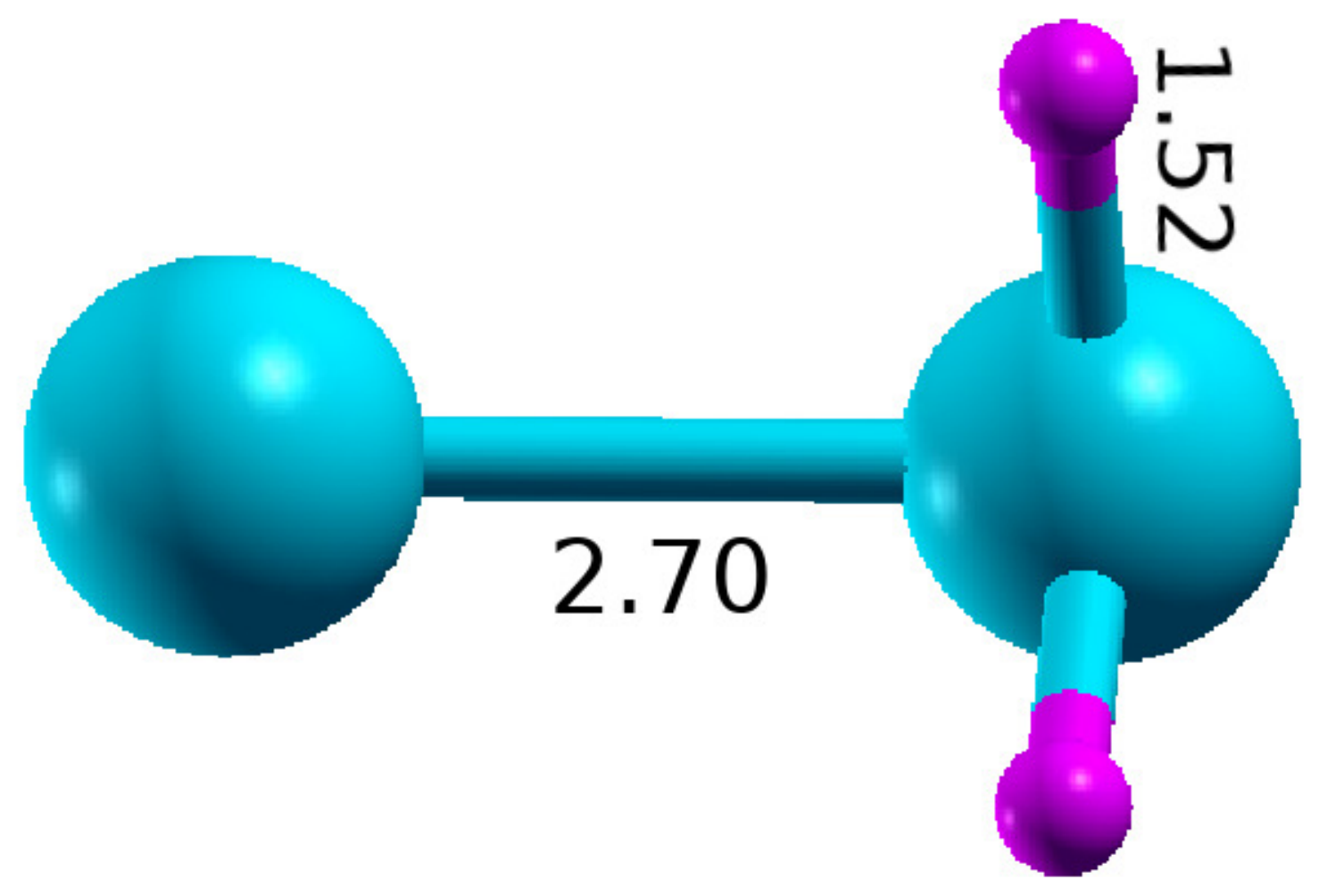}

}\subfloat[Peak-VII]{\includegraphics[scale=0.35]{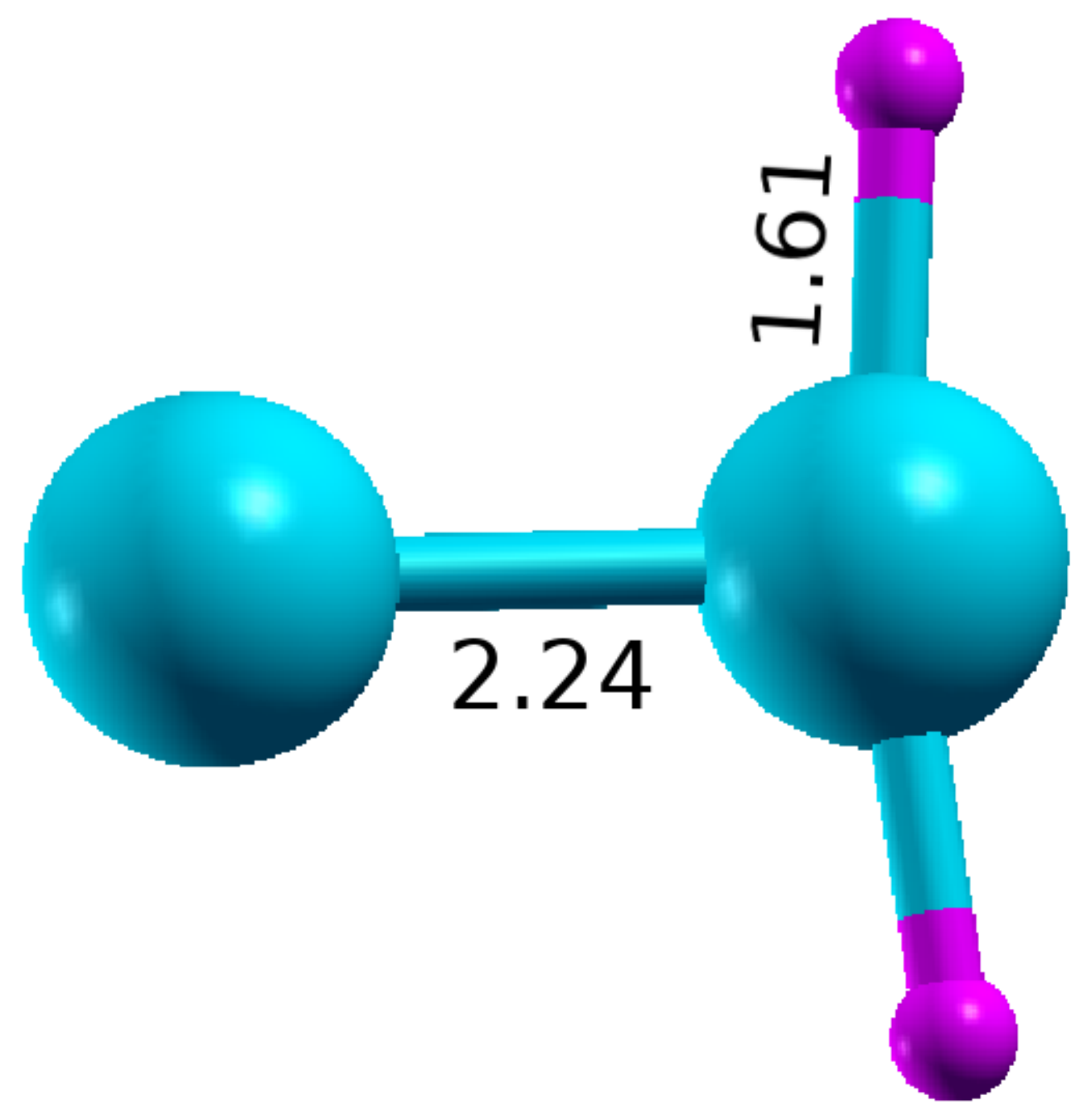}

}
\par\end{centering}
\caption{Optimized excited state geometries corresponding to various peaks
of the absorption spectrum of disilavinylidene (Si-SiH\protect\textsubscript{2})
isomer {[}main article, Fig. 4(c){]}. All the bond lengths are in
$\text{Å}$ unit.}
\end{figure}

\subsection{Trans-bent (HSi-SiH) isomer}

\begin{figure}[H]
\begin{centering}
\subfloat[Peak-I]{\includegraphics[scale=0.35]{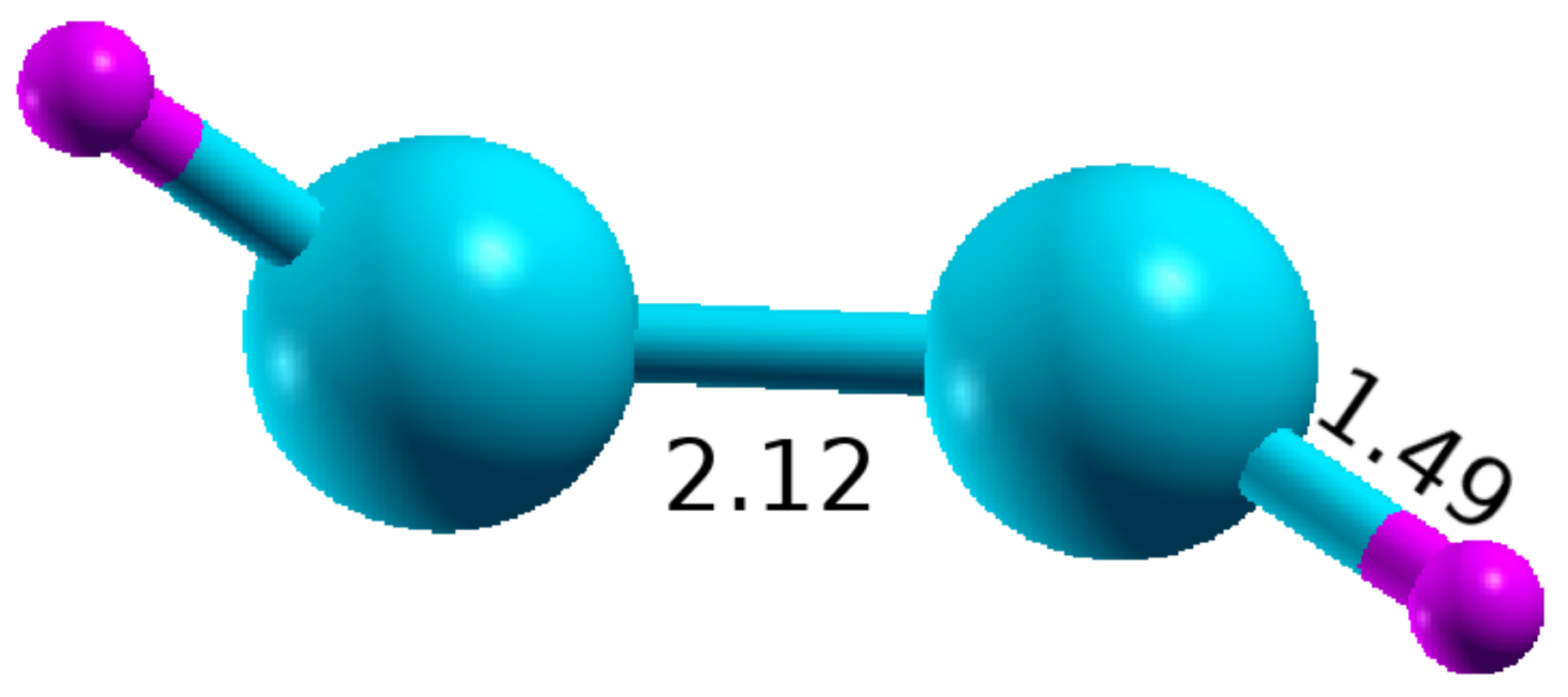}

}\subfloat[Peak-II]{\includegraphics[scale=0.35]{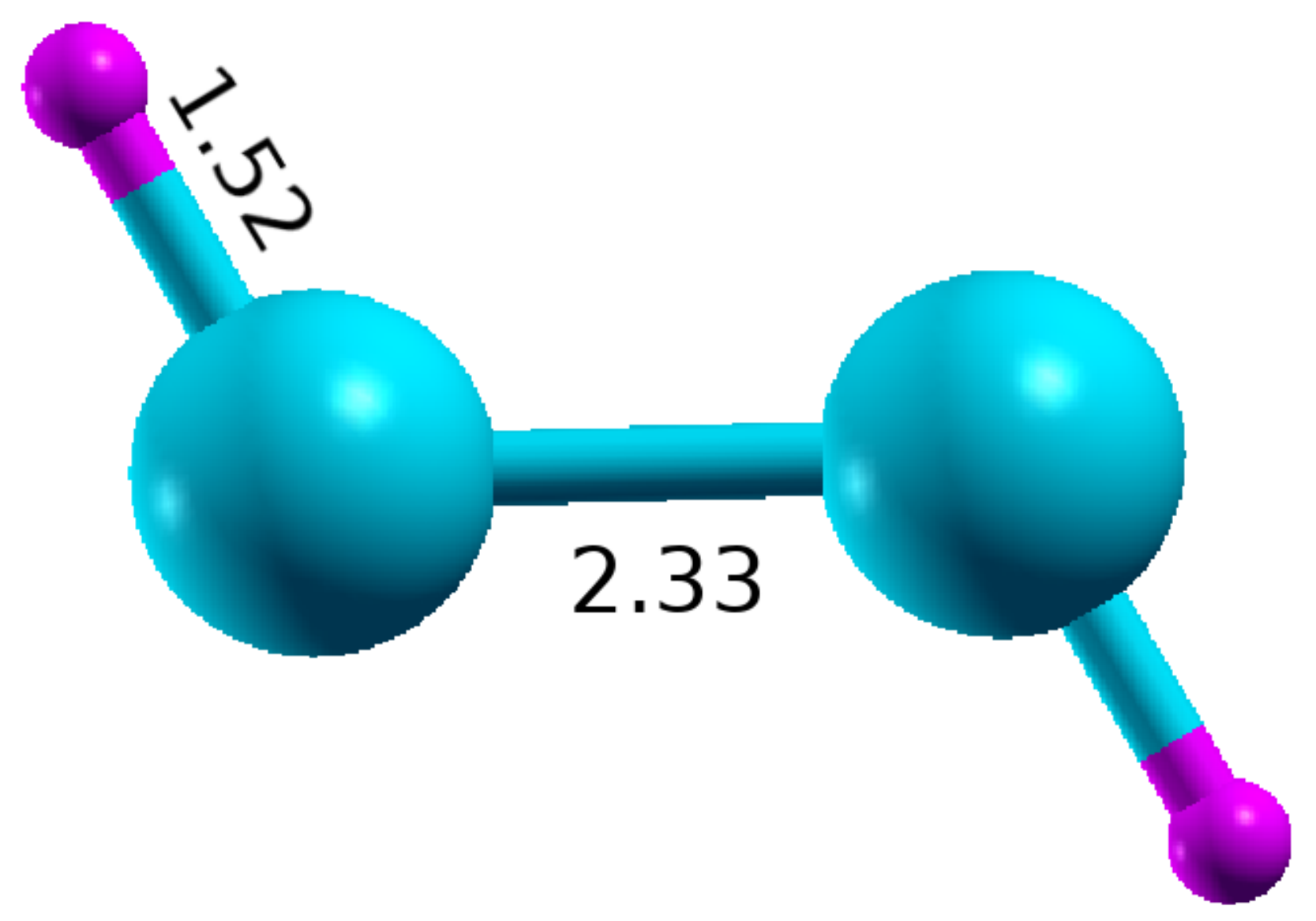}

}
\par\end{centering}
\caption{Optimized geometries of excited states corresponding to various peaks
of the absorption spectrum of trans-bent (HSi-SiH) isomer {[}main
article, Fig. 4(d){]}. All the bond lengths are in $\text{Å}$ unit.}
\end{figure}

\subsection{Disilene (H\protect\textsubscript{\textbf{2}}Si-SiH\protect\textsubscript{\textbf{2}})
isomer}

\begin{figure}[H]
\begin{centering}
\subfloat[Peak-I]{\includegraphics[scale=0.3]{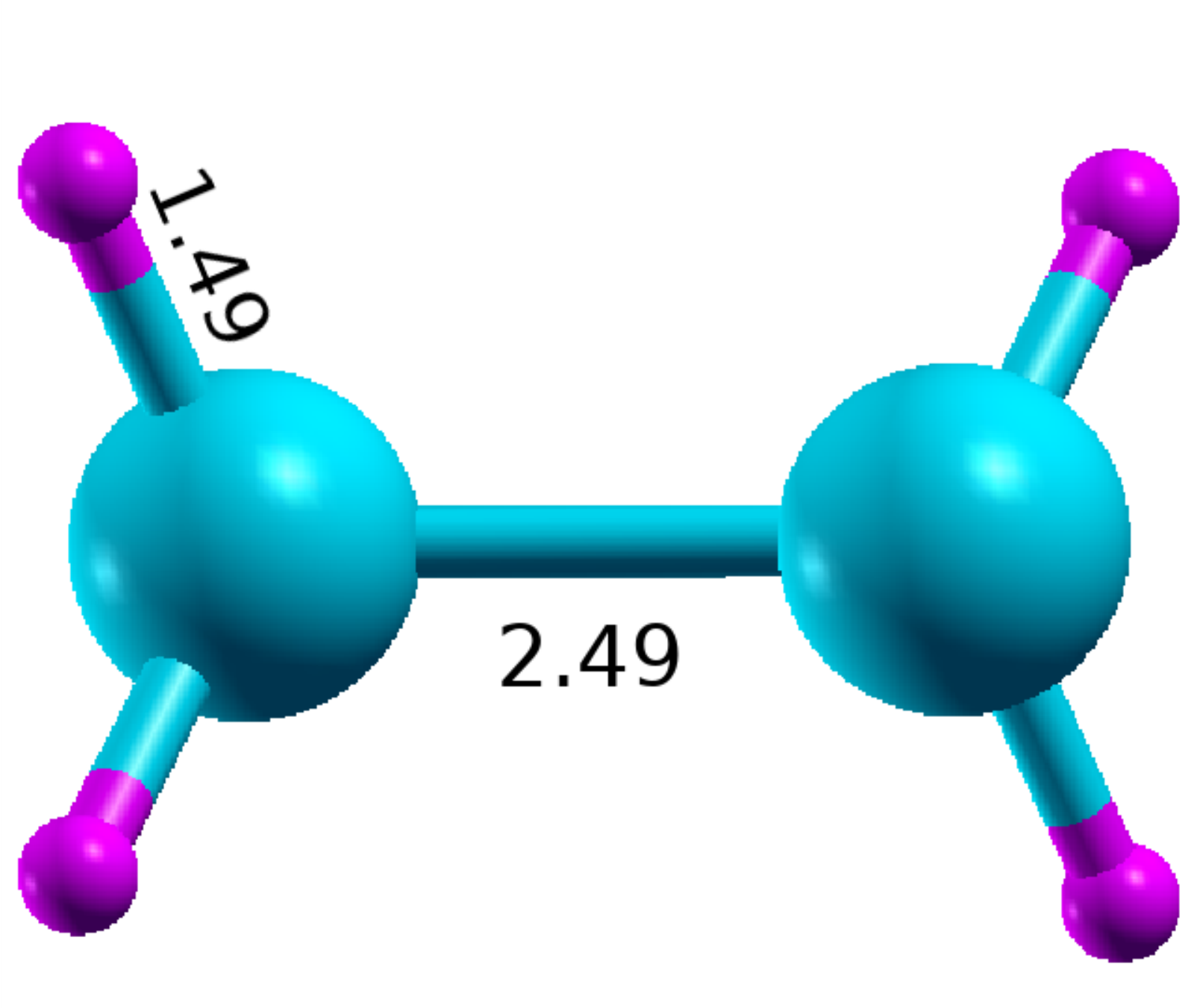}

}\subfloat[Peak-III]{\includegraphics[scale=0.3]{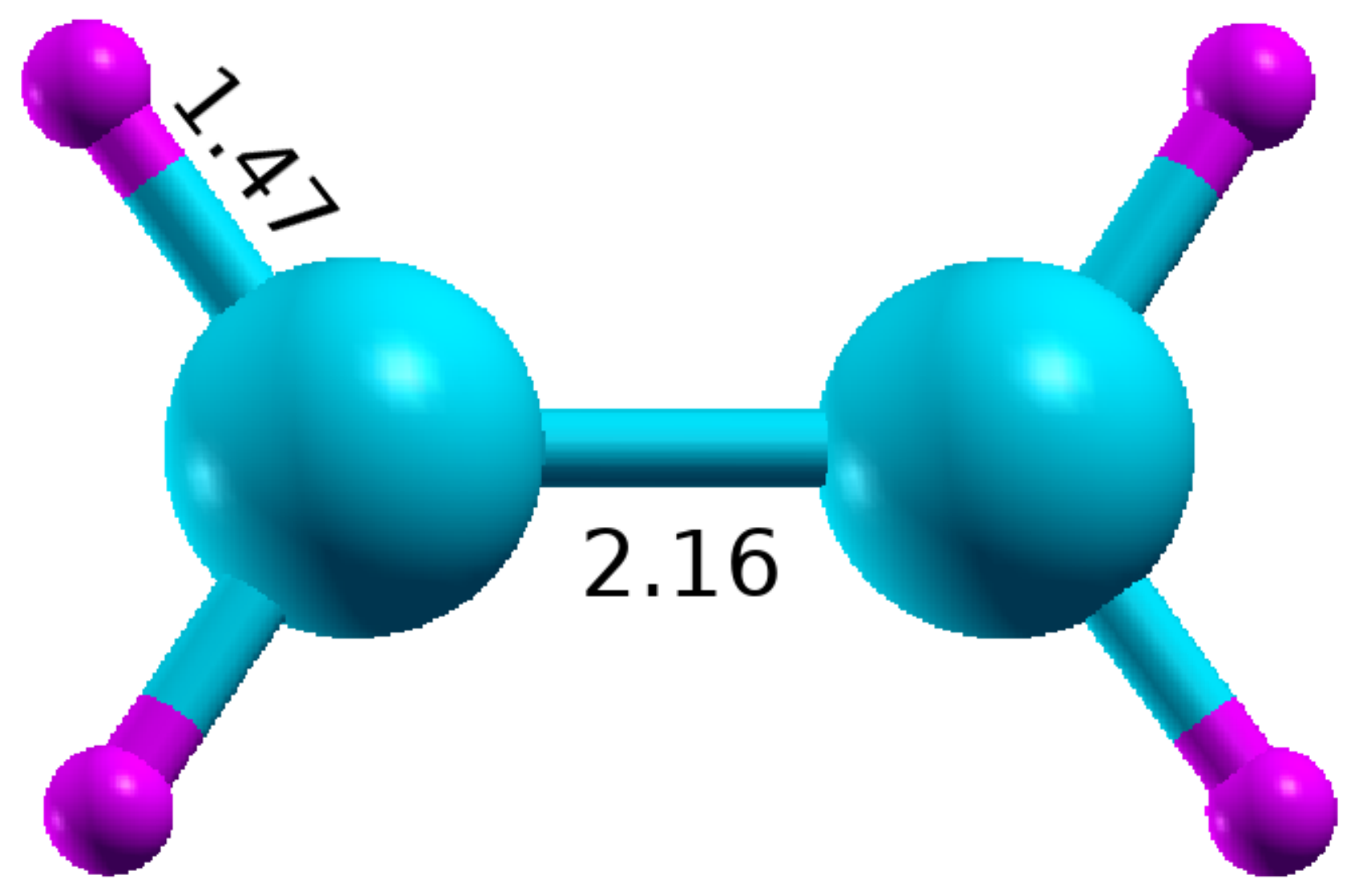}

}
\par\end{centering}
\caption{Optimized excited state geometries corresponding to various peaks
of the absorption spectrum of disilene (H\protect\textsubscript{2}Si-SiH\protect\textsubscript{2})
isomer {[}main article, Fig. 9(a){]}. All the bond lengths are in
$\text{Å}$ unit.}
\end{figure}

\subsection{Monobridged (H\protect\textsubscript{\textbf{2}}Si-H-SiH) isomer}

\begin{figure}[H]
\begin{centering}
\subfloat[Peak-I]{\includegraphics[scale=0.3]{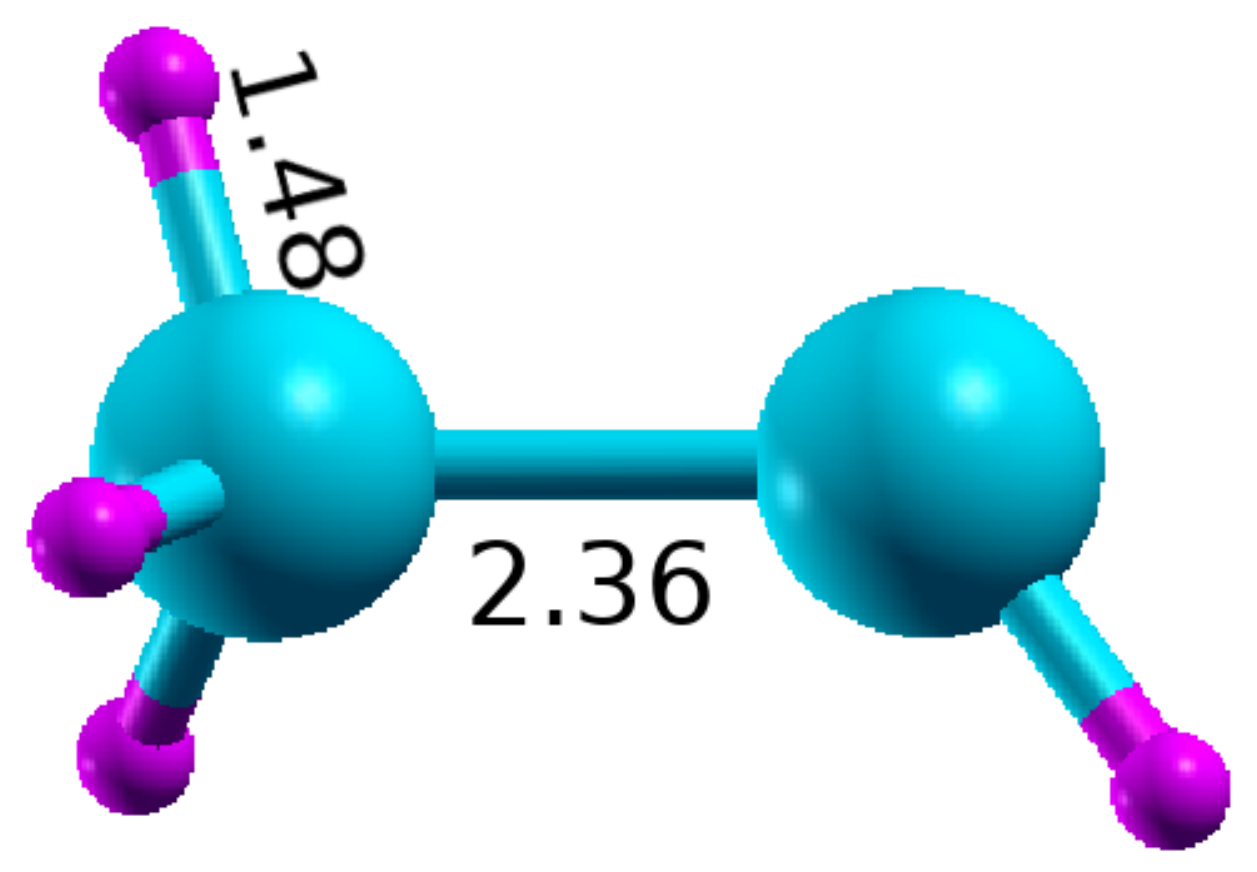}

}\subfloat[Peak-II]{\includegraphics[scale=0.3]{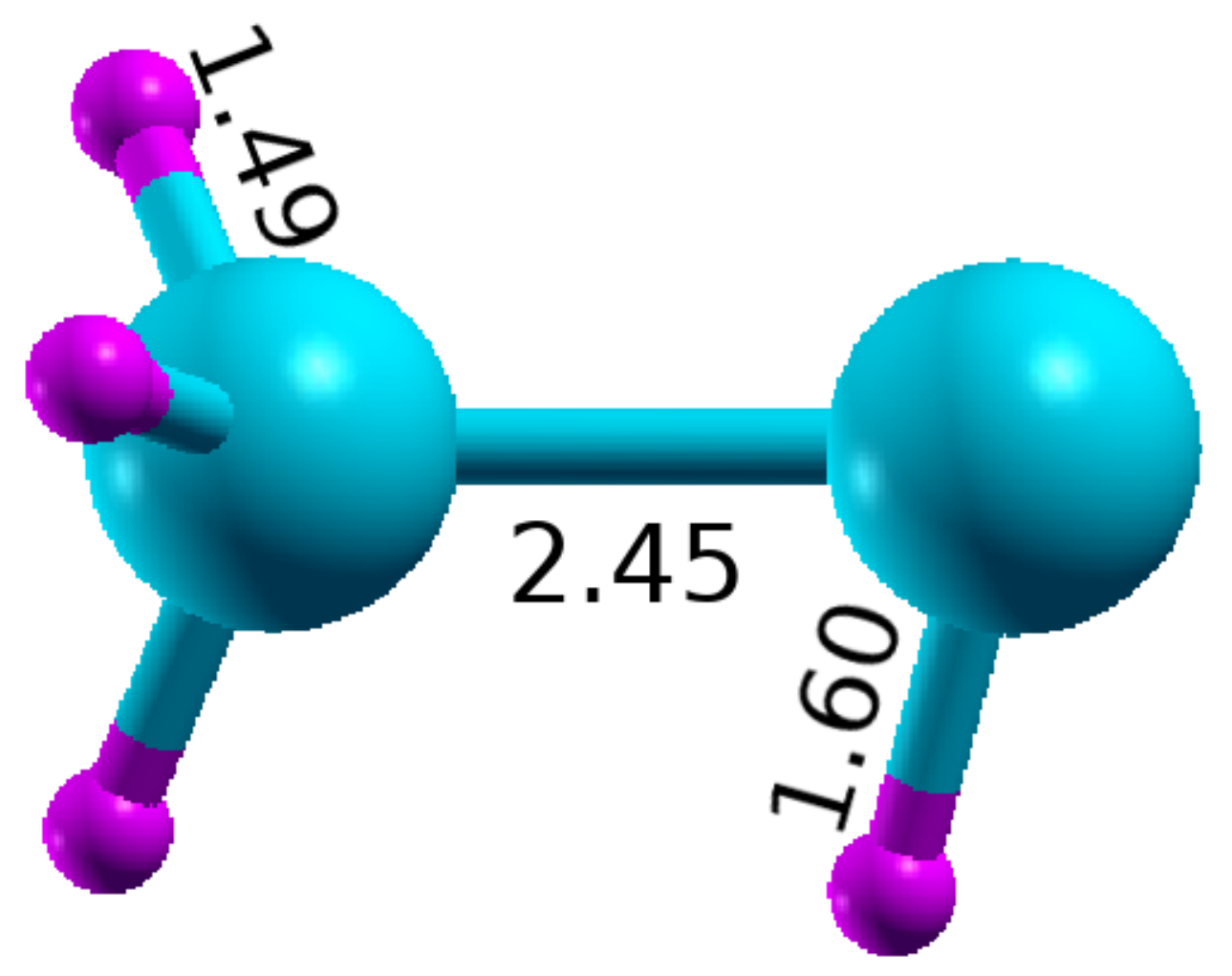}

}\subfloat[Peak-III]{\includegraphics[scale=0.3]{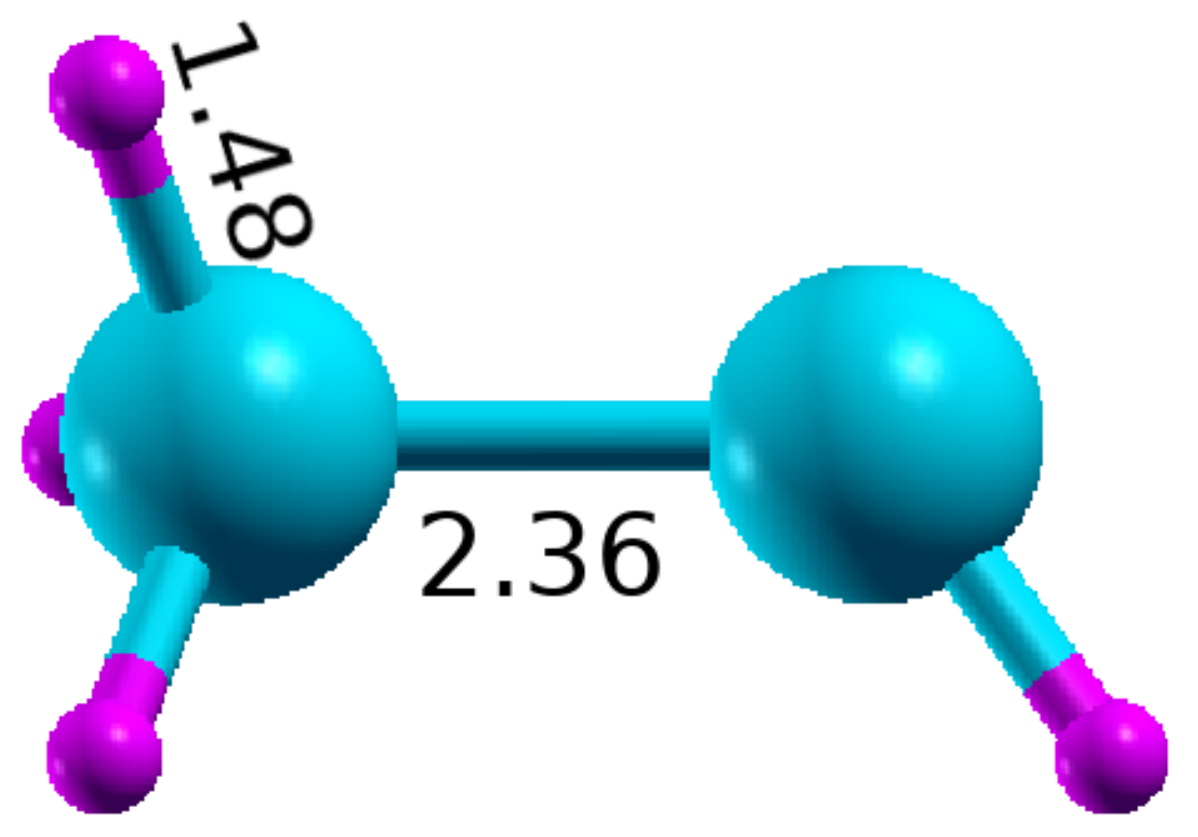}

}
\par\end{centering}
\caption{Optimized excited state geometries corresponding to various peaks
of the absorption spectrum of monobridged (H\protect\textsubscript{2}Si-H-SiH)
isomer {[}main article, Fig. 9(b){]}. All the bond lengths are in
$\text{Å}$ unit.}
\end{figure}

\subsection{Silylsilylene (H\protect\textsubscript{3}Si-SiH) isomer}

\begin{figure}[H]
\begin{centering}
\subfloat[Peak-I]{\includegraphics[scale=0.3]{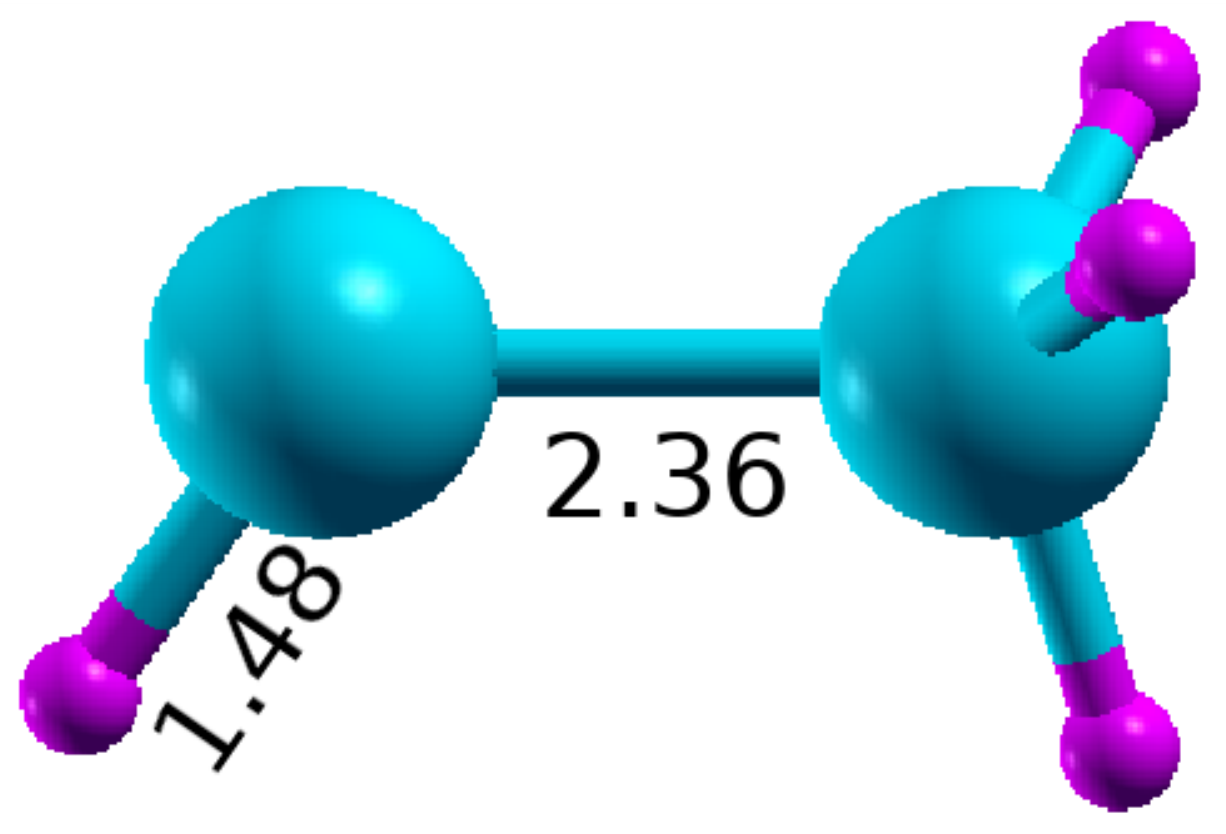}

}\subfloat[Peak-V]{\includegraphics[scale=0.3]{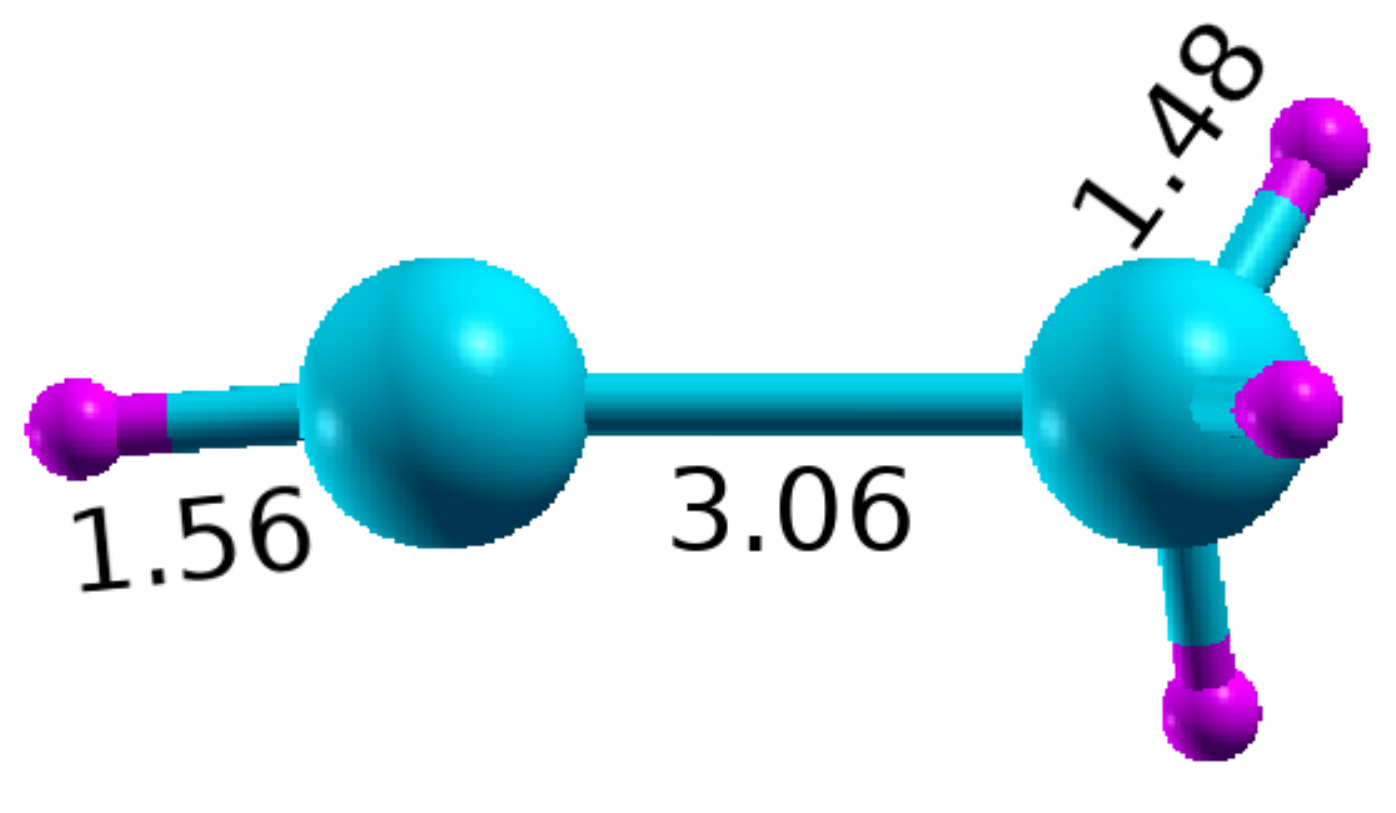}

}
\par\end{centering}
\caption{Optimized excited state geometries corresponding to various peaks
of the absorption spectrum of silylsilylene (H\protect\textsubscript{3}Si-SiH)
isomer {[}main article, Fig. 9(c){]}. All the bond lengths are in
$\text{Å}$ unit.}
\end{figure}

\subsection{Disilane (H\protect\textsubscript{3}Si-SiH\protect\textsubscript{3})
isomer}

\begin{figure}[H]
\begin{centering}
\subfloat[Peak-I]{\includegraphics[scale=0.3]{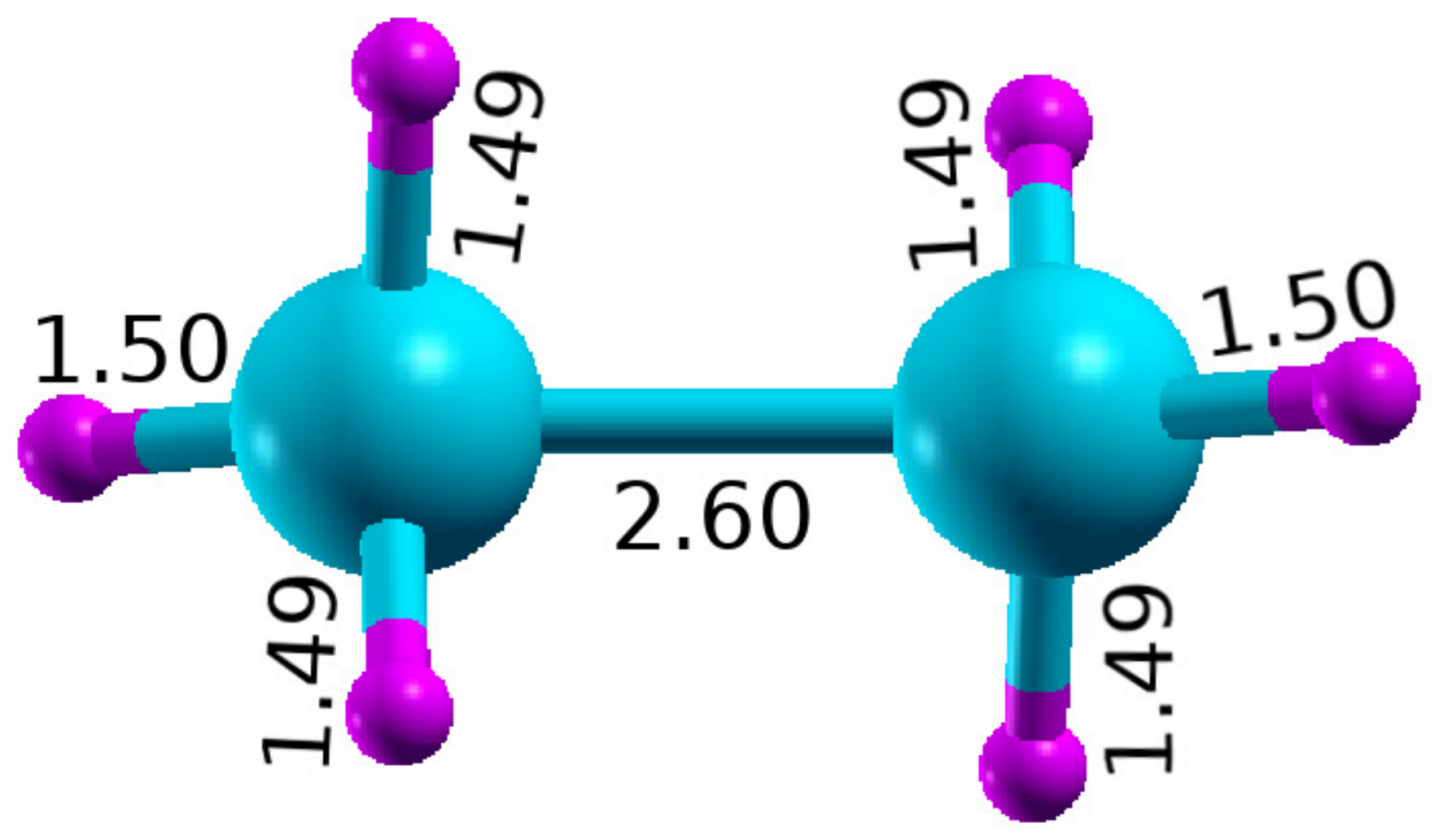}

}\subfloat[Peak-II]{\includegraphics[scale=0.3]{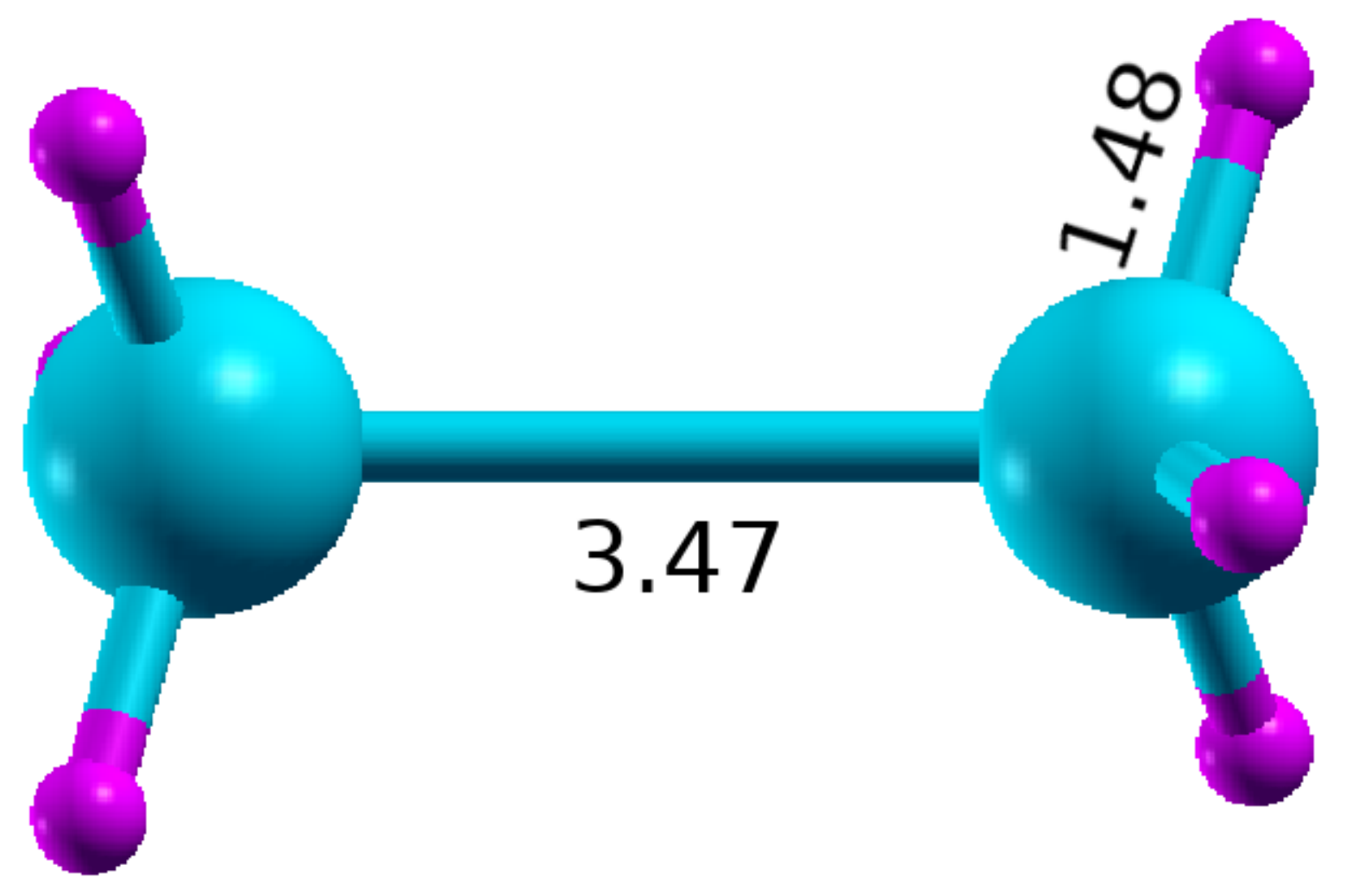}

}
\par\end{centering}
\caption{Optimized excited state geometries corresponding to various peaks
of the absorption spectrum of disilane (H\protect\textsubscript{3}Si-SiH\protect\textsubscript{3})
isomer {[}main article, Fig. 2{]}. All the bond lengths are in $\text{Å}$
unit.}
\end{figure}

\section{Atomic Coordinates Corresponding to the Optimized Ground State Geometries
of Various Clusters}

\begin{table}[H]
\caption{Atomic coordinates (in \AA\  units) corresponding to the optimized
ground state geometry of dibridged disilyne (Si-H$_{2}$-Si)\label{tab:geo-Dibridged-disilyne-si2h2}}

\centering{}%
\begin{tabular}{ccccccc}
\hline 
 & \hspace{0.5cm} & x & \hspace{0.25cm} & y & \hspace{0.25cm} & z\tabularnewline
\hline 
Si &  & 0.000000000000 &  & 1.100740944342 &  & -0.027074095829\tabularnewline
Si &  & 0.000000000000 &  & -1.100740944342 &  & -0.027074095829\tabularnewline
H &  & -0.985364830979 &  & 0.000000000000 &  & 0.751568938885\tabularnewline
H &  & 0.985364830979 &  & 0.000000000000 &  & 0.751568938885\tabularnewline
\hline 
\end{tabular}
\end{table}

\begin{table}[H]
\caption{Atomic coordinates (in \AA\  units) corresponding to the optimized
ground state geometry of monobridged structure (Si-H-SiH)\label{tab:geo-Monobridged-si2h2}}

\centering{}%
\begin{tabular}{ccccccc}
\hline 
 & \hspace{0.5cm} & x & \hspace{0.25cm} & y & \hspace{0.25cm} & z\tabularnewline
\hline 
Si &  & -0.836958729655 &  & 0.017059638406 &  & -0.709155735004\tabularnewline
Si &  & 0.806304271215 &  & -0.014312279108 &  & 0.606308390319\tabularnewline
H &  & -0.772104895994 &  & -0.033803861038 &  & 1.012668921759\tabularnewline
H &  & 1.623063648110 &  & -0.042462025198 &  & 1.842343123724\tabularnewline
\hline 
\end{tabular}
\end{table}

\begin{table}[H]
\caption{Atomic coordinates (in \AA\  units) corresponding to the optimized
ground state geometry of disilavinylidene isomer (Si-SiH$_{2}$)\label{tab:geo-Disilavinylidene-si2h2}}

\centering{}%
\begin{tabular}{ccccccc}
\hline 
 & \hspace{0.5cm} & x & \hspace{0.25cm} & y & \hspace{0.25cm} & z\tabularnewline
\hline 
Si &  & 0.000038658677 &  & 1.165679188613 &  & 0.000000000000\tabularnewline
Si &  & -0.000024520199 &  & -1.031699608620 &  & 0.000000000000\tabularnewline
H &  & 1.227406456834 &  & -1.859865255861 &  & 0.000000000000\tabularnewline
H &  & -1.227798936833 &  & -1.859368487152 &  & 0.000000000000\tabularnewline
\hline 
\end{tabular}
\end{table}

\begin{table}[H]
\caption{Atomic coordinates (in \AA\  units) corresponding to the optimized
ground state geometry of trans-bent isomer (HSi-SiH)\label{tab:geo-Trans-bent-si2h2}}

\centering{}%
\begin{tabular}{ccccccc}
\hline 
 & \hspace{0.5cm} & x & \hspace{0.25cm} & y & \hspace{0.25cm} & z\tabularnewline
\hline 
Si &  & 1.049560638794 &  & -0.012108929052 &  & 0.000000000000\tabularnewline
Si &  & -1.049560638794 &  & 0.012108929052 &  & 0.000000000000\tabularnewline
H &  & 1.901548085640 &  & -1.229661195450 &  & 0.000000000000\tabularnewline
H &  & -1.901548085640 &  & 1.229661195450 &  & 0.000000000000\tabularnewline
\hline 
\end{tabular}
\end{table}

\begin{table}[H]
\caption{Atomic coordinates (in \AA\  units) corresponding to the optimized
ground state geometry of disilene isomer (H$_{2}$Si-SiH$_{2}$)\label{tab:geo-Disilene-si2h4}}

\centering{}%
\begin{tabular}{ccccc}
\hline 
 &  &  &  & \tabularnewline
\hline 
\hline 
 & \hspace{0.5cm} & x & y & z\tabularnewline
\hline 
Si &  & -0.014527084989 & 1.078610931417 & 0.000000000000\tabularnewline
Si &  & 0.014527084989 & -1.078610931417 & 0.000000000000\tabularnewline
H &  & -0.388083856897 & 1.807751814186 & 1.227693520012\tabularnewline
H &  & -0.388083856897 & 1.807751814186 & -1.227693520012\tabularnewline
H &  & 0.388083856897 & -1.807751814186 & -1.227693520012\tabularnewline
H &  & 0.388083856897 & -1.807751814186 & 1.227693520012\tabularnewline
\hline 
\end{tabular}
\end{table}

\begin{table}[H]
\caption{Atomic coordinates (in \AA\  units) corresponding to the optimized
ground state geometry of mono-bridged isomer (H$_{2}$Si-H-SiH)\label{tab:geo-Mono-bridged-si2h4}}

\centering{}%
\begin{tabular}{ccccccc}
\hline 
 & \hspace{0.5cm} & x & \hspace{0.25cm} & y & \hspace{0.25cm} & z\tabularnewline
\hline 
Si &  & 1.172682525006 &  & 0.047991888154 &  & -0.052395704519\tabularnewline
Si &  & -1.070659956950 &  & 0.004587875388 &  & -0.004943355672\tabularnewline
H &  & 1.079592180677 &  & -1.466147217376 &  & -0.026756050259\tabularnewline
H &  & -0.073757438423 &  & -0.034913083199 &  & 1.266626592692\tabularnewline
H &  & -1.881741837422 &  & 1.218385602498 &  & 0.242557859563\tabularnewline
H &  & -1.956209393611 &  & -1.176924076404 &  & 0.109287047982\tabularnewline
\hline 
\end{tabular}
\end{table}

\begin{table}[H]
\caption{Atomic coordinates (in \AA\  units) corresponding to the optimized
ground state geometry of silylsilylene isomer (H$_{3}$Si-SiH)\label{tab:geo-Silylsilylene-si2h4}}

\centering{}%
\begin{tabular}{ccccccc}
\hline 
 & \hspace{0.5cm} & x & \hspace{0.25cm} & y & \hspace{0.25cm} & z\tabularnewline
\hline 
Si &  & 1.126462143734 &  & 0.027035432078 &  & 0.000006113340\tabularnewline
Si &  & -1.258201977707 &  & -0.087127476208 &  & -0.000009612313\tabularnewline
H &  & -1.308985044826 &  & 1.431034864085 &  & 0.000041211523\tabularnewline
H &  & 1.587300210721 &  & 0.767155781338 &  & 1.203027047658\tabularnewline
H &  & 1.791407763335 &  & -1.297349115623 &  & -0.000057042453\tabularnewline
H &  & 1.587336122859 &  & 0.767295944767 &  & -1.202914086260\tabularnewline
\hline 
\end{tabular}
\end{table}

\begin{table}[H]
\caption{Atomic coordinates (in \AA\  units) corresponding to the optimized
ground state geometry of disilane (H$_{3}$Si-SiH$_{3}$)\label{tab:geo-Disilane-si2h6}}

\centering{}%
\begin{tabular}{ccccccc}
\hline 
 & \hspace{0.5cm} & x & \hspace{0.25cm} & y & \hspace{0.25cm} & z\tabularnewline
\hline 
Si &  & 0.000000000915 &  & -1.169370238817 &  & 0.000000000000\tabularnewline
Si &  & -0.000000000915 &  & 1.169370238817 &  & 0.000000000000\tabularnewline
H &  & -1.388422645000 &  & -1.684057891419 &  & 0.000000000000\tabularnewline
H &  & 0.694211326479 &  & -1.684057901803 &  & -1.202409278566\tabularnewline
H &  & 0.694211326479 &  & -1.684057901803 &  & 1.202409278566\tabularnewline
H &  & 1.388422645000 &  & 1.684057891419 &  & 0.000000000000\tabularnewline
H &  & -0.694211326479 &  & 1.684057901803 &  & -1.202409278566\tabularnewline
H &  & -0.694211326479 &  & 1.684057901803 &  & 1.202409278566\tabularnewline
\hline 
\end{tabular}
\end{table}